\documentclass[review]{elsarticle}

\usepackage{lineno,hyperref}
\modulolinenumbers[5]

\usepackage{subcaption}
\usepackage[dvipsnames]{xcolor}
\usepackage{amsmath}
\usepackage{amssymb} % to fix \mathbb issues
\usepackage{bm} % to fix \bm issues
\usepackage{bbold}
\usepackage{float} % to fix figure placement issues
\usepackage[british]{babel}

\journal{Elsevier. Updated to published version.}

\begin{document}

\begin{frontmatter}

\title{Advances in Multi-Variate Analysis Methods for New Physics Searches at the Large Hadron Collider}

\author[y,a]{       Anna Stakia } 
\author[b]{         Tommaso Dorigo\corref{mycorrespondingauthor} }
\cortext[mycorrespondingauthor]{Corresponding author}
\ead{dorigo@pd.infn.it}

\author[ea]{        Giovanni Banelli}           
\author[ox]{        Daniela Bortoletto}         
\author[gm,ac]{     Alessandro Casa}            
\author[b,f]{       Pablo {de Castro}}          
\author[d]{         Christophe Delaere}         
\author[bb]{        Julien Donini}              
\author[lf]{        Livio Finos}                
\author[c]{         Michele Gallinaro}          
\author[d]{         Andrea Giammanco}           
\author[y,z,e]{     Alexander Held}             
\author[bb,g]{      Fabricio Jiménez Morales}   
\author[gm]{        Grzegorz Kotkowski}         
\author[ea]{        Seng Pei Liew}              
\author[d]{         Fabio Maltoni}              
\author[gm]{        Giovanna Menardi}           
\author[h]{         Ioanna Papavergou}          
\author[d,de]{      Alessia Saggio}             
\author[gm,bs]{     Bruno Scarpa}               
\author[c,b,f]{     Giles C. Strong}            
\author[ox,uc]{     Cecilia Tosciri}            
\author[c]{         Jo\~{a}o Varela}            
\author[c,uo,d]{    Pietro Vischia}             
\author[ea]{        Andreas Weiler}

\address[y]{European Organization for Nuclear Research (CERN), Geneva, Switzerland~~~~~~~~~~~~~~~~~~~~~~}
\address[a]{Institute of Nuclear and Particle Physics, National Centre for Scientific Research ~~~~~~~~~~~~~ \\
`Demokritos', Athens, Greece~~~~~~~~~~~~~~~~~~~~~~~~~~~~~~~~~~~~~~~~~~~~~~~~~~~~~~~~~~~~~~~~~~~~~~~~~~~~}
\address[b]{Istituto Nazionale di Fisica Nucleare, Sezione di Padova, Italy~~~~~~~~~~~~~~~~~~~~~~~~~~~~~~~~~~~~~~~~}
\address[ea]{Physics Department, Technical University of Munich, Garching, Germany~~~~~~~~~~~~~~~~~~~~~~~~}
\address[ox]{Department of Physics, University of Oxford, UK~~~~~~~~~~~~~~~~~~~~~~~~~~~~~~~~~~~~~~~~~~~~~~~~~~~~~~~~~}
\address[gm]{Department of Statistical Sciences, University of Padova, Italy~~~~~~~~~~~~~~~~~~~~~~~~~~~~~~~~~~~~~~~~}
\address[ac]{School of Mathematics and Statistics, University College Dublin, Ireland~~~~~~~~~~~~~~~~~~~~~~~~~~}
\address[f]{Department of Physics, University of Padova, Italy~~~~~~~~~~~~~~~~~~~~~~~~~~~~~~~~~~~~~~~~~~~~~~~~~~~~~~~}
\address[d]{Centre for Cosmology, Particle Physics and Phenomenology (CP3), Universit\'e catholique ~~~\\
 de Louvain, Louvain-la-Neuve, Belgium~~~~~~~~~~~~~~~~~~~~~~~~~~~~~~~~~~~~~~~~~~~~~~~~~~~~~~~~~~~~~~~}
\address[bb]{Laboratoire de Physique de Clermont, Universit\'e Clermont Auvergne, IN2P3/CNRS, ~~~~~~~~~\\ 
Clermont-Ferrand, France~~~~~~~~~~~~~~~~~~~~~~~~~~~~~~~~~~~~~~~~~~~~~~~~~~~~~~~~~~~~~~~~~~~~~~~~~~~~~~~}
\address[lf]{Department of Developmental Psychology and Socialization, University of Padova, Italy~~~~~~}
\address[c]{Laborat\'orio de Instrumenta\c c\~ao e F\'isica Experimental de Part\'iculas, LIP Lisbon, Portugal~~~}
\address[z]{Department of Physics, University of British Columbia, Canada~~~~~~~~~~~~~~~~~~~~~~~~~~~~~~~~~~~~~}
\address[e]{Department of Physics, New York University, USA~~~~~~~~~~~~~~~~~~~~~~~~~~~~~~~~~~~~~~~~~~~~~~~~~~~~~~~}
\address[g]{Laboratoire Leprince-Ringuet, CNRS, \'Ecole polytechnique, Institut Polytechnique de Paris, \\ 
Palaiseau, France~~~~~~~~~~~~~~~~~~~~~~~~~~~~~~~~~~~~~~~~~~~~~~~~~~~~~~~~~~~~~~~~~~~~~~~~~~~~~~~~~~~~~~~~~}
\address[h]{Department of Physics, National and Kapodistrian University of Athens, Greece~~~~~~~~~~~~~~~~}
\address[de]{Deutsches Elektronen-Synchrotron (DESY), Hamburg, Germany~~~~~~~~~~~~~~~~~~~~~~~~~~~~~~~~~~~~~}
\address[bs]{Department of Mathematics, University of Padova, Italy~~~~~~~~~~~~~~~~~~~~~~~~~~~~~~~~~~~~~~~~~~~~~~~~}
\address[uc]{Department of Physics, University of Chicago, USA~~~~~~~~~~~~~~~~~~~~~~~~~~~~~~~~~~~~~~~~~~~~~~~~~~~~~~}
\address[uo]{Department of Physics, Universidad de Oviedo, Spain~~~~~~~~~~~~~~~~~~~~~~~~~~~~~~~~~~~~~~~~~~~~~~~~~~~~}

\begin{abstract}
Between the years 2015 and 2019, members of the Horizon 2020-funded Innovative Training Network named ``AMVA4NewPhysics"
studied the customization and application of advanced multivariate analysis methods and statistical learning tools to high-energy physics problems, as well as developed entirely new ones. Many of those methods were successfully used to improve the sensitivity of data analyses performed by the ATLAS and CMS experiments at the CERN Large Hadron Collider; 
several others, still in the testing phase, promise to further improve the precision of measurements of fundamental physics parameters and the reach of searches for new phenomena. In this paper, the most relevant new tools, among those studied and developed, are presented along with the evaluation of their performances.
\end{abstract}


\end{frontmatter}

\tableofcontents
%\linenumbers

\clearpage

\section{Introduction}

\subsection{Background}

\medskip
%\smallskip
\noindent
Over forty quadrillion proton--proton collisions were produced by the CERN Large Hadron Collider (LHC)~\cite{Evans2008} at the centre of the ATLAS~\cite{Aad2008} and CMS~\cite{Chatrchyan2008} detectors since the start of LHC operations in 2009. The data samples produced by the reconstruction of the resulting detector readouts allowed those two experiments to vastly expand our knowledge of matter and interactions at the shortest distance scales. Besides delivering a much awaited discovery of the Higgs boson in 2012~\cite{Aad2012,Chatrchyan2012}, the two giant multi-purpose experiments published hundreds of precision measurements of fundamental physics constants and searches for new phenomena which previous experiments could not be sensitive to~\cite{atlasphysics,cmsphysics}.

\medskip
\smallskip
\noindent
The intrinsic complexity of the collected data and the intent to fully exploit the information they yielded on subnuclear phenomena significantly increased the need of experimentalists to optimize their information extraction procedures by employing the most performant multivariate analysis methods. A concurrent rise in the development of modern machine learning (ML) techniques and the increasing degree of their application to scientific research enabled the LHC experiments to achieve that goal, by squeezing more information from their datasets and improving the quality of their scientific output.

\medskip
\smallskip
\noindent
In the above context are set the activities of AMVA4NewPhysics, an Innovative Training Network funded through the Marie-Skłodowska Curie Actions of the European Union Horizon 2020 program. The network, which operated from September 2015 to August 2019, saw the participation of about fifty researchers and students from nine beneficiary nodes among European research institutes and universities,\footnote{The involved beneficiary nodes were the Italian Institute for Nuclear Physics and the University of Padova (Italy), the University of Oxford (England), the Université catholique de Louvain (Belgium), the Université Clermont Auvergne (France), the Laborat\'orio de Instrumenta\c{c}\~{a}o e F\'isica Experimental de Part\'iculas, Lisbon (Portugal), the CERN laboratories, the Technische Universitat Munchen (Germany), and the Institute for Accelerating Systems and Applications (Greece).} in addition to nine academic and non-academic partners in Europe and the United States.\footnote{ The network included as academic parthers the Universidad de Oviedo (Spain), the University of California Irvine (USA), the \'Ecole Polytechnique Fédérale de Lausanne (Switzerland), the University of British Columbia (Canada), the National and Kapodistrian University of Athens (Greece); and as non-academic partners the Mathworks Company (Massachusetts, USA), SDG group Milano (Italy), B12 (Belgium), and YANDEX (Russia). } The network, while keeping as its primary goal excellence in training of a cohort of Ph.D. students, conducted cutting-edge research and fostered the use of advanced multivariate analysis methods in physics data analysis for the ATLAS and CMS experiments at the CERN LHC~\cite{Stakia2021}.

\medskip
\smallskip
\noindent
The four main pillars, upon which most of the studies performed within\\ AMVA4NewPhysics were based, comprise:

\begin{enumerate} 
\item The customization and optimization of advanced Statistical Learning tools for the precise measurement of Higgs boson properties;

\item The development of new Statistical Learning algorithms to increase the sensitivity of physics analyses targeting model-specific and aspecific searches for new physics;
\item The improvement of the Matrix Element Method through the addition of new tools that extend its applications;
\item The development of new Statistical Learning algorithms for use in high-energy physics (HEP) analyses, ranging from data modelling methods to anomaly detection methods in model-independent searches.
\end{enumerate}

\medskip
\smallskip
\noindent
In this paper we summarize some of the research outcomes that resulted from work performed by AMVA4NewPhysics members in the four pillars defined above, highlighting the importance of the results for future studies at the LHC and beyond.

\subsection{Plan of this document}

\medskip
%\smallskip
\noindent
The structure of this document follows loosely the order of the four pillars defined above; however, new tools belonging to the fourth one are in some cases described earlier, where they find their most relevant research application. We start in Section~\ref{sec:higgsml} where we describe a detailed study of the performance of deep neural networks applied to the complex task of distinguishing a signal of Higgs boson decays to tau lepton pairs from competing backgrounds; the study focuses on the most performant strategies by leveraging information from a competitive effort (the Kaggle ‘HiggsMLChallenge’). This is followed in Section~\ref{sec:mvaHH} by a description of multivariate methods applied to the extraction of the Higgs pair-production signal: an innovative technique for the precise data-driven modelling of multi-jet backgrounds in the search of the $HH \to b \bar{b} b \bar{b}$ process performed by CMS on Run 2 data, and neural-network studies for the extraction of the $HH \to b \bar{b} \tau \tau$ signal in future high-luminosity LHC running conditions.

\medskip
\smallskip
\noindent
In Section~\ref{sec:btagging} we describe the development of a high-performance method for identifying the flavour of the parton originating a hadronic jet in CMS data; the resulting algorithms are now among the crucial ingredients for a wide class of analysis tasks, ranging from high-sensitivity measurements of Higgs boson properties, to top quark precision measurements, and to wide-reach searches for new physics signatures in collider data. Section~\ref{sec:mem} is then devoted to describing improvements achieved on the Matrix-Element Method, which is a complex multi-dimensional calculation that approximates the likelihood function to extract SM parameters from the observed data, and resulting applications to Higgs boson searches in ATLAS and CMS data.

\medskip
\smallskip
\noindent
Section~\ref{sec:anomalyDetection} focuses on the new methods we designed to search for new physics in LHC data in model-aspecific ways through the identification of anomalous regions of the feature space of the observed datasets; the section also includes description of a technique developed to improve inference on the presence of new physics signals in invariant mass distributions, and its expected performance in searches for high-mass resonances decaying to jet pairs. 
Section~\ref{s:caloshowers} describes an innovative study of particle showers in the ATLAS forward electromagnetic calorimeter, aimed at the production of a fast simulation of the complex physics processes detected by that instrument. In Section~\ref{sec:mode} we offer an outlook of future studies targeting the end-to-end optimization of data analyses aimed at the loss-less extraction of information from multi-dimensional datasets such as those common in HEP problems, and describe an algorithm we developed for that task. We finally offer some concluding remarks and summary of our review of AMVA4NewPhysics contributions to LHC data analysis in Section~\ref{sec:summary}.

\section{Supervised Classification Methods for the Search of Higgs Boson Decays to Tau Lepton Pairs}\label{sec:higgsml}

\subsection{Background}

\medskip
%\smallskip
\noindent
The groundbreaking discovery of the Higgs boson in 2012 by the ATLAS and CERN experiments at the LHC~\cite{Aad2012,Chatrchyan2012} led to the award of the 2013 Nobel Prize in Physics to P.W.~Higgs and F.~Englert; the Scottish and Dutch awardees must indeed be commended for their visionary theoretical predictions, which had to wait for almost five decades to be experimentally confirmed. While concluding a long quest for the origin of electroweak symmetry breaking, that scientific milestone initiated a new era of large-scale precision measurement studies and new physics searches related to the Higgs boson and its properties. Curiously, the year 2012 also marks an important milestone for machine learning, as in that year deep neural networks reached paradigm-changing performance in the benchmark problem of image classification~\cite{imagenet}. It is thus not a surprise to observe that, from that year onwards, HEP data analysis withstood a boom of applications of ML-based techniques, aimed at optimizing the experimental output of their measurements and searches.

\medskip
\smallskip
\noindent
A particularly significant activity in that context was represented by the ‘HiggsMLChallenge’ competition on Kaggle~\cite{HiggsMLChallenge}. That competition, held in 2014, brought together thousands of participants, both belonging to the HEP collaborations most interested in the specific application object of the challenge, as well as academic and non academic participants with background in computer science. Besides that success, the challenge managed to achieve the set goal of understanding which were the most performant machine learning techniques in discriminating the Higgs boson decay to a pair of tau leptons from the various background processes, and at the same time allowed to introduce new promising methods and tools to HEP research and to the broader scientific community. The complexity of the classification task, combined with the high expertise behind the best proposed solutions, made the HiggsMLChallenge competition a benchmark against which to evaluate and compare different ML approaches for supervised classification, as well as to gauge their applicability on HEP datasets. Triggered by the interest of the challenge and the derived conclusions, we conducted a thorough study published in Ref.~\cite{Strong2020}, whose results are summarized {\em infra}. In parallel, the new \textsc{Lumin}~\cite{lumin} software package was developed to provide implementations of the investigated methods, using \textsc{PyTorch}~\cite{pytorch} as the underlying tensor library.

\subsection{Challenge details and datasets}

\medskip
%\smallskip
\noindent
The data used in the competition were constituted by information on all particles produced in proton--proton collisions simulated under the 2012 LHC run conditions (a centre-of-mass energy of 8 TeV and a typical instantaneous luminosity of $10^{34} cm^{-2} s^{-1}$), 
which was fed through a simulation of the ATLAS detector, and from which, after applying state-of-the-art reconstruction algorithms, a set of 30 high-level physics observables were derived per simulated collision. The signal was constituted by direct production of a Higgs boson followed by its decay into a pair of $\tau$ leptons, $pp \to H + X \to \tau^+ \tau^- + X$ (where $X$ denotes any additional produced particles) with the subsequent mixed decay of the $\tau$ lepton pair, $\tau^+ \tau^- \to e^+(\mu^+) \nu_{e(\mu)} \overline{\nu_{\tau}}+ \tau_h^- \nu_{\tau}$ (or to the charge-conjugate final state), where $\tau_h$ denotes the hadronic decay products of one of the tau leptons. These simulated collision events thus contained a semi-hadronically decaying tau lepton and in addition a reconstructed muon or electron, plus at least three unobserved neutrinos.
{The background sample consists of three major processes: $Z\rightarrow\tau^+ + \tau^-$, $t\bar{t}$, and $W+\text{jets}$.
The simulated samples include labels identifying the originating process-class (signal or background) of each event, as well as weights normalizing the various contributing processes to the target integrated luminosity.
} 

\subsubsection{Data preprocessing}

\medskip
%\smallskip
\noindent
The training and testing datasets~\cite{higgsml_data} consist of 250,000 and 550,000 events, respectively, each a mixture of signal and backgrounds. At the time of the competition, process-identifying labels were only supplied for the training dataset, and the testing dataset was split into two parts: the public test set, for which scores (see Section~\ref{s:scoringmetric}, {\em infra}) were supplied to the participant after every submission; and the private test set, for which scores were supplied for the solution selected by the participant once the competition ended, and on which the competition was judged. The analysis discussed {\em infra} attempted to reproduce the challenge conditions, by developing the model based on scores on the public dataset, and then checking the performance of the final model on the private dataset.

\medskip
\smallskip
\noindent
Dataset features consist of low-level information and high-level information, the latter calculated via (non)-linear combinations of the low-level features or from hypothesis-based fitting procedures. Additionally, events carry a weight to normalize the datasets to a fixed integrated luminosity. Since this weight carries information about the event process (signal or background), weights were not publicly available for the testing datasets, and they cannot be used as an input feature in developing the models. {Ref.~\cite{higgsml_data} provides a compact summary of the dataset features.}

\subsubsection{Scoring metric \label{s:scoringmetric}}

\medskip
%\smallskip
\noindent
The performance of a solution classifying testing data as belonging to the signal or background classes is measured using the so-called ‘approximate median significance’ (AMS)~\cite{Cowan2011,ams}:
%\vspace{1mm}
\begin{center}
$AMS=\sqrt{2\left((s+b+b_{reg})\ln\left(1+\frac{s}{b+b_{reg}}\right)-s\right)}$,
\label{eq:ams}
\end{center}
%\vspace{1mm}
where $s$ is the sum of weights of true positive events (signal events determined as signal events by the solution), and $b$ is the sum of weights of false positive events (background events determined as signal events by the solution); $b_{reg}$ is a regularization term set to a constant value of 10 for the competition. The AMS provides an approximation of the expected statistical significance of the number of data events selected by the classification procedure, which could be obtained from the $p$-value of observing at least the selected amount of data (signal plus background) under an expectation provided by the background contribution alone in the null hypothesis.

\subsection{Deep Neural Network overview}

\medskip
%\smallskip
\noindent
The model architecture used for the reported study is based on an artificial Deep Neural Network (DNN). A Neural Network (NN) attempts to learn a mathematical function that maps a selection of input features to a target function. This is accomplished by a series of matrix operations involving learned weights, {\em i.e.} parameters that are adjusted throughout the training, and non-linear transformations on the inputs. The NN may be visualized as a set of layers of neurons, each of which receives inputs from the neurons of the previous layer. Layers between the input and output ones are referred to as hidden, and the more hidden layers a NN contains, the ‘deeper’ it is considered to be. The main choices to be made when constructing a NN, including the ones taken into account and tested in the study described here, are: 

\begin{itemize}
\item{The activation function, which provides a (non)-linear response of each neuron based on the weighted sum of their inputs, in order to provide the neuron output;}
\item{The weight initialization, on the basis of which weights are sampled randomly from a (non-uniform) distribution;}
\item{The loss function, which quantifies the performance of the NN,}
\item{The learning rate (LR), which corresponds to the step size the NN makes over the loss surface at each update point;}
\item{The optimization algorithm used for the learning rate adaptation following changes in the loss function;}
\item{The pre-processing step, which refers to the appropriate transformation of the input features towards improving the weight initialization and decreasing the convergence time;}
\item{The cross-validation (k-fold), related to splitting the training sample into k equally sized portions, and repeating the training and testing procedure k times by training each time the NN on all the portions except the one to be eventually used for the NN’s respective testing;}
\item{The ensemble approach, according to which multiple ML algorithms are combined in the direction of improving the performance for a larger range of inputs, and the constituent NN are weighted based on their performance on separate testing sets, for the degree of their respective influence on the output to be regulated.}
\end{itemize}

\subsection{Baseline model and alternative techniques}

\medskip
%\smallskip
\noindent
The baseline model in the study reported here is a fully connected, feed-forward DNN, with 4 hidden layers of 100 neurons each. The activation function used is ReLU (Rectified Linear Unit), and the weight
initialization relies on He’s prescription~\cite{He2015}. The NN output is a single sigmoid neuron with Glorot initialization~\cite{Glorot2010}. The optimization is done via the Stochastic Gradient Descent algorithm with the Adam extension, and the mini-batch size is set to 256 events.

\medskip
\smallskip
\noindent
An 80:20 random split is performed on the original training data into training and validation sets, which are then split into ten folds via random stratified splitting on the event class. The testing data is split into ten folds as
well, but via simple random splitting. During training, each fold is loaded sequentially.

\medskip
\smallskip
\noindent
The tests performed in this study---with a view to comparing different machine learning techniques as to their effect on the performance achieved through the ‘HiggsMLChallenge’ winning solutions---lie on several levels including the following:

\begin{itemize}
\item{{Combining many models in an ensemble by averaging the predictions of each model, which improves the generalization of the final prediction to unseen data;}
\item{{Learning rich, compact, embedding matrices for categorical features~\cite{cat_embed}, as opposed to 1-hit encoding the categories and thereby increasing the number of inputs to the model;}}}
\item{{Choosing a better internal activation function; whilst ReLU is the standard, issues such as ``dead ReLU",\footnote{Random weight initialization, or updates during training push the activation into the far-negative region, meaning that the output (and incoming gradient) are consistently zero and the neuron becomes unusable.} non-zero-centred outputs, and saturated gradient for negative outputs mean that newer activation functions which aim to solve these issues can be beneficial, such as: ReLU, PReLU (Parameterized ReLU)~\cite{He2015}, SELU (Scaled Exponential LU)~\cite{Klambauer2017} 
and Swish~\cite{Ramachandran2018}, the latter defined as Swish(x) = $x$ · Sigmoid(x);}}
\item{{Using learning rate and momentum scheduling to improve both performance and convergence time, e.g.~cosine-annealed LR~\cite{Loshchilov2017} and 1-cycle annealing of LR and momentum~\cite{Smith2018};}}
\item{{Employing domain-specific data-augmentation techniques (see {\em e.g.}~\cite{Krizhevsky2017}) to improve the performance and generalization of the model; for LHC collisions in symmetrical detectors like ATLAS and CMS, events can be rotated in azimuth and flipped in the transverse and longitudinal plane to create new, valid inputs without affecting the class of the events. This may be applied during training time to artificially increase the available training data and during testing time to reduce residual biases in predictions by averaging over a set of augmented inputs;}}
\item{{Performing advanced ensembling via: Snapshot Ensembling (SSE)~\cite{snapshot_ensemble}, Fast Geometric Ensembling (FGE)~\cite{FGE,FGE_simul}, and Stochastic Weight Averaging (SWA)~\cite{Izmailov2018}; These methods can either reduce the time required to train and apply ensembles, or allow a larger ensemble to be trained in a similar time as traditional ensembling;}}
\item{{Using densely connected hidden layers~\cite{Strong2020}; similar to \textsc{DenseNet}~\cite{densenet}, each hidden layer receives as input the output of all previous layers, meaning that prior representations of the data are never lost (thereby protecting against over-parameterization of the model), and parameters have a more direct connection to the output and subsequent back-propagated gradient.}}
\end{itemize}

%\medskip
%\smallskip
\noindent
Based on the above options, which are studied separately, this investigation manages to carry out several comparisons with the baseline model, and in so doing reaches important conclusions on the set of choices that is found to benefit the performance most.

\subsection{Performance tests}

\medskip
%\smallskip
\noindent
After evaluating the proposed changes (some of them being mutually exclusive), the final model was decided upon. This consists of an ensemble of 10 DNNs, in which predictions are weighted according to their performance on validation data during training (the reciprocal of their loss). The use of ensembling resulted in the largest improvement in performance that was seen in the study. The single categorical feature of the data was passed through an embedding matrix, which offered a minor performance boost. The DNNs were trained using HEP-specific data augmentation, in which the final state particles measured in an event were randomly flipped and rotated in a class-preserving and physics-invariant manner.\footnote{For example, all particles pseudo-rapidities may be simultaneously changed of sign, given the symmetry in the initial state of the collision; similarly, if one observed particle is taken as a reference, all others may be subjected to a mirroring of their azimuthal angles about the axis of the reference particle, $\phi_{ref}$ as $\phi_i' = 2\phi_{ref}-\phi_{i}$} This was also applied at testing time by computing the average prediction on a set of data transformations. This procedure resulted in the second largest improvement observed in the scoring metric. 

\medskip
\smallskip
\noindent
Among the activation functions tested, Swish offered the largest improvement in performance. 1-cycle scheduling of the LR and momentum of the optimizer were found to allow DNNs to be trained to a higher level of performance in half of the required time. Finally, by changing the DNNs to be thinner and deeper (six layers of 33 neurons each), and then passing all the outputs of previous layers to the inputs of subsequent layers (dense connections for non-convolutional layers), a moderate improvement in performance was found (along with potential resilience to the exact settings of architecture hyper-parameters).

{\setlength{\tabcolsep}{9.5pt}
\begin{table}[ht]
\captionsetup{font=scriptsize}
\caption{Comparison between our solution (``This Work") and the top three solutions during the HiggsML Challenge. Performance is measured in AMS (see Section~\ref{s:scoringmetric}). We report the average performance of our model (evaluated via repeated trainings), but for the challenge entrants' solutions we report the AMS of their submitted predictions. The inference times are the recorded time taken to evaluate every event in the full testing dataset (public and private).
Accounting for the differences in precision and throughput between the Nvidia Titan and 1080 Ti GPUs, it is estimated that the $1^{st}$-place solution may be trained in 100 minutes and applied to the full testing set in 8.5 minutes on a 1080 Ti machine. The developed solution therefore provides an effective speedup of: 92\% on GPU or 86\% on CPU, for training; and 97\% on GPU or 65\% on CPU, for inference. \vspace{0.1mm}}
\label{table:1}
    \scriptsize 
    \centering
    \begin{tabular}{@{}l l l l l@{}}
%        \thickhline
        \cline{1-5}
        \cline{1-5}
%                            &   Ours            &   $1^{\textrm{st}}$   &   $2^{\textrm{nd}}$   &   $3^{\textrm{rd}}$   \\
                            &   This Work            &   1st    &   2nd    &   3rd    \\
        \cline{1-5}            
            Method          &   10 DNNs         &   70 DNNs             &   Many BDTs           &   108 DNNs            \\
            Train time (GPU)     &   8 min      &   12 h                &   Unknown             &   Unknown             \\
            Train time (CPU)     &   14 min     &   35 h                &   46 h                &   3 h                 \\
            Inference time (GPU) &   15 s       &   1 h                 &   Unknown             &   Unknown             \\
            Inference time (CPU) &   3 min      &   Unknown             &   Unknown             &   20 min              \\
            Private AMS     &   3.806$\pm$0.005      &   3.806             &   3.789              &   3.787               \\
        \cline{1-5}
        \cline{1-5}
%        \thickhline
%    \vspace{0.05mm}
    \end{tabular}
\end{table}
}

\subsection{Solutions comparison and outlook}

\begin{figure}[t]
    \centering
        \includegraphics[scale=.40]{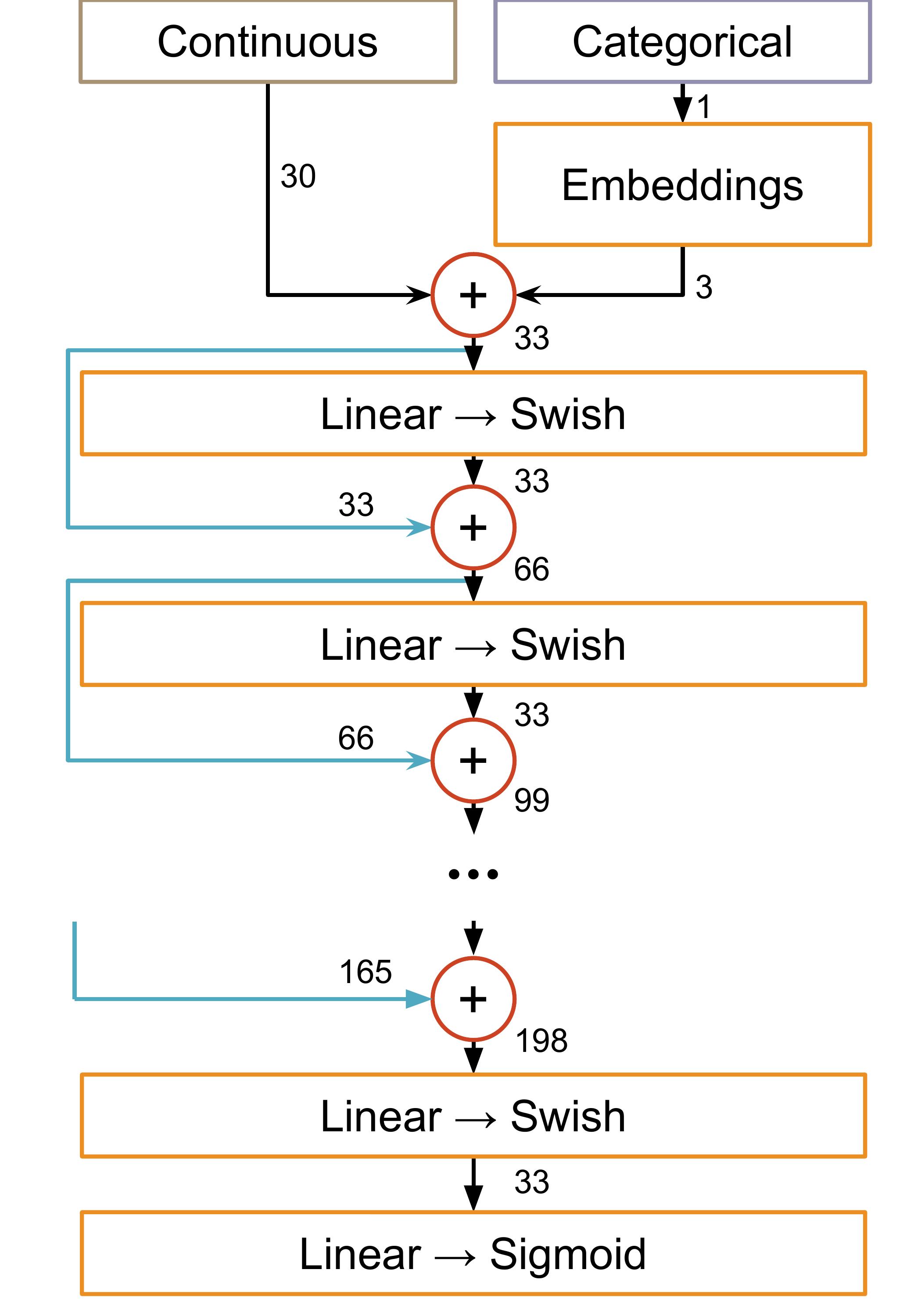}
    \caption{Illustration of the layout of the network used to build the final model. $\oplus$ indicates a feature-wise concatenation, providing dense skip-connections to the hidden layers. Six hidden layers (each with 33 neurons) are used in total. The single categorical feature is encoded via an embedding matrix prior to concatenation with the continuous features. The final model consists of 10 such networks in a weighted ensemble. Image source~\cite{Strong2020}.}
    \label{fig:higgsml:model}
\end{figure}

\medskip
%\smallskip
\noindent
Considering the set of choices regarding the model structure and characteristics that gives the most performant outcome, the study proceeds to a comparison with the winning solution. As shown in Table~\ref{table:1}, the model proposed by the study can match the score of the winning solution of the HiggsML Challenge, whilst being more lightweight, allowing it to be trained and applied much more quickly, even on a typical laptop. The relative contributions to overall improvement in metric score of each of the changes included in the final model are illustrated in Table~\ref{tab:higgsml:summary}. Similar to the other solutions, ensembling was found to provide a large improvement, however we can see that domain-specific data augmentation also provides a significant benefit. Whilst smaller in terms of score improvement, the 1-cycle training schedule allows to halve the required training time, and the dense connections provide some resilience to poor choices of network hyper-parameters. An illustration of the final model is shown in Fig.~\ref{fig:higgsml:model}.

\begin{table}
\caption{Breakdown of sources of improvement in metric score for the final ensembled model compared to the starting baseline model.}
\label{tab:higgsml:summary}
    \scriptsize
    \centering
    \begin{tabular}{ll}
    \hline
    Technique & Improvement contribution \\
    \hline
    Ensembling & 61.3\% \\
    Data augmentation & 32.1\% \\
    Dense connections & 3.6\% \\
    Swish + 1cycle & 3.0\% \\
    Entity embedding & 0.1\% \\
    \hline
    \end{tabular}
\end{table}

Code and hardware details of the HiggsML solutions may be found below:
\begin{itemize}
    \item This Work (\url{https://github.com/GilesStrong/HiggsML_Lumin}):
        \begin{itemize}
            \item GPU: NVidia 1080 Ti, $<$1GB VRAM, $<$1GB RAM
            \item CPU (2018 MacBook Pro): Intel i7-6500U 4-core CPU, $<$1 GB RAM;
        \end{itemize}
    \item 1st place: Melis (\url{https://github.com/melisgl/higgsml})
        \begin{itemize}
            \item GPU: NVidia Titan, $<$24 GB RAM
            \item CPU (AWS m2.4.xlarge): 8vCPU, $<$24 GB RAM;
        \end{itemize}
    \item 2nd place: Salimans (\url{https://github.com/TimSalimans/HiggsML}), 8-core CPU, 64 GB RAM;
    \item 3rd place: Pierre (\url{https://www.kaggle.com/c/higgs-boson/discussion/10481}), 4-core CPU (2012 laptop).
\end{itemize}

\medskip
%\smallskip
\noindent
This result highlights the need of optimizing the architecture and training of the ML algorithms, as this has been shown to lead to significant improvement over common default choices for both classification and regression problems.
The improvement in timing and hardware requirements are also of great importance, given that most analysers at the LHC do not have on-demand access to high-performance GPUs and must rely solely on their laptops and CPU farms.

\medskip
\smallskip
\noindent
The results of this study were verified in a partially independent study of the projected sensitivity to Higgs pair production of the CMS experiment in the High-Luminosity LHC (HL-LHC) scenario~\cite{CYRM_main,CYRM_notes}, { discussed further in \autoref{sec:hl_lhc_hh},} in which using some of the solution developments discussed above resulted in a 30\% improvement in { AMS}, compared to a baseline DNN { under otherwise similar circumstances}. This investigation is briefly summarized in the following section.

\section{Multi-Variate Techniques for Higgs Pair Production Studies}\label{sec:mvaHH}

\subsection {Introduction}

\medskip
%\smallskip
\noindent
With the mass of the Higgs boson ($m_H$) now experimentally measured with sub-GeV precision~\cite{higgs_mass1}, the structure of the Higgs scalar field potential and the intensity of the Higgs boson self-couplings are precisely predicted in the SM. While measured properties are so far consistent with the expectations from the SM predictions, measuring the Higgs boson self-couplings provides an independent test of the SM and allows a direct measurement of the scalar sector properties. In particular its self-coupling strength $\lambda_{hhh}$, can directly be measured through the study of particle collisions in which two Higgs bosons are produced by the same hard subprocess. In LHC proton--proton collisions this process occurs mainly via gluon fusion (ggF) and it involves either couplings to a loop of virtual fermions, or the $\lambda_{hhh}$ coupling itself. The SM prediction for the double Higgs production cross section is small, {\em i.e.} approximately $\sigma$(ggF)=31~fb at NNLO for a Higgs boson mass $m_H$=125 GeV at 13 TeV~\cite{Grazzini:2018bsd}. Beyond Standard Model (BSM) physics effects in the non-resonant case may appear via anomalous couplings of the Higgs boson; the experimental signature of those effects would be a modification of the Higgs boson pair production cross section and of the event kinematics.

\medskip
\smallskip
\noindent
If the value of the Higgs self-coupling were significantly different from the value predicted by the SM, it could well be an indication of new physics. Possible scenarios that might give rise to such a deviation include resonant production via a ‘heavy Higgs’ such as those predicted by BSM theories like the Minimal Super-Symmetric Standard Model, and BSM particles being produced virtually within the coupling loops~\cite{MarchRussell:2008yu,Khachatryan:2015tha,Aaboud:2017cxo}.
Unfortunately, the expected cross section for Higgs pair production is several orders of magnitude below other particle processes that form an irreducible background. Because of this, detecting a statistically significant presence of Higgs pair production events at the LHC requires a much larger number of recorded collisions than the LHC in its current form is expected to produce.

\medskip
\smallskip
\noindent
Higgs bosons have a wide range of possible decay channels~\cite{pdg}. The most probable decay for a Higgs boson is to a pair of b quarks, but this decay mode incurs in a large background from Quantum Chromo-Dynamics (QCD) processes yielding multiple hadronic jets. Decay to a pair of $\tau$ leptons is the next most probable fermionic decay mode, where $\tau$ leptons can decay to light leptons (electrons and muons, with a branching ratio BR of approximately 35\%) or hadronically (with a few charged particles in 1- or 3-prongs, with a BR of approximately 65\%) in their secondary decays; these decay modes offer a powerful handle on suppressing QCD backgrounds. The $HH\rightarrow\tau\tau bb$ decay mode therefore offers a favourable compromise due to the high branching ratio of $H\rightarrow bb$ and a source of high-purity leptons from the $H\rightarrow\tau\tau$ decay. On the other hand, the $b \bar{b} b \bar{b}$ final state of $HH$ pairs is the most frequent one, yet it has to fight a very large background of QCD production of multiple b-quark pairs. In the following we describe a multivariate study aimed at modelling with high precision the QCD background in the four-b-quark final state, and a study of the potential of the $b \bar{b} \tau \tau$ final state in the HL-LHC data-taking scenario.

\subsection{Modelling QCD backgrounds for the \texorpdfstring{$HH\to b\bar{b}b\bar{b}$}{HH->bbbb} search}

\medskip
%\smallskip
\noindent
The QCD background is the omnipresent problem of searches for rare processes in hadron collisions. While Monte Carlo generators can today accurately model processes with several hadronic jets in the final state, their reliability remains limited in regions of phase space populated by a large number of energetic jets, which are contributed by radiative sub-leading processes that MC generators can only handle with limited precision. In addition, the computational requirements of a thorough simulation of events with a large number of jets makes reliance on MC simulation not always easily practicable. Finally, in specific applications where one needs a precise modelling of not just one single variable, but of the full multi-dimensional density of the multi-jet kinematics, the problem becomes intractable by parameterizations.

\medskip
\smallskip
\noindent
Having in mind an application to the search of Higgs boson pair production in CMS data using the four b-jets final state, we devised a precise modelling of the QCD background employing exclusively experimental data. The novel technique we designed, called ‘hemisphere mixing’, allows for the generation of a high-statistics, multi-dimensional model of QCD events, such that we can base on its properties the training of a multivariate classifier capable of effectively discriminating the Higgs pair production signal. 

\medskip
\smallskip
\noindent
Event mixing techniques are not a novelty in HEP applications; they have been used extensively in electron--positron collider experiments. Applications to hadron collider physics analysis also exist~\cite{mixing1,mixing2,mixing3,mixing4,mixing5}, but they are limited to the mixing of individual particles or jets, while our technique for the first time employs hemispheres of jets as the individual objects subjected to a mixing procedure. As jets are direct messengers of the hard subprocess, the creation of artificial events through the mixing of entire jet collections is considerably more complex than the mixing of single particles or jets.

\subsubsection{The hemisphere mixing technique}

\medskip
%\smallskip
\noindent
The idea on which hemisphere mixing is based stems from the observation that QCD events, while extremely complex to model and interpret from the final state point of view, may be thought to originate from a simple tree-level two-body reaction, whereby two partons scatter off one another. The reaction then produces the complex kinematics of a multi-jet event by means of the intervention of second-order effects involving initial and final state QCD radiation, in addition to  pile-up collisions in the same proton--proton bunch crossing or even multiple parton scattering of the same colliding protons. Still, the kinematics of the two leading partons emitted in the final state of the hard subprocess offers itself, if properly estimated, as a basis of a similarity measure, which can be exploited to create artificial events based on the individual parton properties.

We consider a dataset of QCD events that feature at least four jets in the final state, and construct in each event a ‘transverse thrust axis’ using the transverse momentum of the jets. This axis can be defined by the azimuthal angle $\phi_T$ such that
\begin{equation}
    \phi_T = arg \, max_{\phi_{T'}} \left[ \sum_j p_{T,j} |\cos (\phi_j - \phi_{T'}) | \right]\,,
\end{equation}

\medskip
%\smallskip
\noindent
where the sum runs over all observed jets $j$; alongside with $T$ we may define the related variable $T_a = \sum_j p_{T,j} | \sin(\phi_j-\phi_T)|$. Once $T$ is defined, the event can be ideally split in two hemispheres by the plane orthogonal to $T$. The two hemispheres contain two sub-lists of the jets, characterized by the different signs of $\cos(\phi_j - \phi_T)$. We may describe each hemisphere by a number of observable features, $h(N_j, N_b, T,M,T_a,P_z)$. In this expression $N_j$ is the number of contained jets, $N_b$ is the number of b-tagged jets, $T$ ($T_a$) is their transverse momentum sum along the thrust axis (orthogonal to it), $M$ is their combined invariant mass, and $P_z$ is the sum of longitudinal components of the jets. Using the $2N$ hemispheres that can be obtained from $N$ observed data events in the same sample where we wish to search for a small signal we may build a library ${h_i}$ $(i=1,...,2N)$, which is the basis of the construction of synthetic events. If $h_l$ and $h_m$ are the hemispheres obtained from the splitting of a real event, we may construct an artificial replica of that event by identifying the two hemispheres $h_p$, $h_q$ in the library that are the most similar to $h_l$ and $h_m$ (subjected to the condition that none of the indices $l,m,p,q$ are identical). Similarity is defined within the sub-classes of hemispheres with the same value of $N_j$ and $N_b$ by a normalized Euclidean distance in the space of continuous parameters $(T, M, T_a, P_z)$ describing the hemispheres. More detail on the procedure is available in~\cite{hemispheremixing}.

\begin{figure}[ht]
\includegraphics[width=12cm]{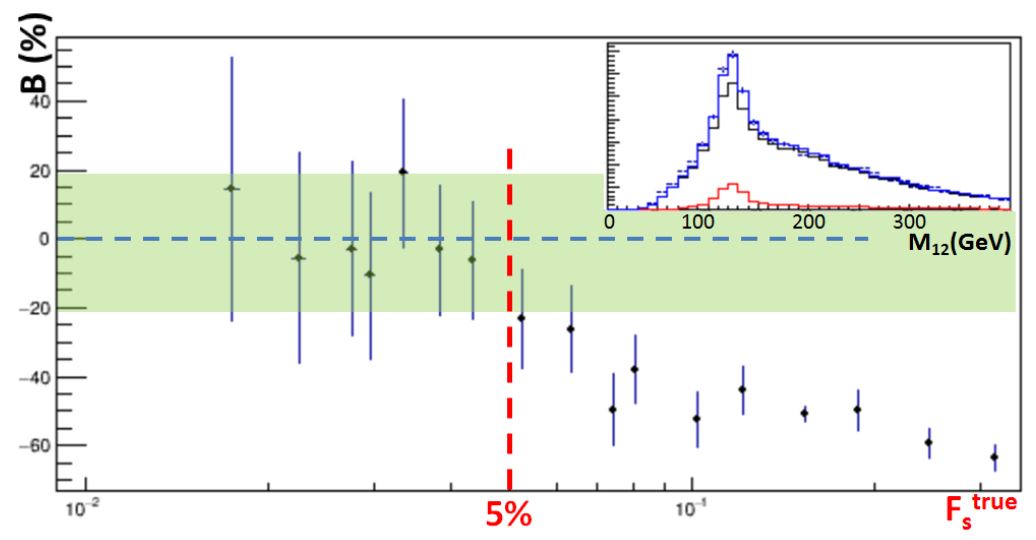}

\caption{ Size of the bias B, in percentage, on the signal fraction extracted by a fit to the reconstructed Higgs boson mass distribution in data containing a small $HH \to b \bar{b} b \bar{b}$ signal and a majority of QCD background events, as a function of the signal fraction $f_s$. The green band shows the level of bias considered acceptable in searches of new signals in hadron collider data. The upper-right inset shows the distribution of the reconstructed Higgs boson mass ($M_{12}$) for simulations of QCD (black), HH signal (red), and sum of the two components (blue), with overlaid the fit results (point with uncertainty bars) for a 5\% signal contamination, which is close to the maximum contamination that still allows a successful modelling of the multi-dimensional distributions~\cite{hemispheremixing}. }
\label{f:hemispheremixingresults}
\end{figure}

\subsubsection{Results}

\medskip
%\smallskip
\noindent
Multi-dimensional statistical tests that employ a complete set of kinematic variables describing the event features prove that the produced artificial events model the multi-dimensional distribution of the original features to sub-percent accuracy, if the dataset is constituted by QCD events. Furthermore, when the data contain a small fraction of events originated by Higgs pair production events the mixing procedure washes out the features of that minority class, such that the artificial data sample still retains accuracy in modelling the QCD properties (see Fig.~\ref{f:hemispheremixingresults}, which shows the effect of different signal fractions in the modelling). This happens because the probability that two hemispheres in the library, chosen to model a signal event, be both originated from other signal events (and thus retain memory of the peculiarities of the multi-dimensional density of signal in the event feature space) scales with the square of the signal fraction in the original data. Hence a small signal contamination present in the dataset will not impair the validity of the model. This property makes the hemisphere mixing method an attractive option for the search of rare signals in QCD-dominated datasets. It is of special interest the fact that the user does not need to identify a control sample of data where to perform modelling studies: the method can be directly applied to the same sample where the signal is sought for. This is a considerable simplification of the experimental analysis, which also reduces modelling systematics. Finally, by searching for multiple similar hemispheres to the two that make up the event to be modelled, one may construct a synthetic dataset much larger than the original one, reducing the statistical uncertainties without introducing appreciable systematic biases.

\medskip
\smallskip
\noindent
The hemisphere mixing procedure has been successfully used for the first search of Higgs pair production in the four b-jets final state performed by the CMS experiment~\cite{hhbbbbcms}. In that analysis the technique enabled the training of a multivariate classifier on a large synthetic dataset of hemisphere-mixed events, as well as provided the background model from which a limit on the Higgs pair production signal was extracted.

\subsection {Prospects of the \texorpdfstring{$HH \to b \bar{b} \tau \tau$}{HH->bbtautau} channel at the HL-LHC}\label{sec:hl_lhc_hh}

\medskip
%\smallskip
\noindent
A study of the sensitivity of the HL-LHC to the Higgs self-coupling using advanced analysis techniques was performed by considering SM Higgs pair production in proton--proton collisions at $\sqrt{s}=14$~TeV. DNNs were trained for the task of separating the signal from background contributions. Details can be found in~\cite{CYRM_main, CYRM_notes}. The study was pursued on the $HH\rightarrow\tau\tau bb$ decay mode.

\medskip
\smallskip
\noindent
The decay of a Higgs boson to $\tau\tau$ pairs gives rise to six possible combinations of final state signatures for the signal: $e\tau_h$, $\mu\tau_h$, $\tau_h\tau_h$, $ee$, $\mu\mu$, and $e\mu$, where $\tau_h$ indicates a hadronically decaying $\tau$ lepton. For this investigation, we only consider the three most frequent final states, {\em i.e.} those involving at least one $\tau_h$. From the event selection, a total of 52 features are used in the study. They are split into ``basic" (27), ``high-level/reconstructed" (21), and ``high-level/global" (4) features. These proved to give the best performance.

\begin{figure}
    \centering
    \begin{subfigure}[H]{0.49\textwidth}
     \includegraphics[width=\textwidth]{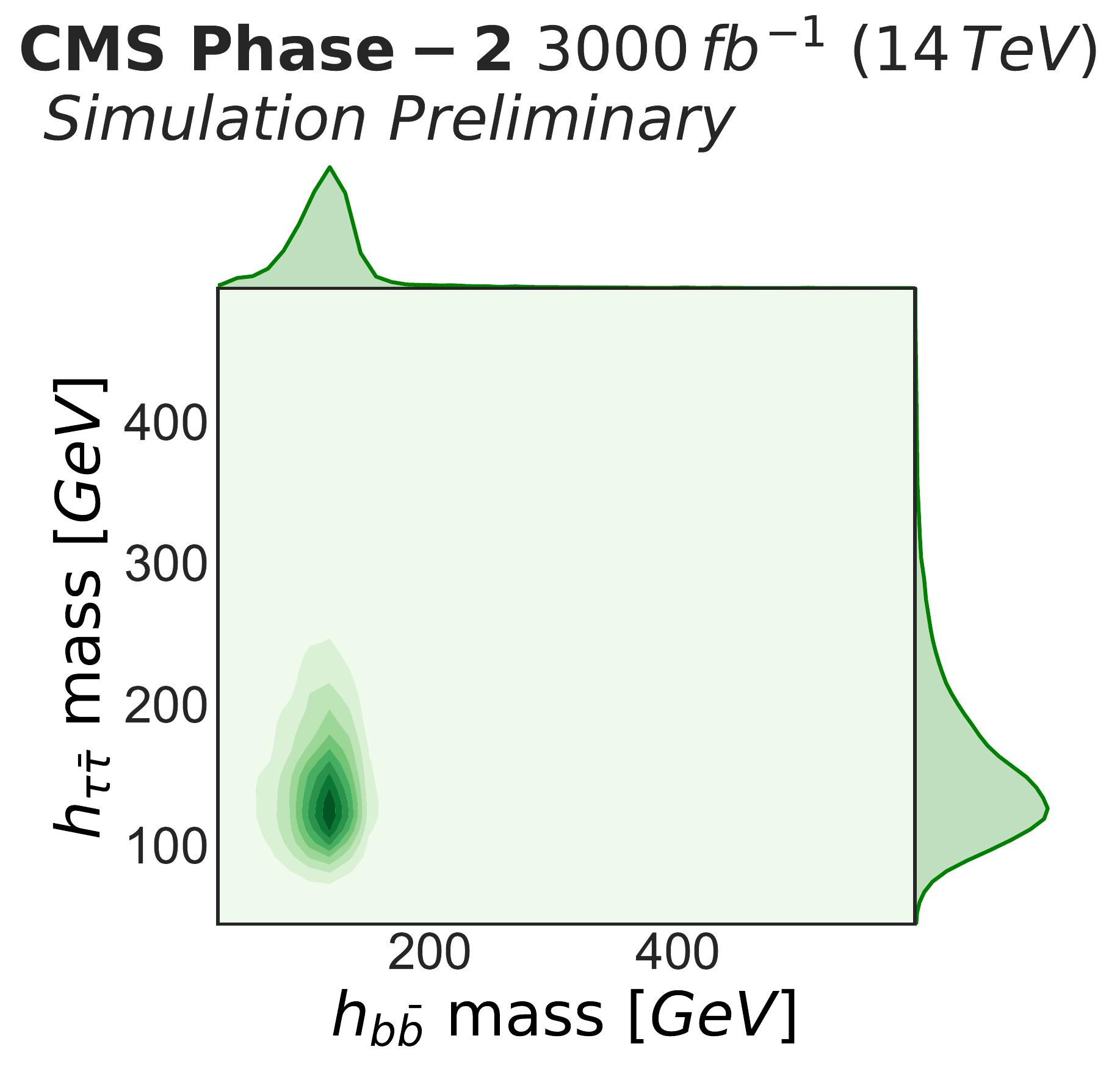}
    \end{subfigure}
    \begin{subfigure}[H]{0.49\textwidth}
\includegraphics[width=\textwidth]{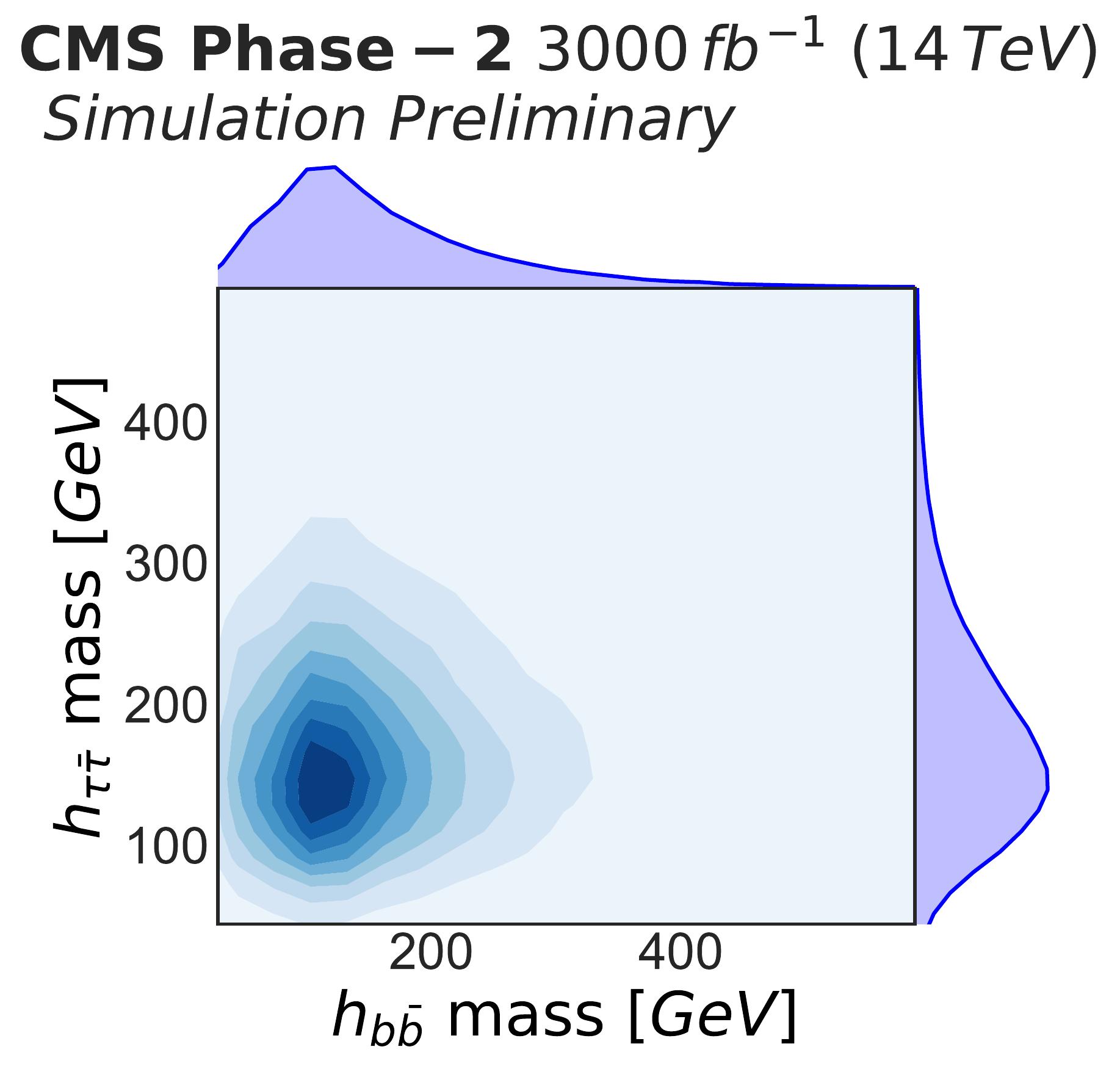}
    \end{subfigure}
    \begin{subfigure}[H]{0.49\textwidth}
\includegraphics[width=\textwidth]{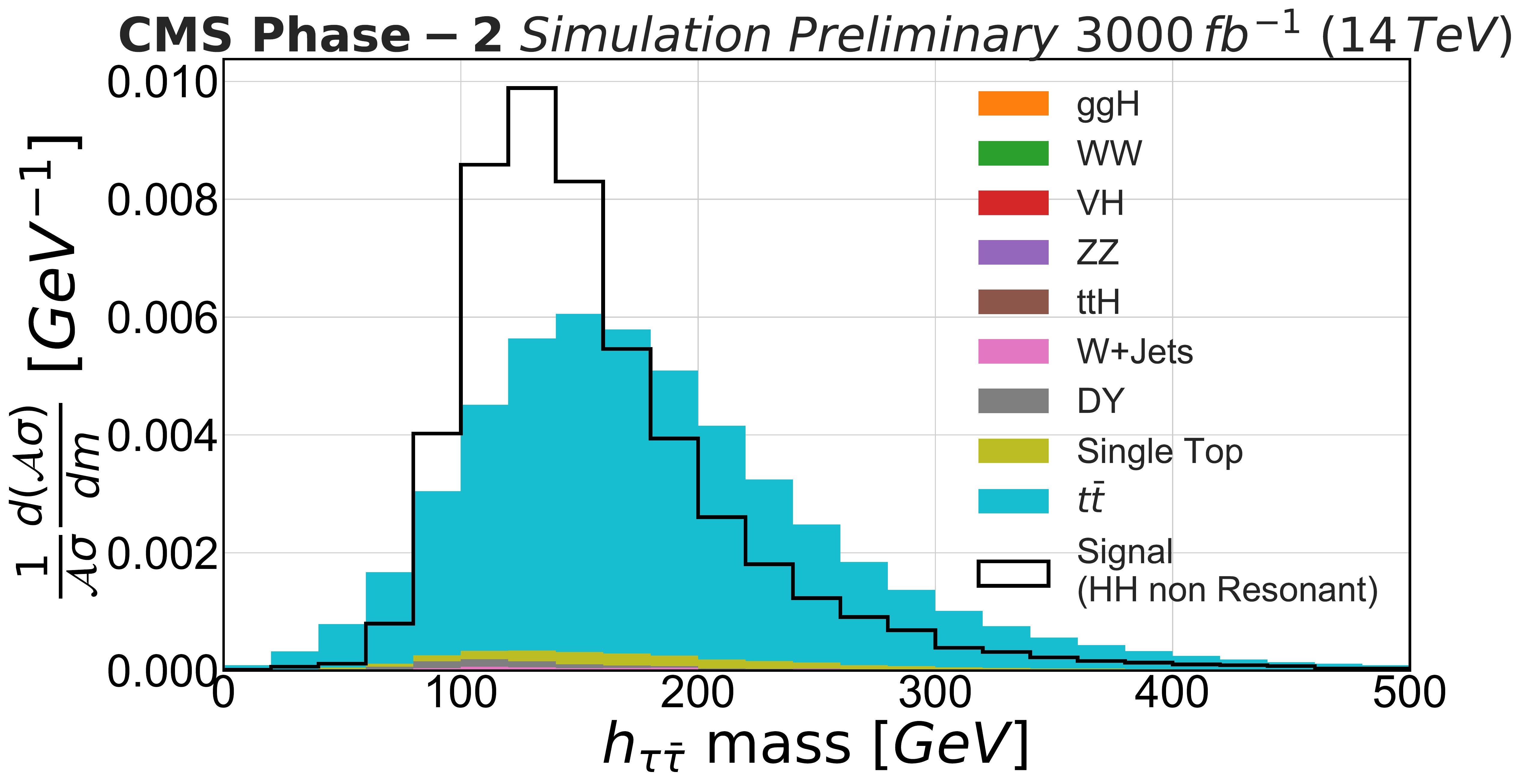}
    \end{subfigure}
    \begin{subfigure}[H]{0.49\textwidth}
\includegraphics[width=\textwidth]{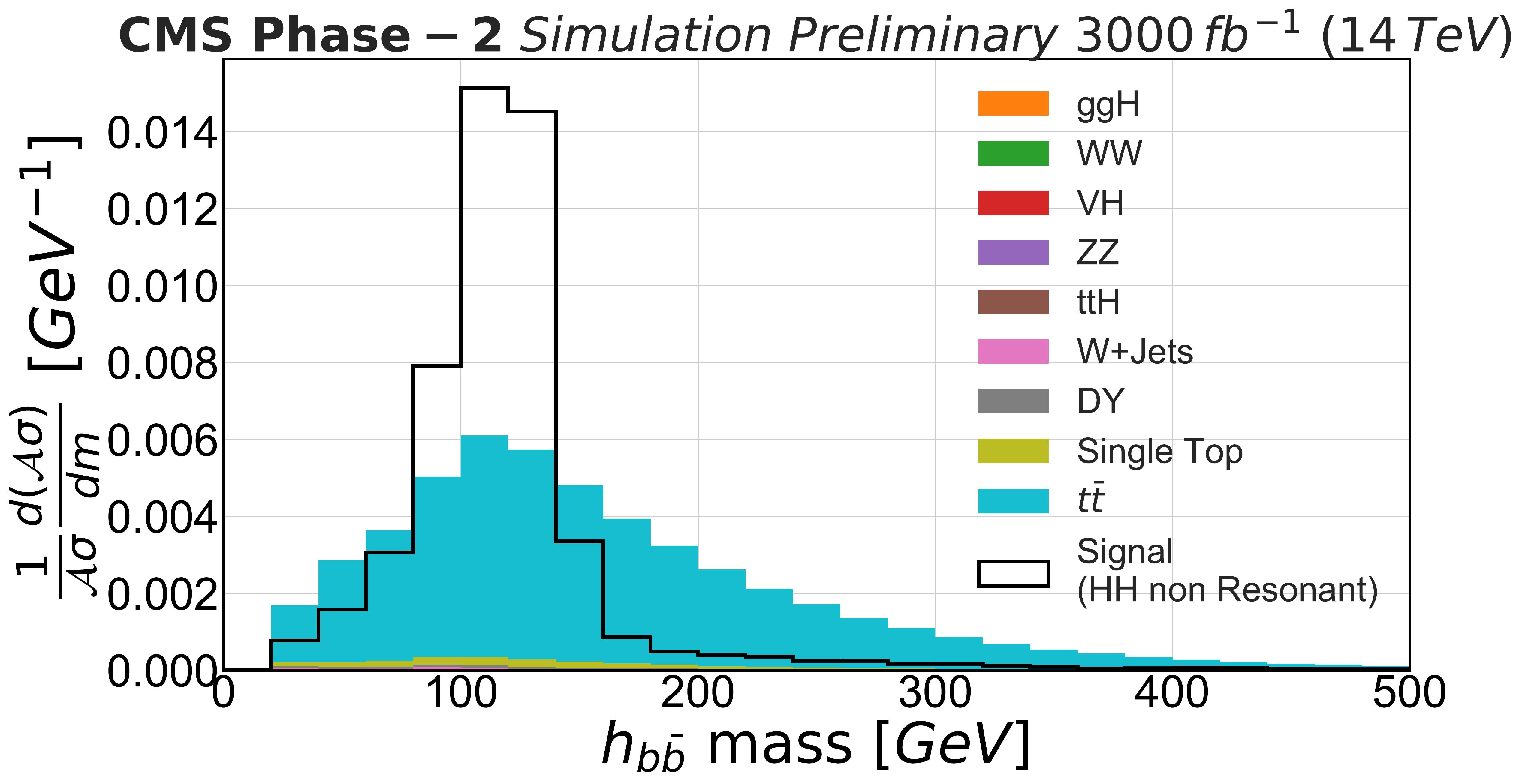}
    \end{subfigure}
    \caption{Two-dimensional distributions of the invariant masses of the reconstructed Higgs bosons ($h_{\tau\tau}$ and $h_{bb}$) in signal (left) and background processes (right). One-dimensional invariant-mass distributions are projected on the corresponding axes; Bottom: One-dimensional distribution of the $h_{\tau\tau}$ (left) and $h_{bb}$ (right) invariant masses for signal and background process for all final states together. Distributions of signal and background are separately normalized to unit area. Images source~\cite{hl-lhc_FTR} (supplementary material).}
    \label{fig:dihiggs:features}
\end{figure}

\medskip
\smallskip
\noindent
From each selected event the two Higgs bosons are reconstructed making use of the considered final states. Distributions of the invariant masses of the reconstructed Higgs bosons used as inputs for the DNN are shown in Fig.~\ref{fig:dihiggs:features}.

\medskip
%\smallskip
\noindent
Simulated data were pre-processed with a 50--50 split into training and testing sets, and were used to train a DNN with several optimizations. { Models were compared using the Approximate Median Significance~\cite{Cowan2011, ams}, however in order to include the presence of uncertainties on the background, an extended version is used, compared to \autoref{eq:ams}}:
\begin{equation}
    \mathrm{AMS}=\sqrt{2\left(s+b\right)\ln\left(\frac{\left(s+b\right)\left(b+\sigma^2_b\right)}{b^2+\left(s+b\right)\sigma^2_b}\right)-\frac{b^2}{\sigma^2_b}\ln\left(1+\frac{\sigma^2_bs}{b\left(b+\sigma^2_b\right)}\right)}\,,
    \label{eq:ams_full}
\end{equation}

\medskip
%\smallskip
\noindent
where $\sigma_b$ is the uncertainty on the number of expected background events. For model development, a 10\% systematic uncertainty on the background normalization was assumed, in addition to statistical uncertainties. The final computation of the analysis sensitivity used appropriately estimated systematic uncertainties from a variety of contributions, accounting for their correlations.

\medskip
\smallskip
\noindent
We studied the performance of NN classifiers in terms of how well they can classify signal and background events. This involved reconfirming the benefits of several of the techniques already studied in the work summarized in Section~\ref{sec:higgsml}. The final ensembled model uses SELU activation~\cite{Klambauer2017}, a cosine-annealed learning rate~\cite{Loshchilov2017}, and the data augmentation described in Section~\ref{sec:higgsml}. These techniques resulted in a 30\% improvement in AMS over the performance of a baseline ReLU model.

\begin{figure}[htp]
\centering
\includegraphics[width=1\textwidth]{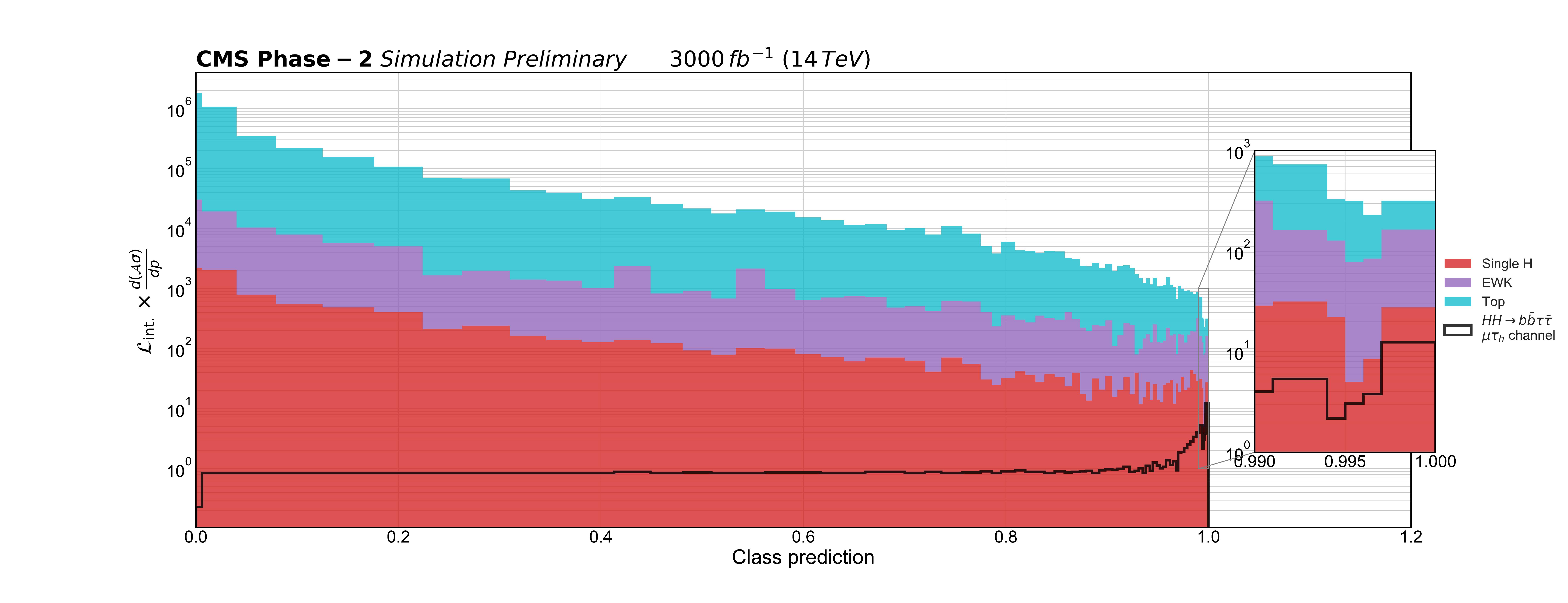}

\caption{Predictions of the classifier evaluated on the test dataset for the $\mu\tau_h bb$ final state. Both signal and background are normalized to the expected yields. Images source~\cite{hl-lhc_FTR} (supplementary material).}
  \label{fig:dihiggs:score}
\end{figure}

\medskip
\smallskip
\noindent
Signal and background events are binned in the distributions of the classifier prediction for signal and background in each channel (Fig.~\ref{fig:dihiggs:score}).
A simultaneous fit is performed on the expected event distributions for the three final states considered.
Including systematic uncertainties, an upper limit on the HH cross section times branching fraction of 1.4 times the SM prediction is obtained, corresponding to a significance of 1.4~$\sigma$ in this final state alone. When results are combined with the other final states, a significance of 2.6~$\sigma$ is achieved, and 4~$\sigma$ when combining the two experiments, ATLAS and CMS.

\medskip
%\smallskip
\noindent
The precise characterization of the Higgs boson will be one of the highest priorities of the HL-LHC physics program.  The  MIP Timing  Detector  (MTD)~\cite{cms_hl-mtd} is a new detector planned for the CMS experiment during the HL-LHC era and it will enhance the physics reach capabilities.
In this context, the improved object reconstruction and the related effects were quantified. In particular, the HH study discussed {\em supra} was repeated to account for the new MTD showing a further improvement. For details see~\cite{mtd-tdr}.

\section{Jet Flavour Classification}\label{sec:btagging}

\subsection{Overview}

\medskip
%\smallskip
\noindent
The correct reconstruction and identification of all particles interacting with the different types of detector material constitutes a fundamental prerequisite to extract information from detected particle collisions at the LHC experiments. 
Here we focus on the reconstruction of hadronic jets, which is more challenging than that of other measurable physics objects because of the complexity of the physics processes of relevance, and which is an important ingredient to the vast majority of measurements and searches carried out with the ATLAS and CMS experiments.

\medskip
\smallskip
\noindent
Hadronic jets can be defined as collimated sprays of particles emerging during the hadronisation process of a parton (quark or gluon) emitted with high energy from the collision point. Jets may originate from b quarks, c quarks, so-called ‘light quarks’ (u, d, s), and gluons. Due to their larger mass than all other partons, and to other specific properties of the hadrons they produce in their hadronisation, b and c quarks (usually denoted as ‘heavy flavour’ quarks) yield jets that may be distinguished from the rest. The identification of the type/flavour of the initial parton that is associated with the jet, referred to as jet tagging, constitutes an essential stage of the jet reconstruction process. In particular, the efficient identification of heavy flavour jets is a subject of paramount importance for a number of measurements and searches, due to the possible connection of their production with physics processes preferentially coupling to the second and third generation of matter fermions {-- as is the case for the search in~\cite{search4}}, while it is also critical for Higgs boson measurements because of the large branching fraction of the Higgs to $b \bar{b}$ and $c \bar{c}$ quark pairs.

\medskip
\smallskip
\noindent
The potential of new ML tools for heavy-flavour tagging must therefore be investigated thoroughly by HEP experiments. In this context, the degree to which the use of novel deep learning techniques may improve heavy flavour jet identification in CMS has been clarified by developing the \textsc{DeepCSV}, \textsc{DeepFlavour}, and \textsc{DeepJet} taggers~\cite{Bols2020,Stakia:ICNFP2017,Stoye:DLPS2017}, a task that received a significant contribution by AMVA4NewPhysics members. Extension of the applicability of the above mentioned taggers for the distinctive case of quark/gluon discrimination has also been examined within this study. On top of the development and evaluation of the jet tagging models, AMVA4NewPhysics researchers have also considerably contributed to the integration of these taggers into the CMS reconstruction software~\cite{cmssw_git, DeepFlavour_integration}, meeting the strict computational and performance requirements set out by this particular task. The architectures developed have been the first advanced deep-learning architectures to be integrated within the CMS reconstruction pipeline, and the core integration implemented for this purpose has been re-used for additional models and tasks thereafter.

\subsection{Particle-flow jets and B hadrons identification}

\medskip
%\smallskip
\noindent
The so-called ‘particle-flow’ jets consist of a list of particles reconstructed via the particle-flow algorithm~\cite{Sirunyan2017}, which is commonly used in CMS reconstruction, and clustered with the anti-$k_T$ clustering algorithm~\cite{Cacciari2008}. The particle-flow algorithm aims at identifying all observable particles in the event by combining information from all CMS sub-detectors. What distinguishes a b-quark-initiated jet from other jets at reconstruction level are several detectable particularities stemming from its typical features. {At the jet-formation stage, }while light quarks and gluons hadronize by predominantly producing short-lived hadrons whose decay products yield tracks that originate directly at the collision point (primary vertex), the B hadron created by the hadronisation of the b-quark has a comparatively long lifetime (of the order of a picosecond), which leads to the creation of a secondary vertex at the point of their decay, significantly displaced from the primary vertex; the secondary vertex can be reconstructed from the measured trajectories of charged tracks possessing a significant impact parameter\footnote{Impact parameter is the distance of closest approach of the back-extrapolated particle trajectory to the primary vertex.} with respect to the primary event vertex. Other detectable characteristics of the b-jets include a relatively large opening angle of decay products of the heavy B hadron, the possible presence of an electron or muon produced by the semi-leptonic B hadron decay, and a different track multiplicity distribution and fragmentation function with respect to those of jets originated by other partons. All the above information is used by the b-jet taggers in CMS and ATLAS, so as to identify b-quark-originated jets with the best possible accuracy.

\subsection{The \textsc{DeepCSV} tagger}

\medskip
%\smallskip
\noindent
The \textsc{DeepCSV} algorithm~\cite{CMS-DP-2017-005}, compared to the previously-standard b-tag classifier \textsc{CSVv2}~\cite{CMS-PAS-BTV-15-001}, has the same input in terms of observable event features, but processes a larger number of charged tracks. \textsc{DeepCSV} is also a deeper NN, and it is trained for multi-class classification. More specifically, the network is composed of five dense layers of 100 nodes each, and its input can be in total of around 70 variables. After selecting charged tracks passing quality criteria, eight features are used from up to six tracks with the highest impact parameter as part of the input set. Eight additional features summarize information from the most displaced secondary vertex, and finally, 12 features are constructed with jet-related observables (global variables). In general, the features used in \textsc{DeepCSV} are similar to the ones traditionally used in CMS for b-tagging~\cite{CMS-PAS-BTV-15-001}. The multi-class classification property of \textsc{DeepCSV} allows the use of four output classes instead of two of binary classifiers. The classes include the following cases describing the originating parton: a b-quark, a c-quark, a light quark (u/d/s), or a gluon. Also, the case that two B hadrons happen to co-exist inside the same jet is studied through a separate class.

\subsection{The \textsc{DeepFlavour} and \textsc{DeepJet} taggers}

\medskip
%\smallskip
\noindent
\textsc{DeepFlavour}~\cite{CMS-DP-2017-013}, compared to \textsc{DeepCSV}, has a much larger input (around 700 variables at most), is a deeper NN (eight fully connected layers, the first one being of 350 nodes, while the rest of 100 nodes each), and it includes convolutional layers. Besides the charged jet constituents, whose number is now increased (up to 25), along with the number (16) of features extracted from their kinematical properties, the input includes also information from identified neutral jet constituents (up to 25), with which six additional features are constructed. Moreover, there are now up to four secondary vertices considered as additional input, upon which 12 features are used. Finally, the input information includes six global variables describing the jet. In order to extract and engineer features per object, particle or vertex, several 1$\times$1 convolutional layers are used on the input lists of objects. For charged particles and secondary vertices, four layers of 64, 32, 32, and 8 filters are applied. For the neutral particles, which carry considerably less information, only three layers are used, with 32, 16, and 4 filters.

\medskip
\smallskip
\noindent
The \textsc{DeepJet} tagger~\cite{CMS-DP-2017-027} was introduced as an updated version of \textsc{DeepFlavour} intended for additional discrimination power between jets originating from gluons and jets originating from light quarks: both categories were expected to fall into the same ‘udsg’ output class in the case of \textsc{DeepFlavour}. In \textsc{DeepJet} the output of the convolutional layers is given to Long Short-Term Memory (LSTM)~\cite{Hochreiter1997} recurrent layers of 150, 50, and 50 nodes, which respectively correspond to the charged particles, the neutral particles, and the secondary vertices. These intermediate features are concatenated with the six global features, and then given to the dense NN, whose first layer has 200 nodes for \textsc{DeepJet} instead of the 350 nodes of \textsc{DeepFlavour}~\cite{Zenodo_DeepJet,Zenodo_DeepJetCore,Zenodo_DeepNTuples}.

\begin{figure}
    \centering
    \begin{subfigure}[H]{0.49\textwidth}
        \includegraphics[width=\textwidth]{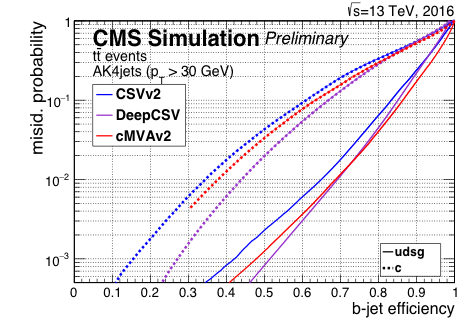}
        \par\medskip
        \caption{Performance comparison of \textsc{DeepCSV}, \textsc{CSVv2}, and \textsc{cMVAv2}. \cite{CMS-DP-2017-005}}
        \label{fig:1_top}
    \end{subfigure}~~
\hspace{1em}
    \begin{subfigure}[H]{0.46\textwidth}
        \includegraphics[width=\textwidth]{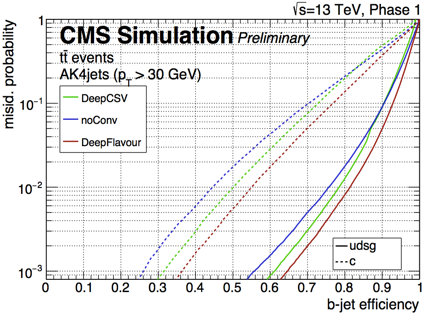}
        \par\medskip
        \caption{Performance comparison of \textsc{DeepFlavour}, \\\textsc{DeepCSV}, and \textsc{noConv}. \cite{CMS-DP-2017-013}}
        \label{fig:1_bottom}
    \end{subfigure}
    \caption{b-jet identification: True positive rate (b-jet efficiency) versus false positive rate (misidentification probability) for c-jets, and uds- and gluon-jets of simulated events {of top quark pair production}, requiring a minimal transverse momentum of 30 GeV to the considered jets.} %~\cite{btagrocsource}.}
    \label{fig:1}
\end{figure}

\subsection{Performance comparison}

\medskip
%\smallskip
\noindent
Fig.~\ref{fig:1_top} shows a significant improvement in performance between \textsc{DeepCSV} and \textsc{CSVv2}. For example, for the same true positive rate (b-jet efficiency) of 65\%, \textsc{DeepCSV} offers a 40\% reduction in false positive rate (misidentification probability) for light jets (uds- and gluon-jets). The 1\% false positive rate is a typical working point used for the classification. This result refers to simulated event samples {of top quark pair production}; however, this gain in the performance has been validated in real collision data. {The latter was made possible via the comparison of the data-to-simulation scale factors---that refer to the b-tagging efficiency measurement with the use of several different methods---between \textsc{DeepCSV} and \textsc{CSVv2}, which shows agreement~\cite{CMS-DP-2017-005}, thus implying that the observed improvement in performance in simulation is additionally reflected on the data part.}

\medskip
\smallskip
\noindent
Fig.~\ref{fig:1_bottom} demonstrates the further significant gain in the performance of \textsc{DeepFlavour} with respect to \textsc{DeepCSV}. For example, for the same true positive rate of 78\%, \textsc{DeepFlavour} offers an almost 40\% reduction in false positive rate for light jets. We also see that \textsc{noConv}, which is an algorithm with the same structure and input as \textsc{DeepFlavour}, but trained without the convolutional layers (only for comparison), provides an even worse result than \textsc{DeepCSV}. This indicates that a larger input set of features alone is not able to increase the performance of the NN; on the contrary, it can even degrade the overall discrimination. The choice of a sophisticated architecture, {\em i.e.} the addition of convolutional layers in this case, which help exploiting the structure of the input (jet), is what provides sufficient information for the NN to perform as expected, when combined with a larger number of input variables.

%\iffalse
\begin{figure}
    \centering
    \begin{subfigure}[H]{0.44\textwidth}
        \includegraphics[width=\textwidth]{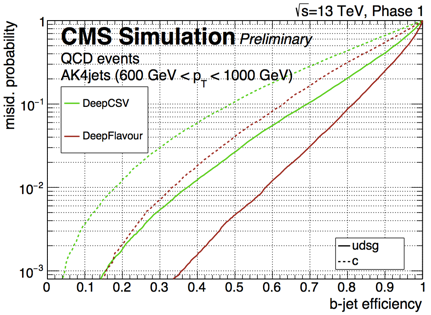}
%        \par\medskip
        \par\medskip
        \caption{
        %Comparison of the true positive rate (b-jet efficiency) versus false positive rate of DeepCSV and DeepFlavour for c-jets, and uds- and gluon-jets in simulated events, requiring a transverse momentum between 600 and 1000 GeV.
        b-jet identification: True positive rate (b-jet efficiency) versus false positive rate for c-jets, and uds- and gluon-jets of simulated events {of QCD production}, requiring a transverse momentum between 600 and 1000 GeV. Performance comparison between \textsc{DeepCSV} and \textsc{DeepFlavour}. \cite{CMS-DP-2017-013}
        }
        \label{fig:2_top}
    \end{subfigure}~~
\hspace{1em}
    \begin{subfigure}[H]{0.47\textwidth}
    \captionsetup{skip=5pt}
        \includegraphics[width=\textwidth]{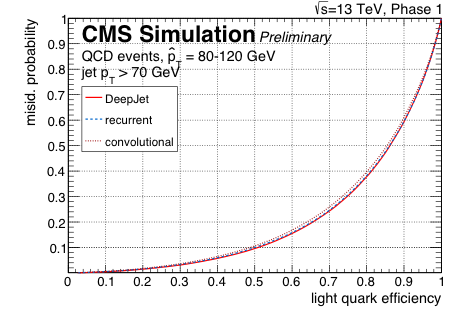}
%        \par\medskip
      \par\medskip
        \caption{
        %Comparison of the true positive rate (light-jet efficiency) versus false positive rate of DeepJet and recurrent and convolutional NNs, for gluon-jets in simulated events requiring a minimal transverse momentum of 70 GeV. 
        Quark/gluon discrimination: True positive rate (light-jet efficiency) versus false positive rate for gluon-jets of simulated events {of QCD production}, requiring a minimal transverse momentum of 70 GeV. Performance comparison between \textsc{DeepJet} and ‘recurrent’, ‘convolutional’. \cite{CMS-DP-2017-027}}
        \label{fig:2_bottom}
    \end{subfigure}
  %  \newline
%\hfill \break
%\par\bigskip
%\par\medskip
    \caption{
    %Comparisons of the performance of different NN for b-jet identification at high transverse momentum.
    Comparison of the performance of different algorithms for b-jet identification and quark/gluon discrimination}
    \label{fig:2}
\end{figure}
%\fi

\medskip
\smallskip
\noindent
Fig.~\ref{fig:2_top} shows the gain in performance of \textsc{DeepFlavour} over \textsc{DeepCSV} at very high values of transverse momentum of the b-jets, which implies a major gain in the sensitivity of physics analyses targeting highly energetic b-jets in the final state. For example, for a true positive rate of 37\%, there is an almost 90\% reduction in false positive rate when using the \textsc{DeepFlavour} tagger.

\medskip
\smallskip
\noindent
Finally, as part of the investigation of \textsc{DeepJet}'s capability to perform quark/ gluon discrimination, Fig.~\ref{fig:2_bottom} shows the comparison between \textsc{DeepJet} and each one of two reference approaches, namely the ‘convolutional’ and the ‘recurrent’ one. The ‘convolutional’ approach, which involves the use of 2D convolutional layers working on ‘jet images’, as in~\cite{Komiske2017}, stems from the idea of considering the calorimeter cells as image pixels so as to be able to apply techniques already implemented within research in computer vision. More specifically, the jet is treated as an image in the $\eta - \phi$ plane of the detector; the continuous particle positions are pixelised, and for each pixel the intensity is provided by the corresponding energy deposits in the calorimeter; additionally, the RGB colour is determined by the relative transverse momenta of the charged and neutral jet constituents and by the charged particle multiplicity. The ‘recurrent’ approach, inspired by~\cite{Louppe2019}, is a slimmed-down version of \textsc{DeepJet}, given that for light quark/gluon discrimination only a fraction of the initial input is relevant. Therefore only four features are used per particle (relative transverse momentum, $\eta$, $\phi$, and the so-called ‘pile-up per particle identification’ weight~\cite{Bertolini2014}); secondary vertex information is removed, and there are no 1$\times$1 convolutional layers. After training all three NN with the same samples, we observe that \textsc{DeepJet} and recurrent NN perform similarly well and marginally better than the convolutional NN. The convolutional NN is already expected to be performant in this case, as this kind of discrimination mostly relies on particle and energy densities, which may be well represented through an image related approach, in contrast to the significantly more complex case of heavy flavour tagging. \textsc{DeepJet}'s capability to achieve quark/gluon discrimination is further established by the considerable gain in performance it offers when compared to the ``quark/gluon likelihood'' discriminator~\cite{CMS-PAS-JME-16-003}, a binary quark/gluon classifier that is included in the CMS reconstruction framework, as described in~\cite{Bols2020}.

\subsection{Summary}

\medskip
%\smallskip
\noindent
The above-described taggers \textsc{DeepCSV} and \textsc{DeepFlavour}/\textsc{DeepJet} significantly outperform the standard b-jet tagger previously used in CMS, thus offering a major gain in the sensitivity for physics analyses involving b-quark jets, including new physics searches and precision measurements. At the same time, their ability to implement multi-class classification extends their use to generic heavy-flavour (b- or c-quark jet) tagging, and also to quark/gluon discrimination, in which regard \textsc{DeepJet} exhibits satisfying performance when compared to approaches sharing the same goal. Broadening the feature selection by increasing the number of input variables that describe the jet constituents, applying a new machine learning algorithm with a deeper NN that also exploits the jet structure as it being an image, and using a larger and more diverse training sample that prevents building a process-specific tagger, constitute the main factors responsible for the observed advantage in performance. This result was validated on real collision data, indicating an equivalent gain in performance to the one estimated in the simulated samples. These taggers are the currently recommended ones for multiple studies carried out within the CMS Collaboration, and have already been used to produce a number of competitive physics results. CMS analyses that made use of the \textsc{DeepCSV} and \textsc{DeepJet} taggers include not only ones involving new physics searches, but also ones performing precision measurements, with~\cite{search1,search2,search3,search4,search5,search6,search7} and~\cite{measurement1,measurement2,measurement3,measurement4,measurement5,measurement6,measurement7}, respectively, serving as a few such examples.

\section{Improvements and Applications of the Matrix Element Method}\label{sec:mem}

\medskip
%\smallskip
\noindent
Machine learning techniques employed at the LHC typically rely on the presence of large sets of training data for optimization purposes. The Matrix Element Method (MEM) takes a different approach, and provides a way to approximate the likelihood function for parameters in the SM given observed data. This calculation is performed from first principles, without the need for training. It was first used by the D0 Collaboration for a top quark measurement~\cite{Abazov:2004cs}, with the original proposal provided by Kunitaka Kondo~\cite{Kondo:1988yd}. The method can also be used to discriminate between different collision processes by providing powerful observables in searches for rare signals. Ratios of likelihoods describing the probabilities that observed events be consistent with signal or background processes are used in this context.
The \textsc{MoMEMta} software package~\cite{Brochet2019,Deliverable_3_1,Deliverable_3_2,Deliverable_3_3}  was developed with contributions from the AMVA4NewPhysics Network to provide a convenient framework for calculating MEM likelihoods for LHC applications.

\subsection{The Matrix Element Method}

\medskip
%\smallskip
\noindent
Let $q_1$, $q_2$ be the momentum fractions of the initial state partons, and $y$ the kinematics of the partonic final state. The differential cross section d$\sigma_{\alpha}(y)$, a function of the parton configuration $y$, is obtained by integrating the differential cross section d$\sigma_{\alpha}(q_1, q_2, y)$ for the process $\alpha$ over the possible initial state parton configurations, weighted by the parton distribution functions (PDFs) of the colliding partons. The so-called transfer function $T(x \vert y)$ is the probability density for reconstructed event kinematics $x$, given a parton configuration $y$. It provides an approximate expression to capture effects from parton shower, hadronization, and detector reconstruction. Reconstruction efficiency effects can also be modelled with an appropriate efficiency $\epsilon(y)$. The probability density for observing an event $x$, given a hypothesis with parameters $\alpha$, is given by 
%\vspace{1mm}
\begin{center}
    $P(x \vert \alpha)=\frac{1}{ \sigma_{\alpha}^{\tiny {\textrm{vis}} } } \int\limits_{q_1, q_2} \sum\limits_{a_1, a_2}\int\limits_y \textrm{d}\Phi(y) \,\, \textrm{d}q_1 \,\, \textrm{d}q_2 \,\, T(x \vert y) \,\, \epsilon(y) \, \times$
\end{center}
\begin{equation}\label{eq_mem_probability}
    \times \, f_{a_1}(q_1) \,\, f_{a_2}(q_2) \,\, {\vert M_{\alpha}(q_1,q_2,y) \vert}^2\,.
\end{equation}
%\vspace{1mm}
In the above expression $f_{a_i}(q_i)$ are the PDFs for a given flavour $a_i$ and momentum fraction $q_i$, $i=1, 2$, ${\vert M_{\alpha}(q_1,q_2,y) \vert}^2$ is the squared matrix element for the process $\alpha$, d$\Phi(y)$ the n-body phase space of $y$, and $\sigma_{\alpha}^{\tiny {\textrm{vis}} }$ is 
a normalization factor. The integral result alone, namely the above quantity without the normalization factor, is referred to as the matrix element weight, $W(x \vert \alpha)$. It is also commonly used.

\medskip
\smallskip
\noindent
Eq.~\ref{eq_mem_probability}, from which the most probable value of theory parameters can be estimated through likelihood maximization, involves an integration typically performed with Monte Carlo methods.
This requires the evaluation of the matrix element ${\vert M_{\alpha}(q_1,q_2,y) \vert}^2$, which contains the (theoretical) information on the hard scattering, and can be computationally expensive to evaluate.
The integrand can vary by many orders of magnitude in different regions of the phase space, necessitating appropriate choices for the parameterization of the integral to ensure computational efficiency.
The design of the \textsc{MoMEMta} framework makes it easy to find suitable parameterizations, which can help overcome the computational hurdle traditionally associated with the MEM.

\subsection{MoMEMta framework}

\subsubsection{Main implementation aspects}

\medskip
%\smallskip
\noindent
The \textsc{MadWeight} package~\cite{Artoisenet2010} introduced a general way to approach the problem of finding an efficient phase space parameterization for integration purposes. This includes the removal of sharp peaks in the integrand, for example due to resonances.
\textsc{MadWeight} is no longer supported, and its lack of flexibility hinders its application. \textsc{MoMEMta}~\cite{Brochet2019} has been designed to build upon the ideas of \textsc{MadWeight}. It is a modular C++ software package, introduced to compute the convolution integrals at the core of the method. Its particular modular structure provides the required flexibility, allowing it not only to extend its applicability beyond its use in smaller programs so as to cover the needs of the complex LHC analysis workflows that handle large amounts of data, but also to be open to specific optimizations in the integration structure or engine. At the same time, since the MEM may be used in both theoretical and experimental high-energy physics problems, with different purposes and use cases in each field, \textsc{MoMEMta}’s modular structure constitutes a significant advancement.
In terms of accuracy and CPU time, the performance of \textsc{MoMEMta} is similar to that of \textsc{MadWeight}, because they rely on the same algorithmic approach of phase-space parameterization.
\textsc{MoMEMta} is however further designed to adapt to any process and allows for implementation in any C++ or Python environment, offering more freedom to the user, who can therefore wrap new modules that handle specific tasks, while benefiting from the existing
features of the framework.

\subsubsection{Modules and blocks}

\medskip
%\smallskip
\noindent
The functionality of modules provided in \textsc{MoMEMta} include representing and evaluating the matrix element and parton density functions, as well as the transfer functions, performing changes of variables, and handling the combinatorics of the final state. This implies that when calculating the probability to be assigned to the experimental events, every term in Eq.~\ref{eq_mem_probability} may be treated as a separate, user-configured module within this framework. The weights for a given process $W(x \vert \alpha)$ are computed by calling and linking the proper set of modules in a configuration file. This computation usually requires the evaluation of multi-dimensional integrals via adaptive Monte Carlo techniques, whose efficiency depends on the phase-space mapping that is used. Such parameterization can be optimized by using a finite number of analytic transformations over subsets of the integration variables, called ‘blocks’. Some of these blocks are also responsible for removing degrees of freedom by imposing momentum conservation, while the rest merely constitute the corresponding change of variables. Due to the potentially large combinatorial ambiguity in the assignment between reconstructed objects and partons, there exists a dedicated module that provides for the averaging over all possible permutations of a given set of particles. The functionality of this module, as opposed to a simple averaging over the possible assignments, allows the adaptive integration algorithms to focus on the assignments contributing most to the final result, thus increasing the precision of the result. This novel feature can potentially significantly speed up the computation, as the evaluation of the matrix element is what actually dominates the computation time.

\subsection{MEM use cases and MoMEMta application}

\medskip
%\smallskip
\noindent
The MEM has proven to be an excellent technique to address two of today's main tasks in HEP analysis: signal-background discrimination and parameter estimate. In the former, the weights $W(x \vert \alpha)$ computed under different hypotheses are used to build a discriminating variable; in the latter, the MEM weights are instead used to build a likelihood function, which is then maximized in order to estimate the parameters of interest. Given its ability to efficiently compute the integral in Eq.~\ref{eq_mem_probability}, \textsc{MoMEMta} meets the needs for both these MEM use-case categories. Examples of signal-background discrimination using the \textsc{MoMEMta} framework can be found in~\cite{Brochet2019}, ranging from cases with low level of complexity (where the final state is precisely reconstructed with detectable particles) to cases with a high-multiplicity final state containing unobserved objects, where a careful consideration of the several degrees of freedom involved is required. In this section, a proof of principle for performing parameter estimation using \textsc{MoMEMta} and an example of signal extraction with the MEM in a LHC analysis are reported.

\subsubsection{Statistical inference in SMEFT using MoMEMta}

\begin{figure}[t]
    \centering
    \includegraphics[scale=.28]{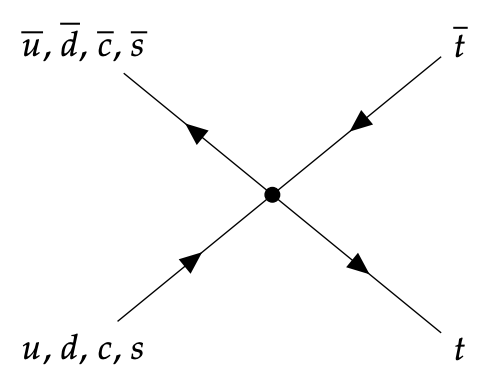}
    \includegraphics[scale=.42]{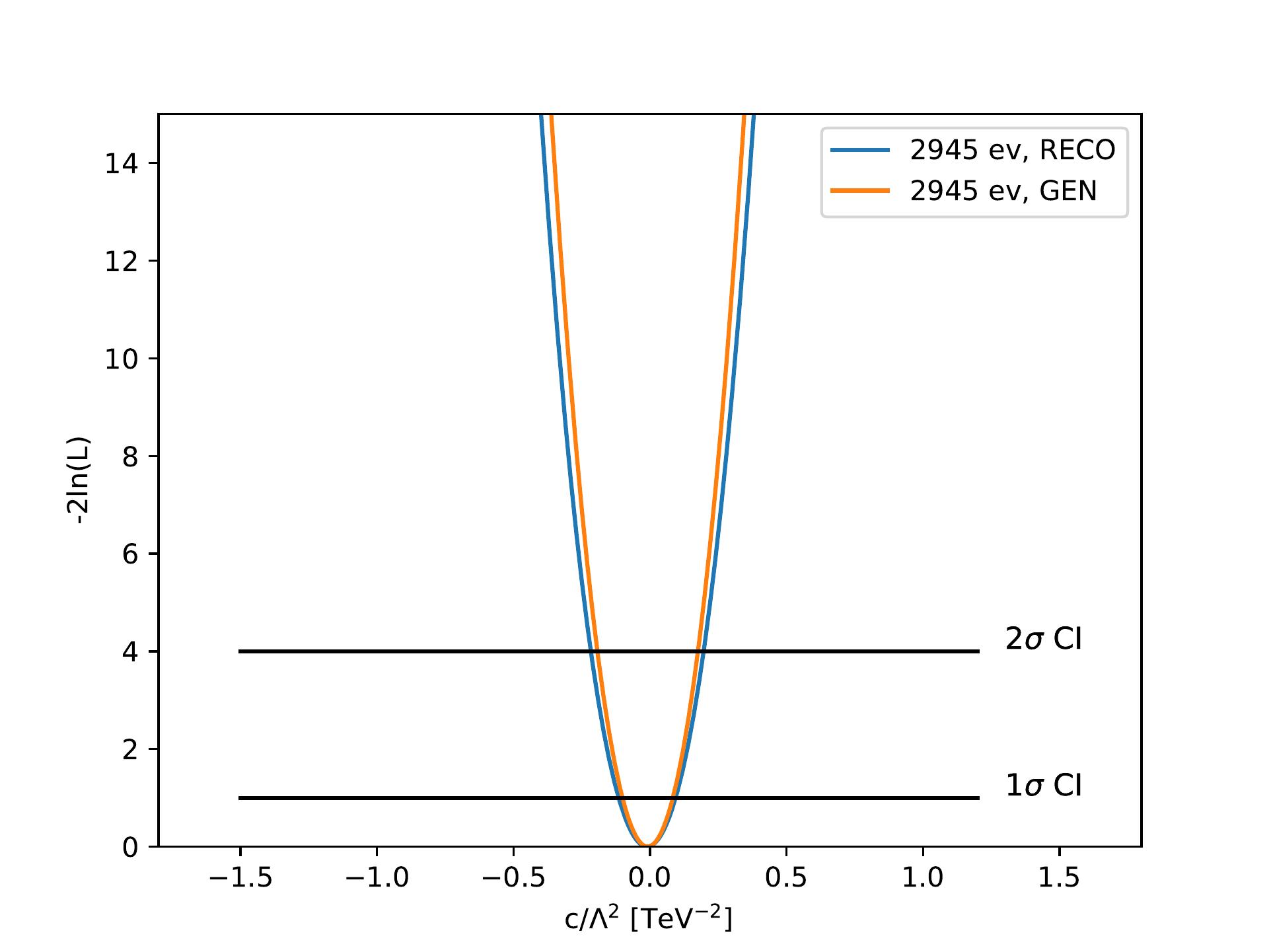}
    \caption{Left: Example of $t\bar{t}$ process production at LO, where the vertex between the quark--antiquark pair and the top pair is described by a dimension-6 effective operator. Right: Negative log-likelihood as a function of $c/\Lambda^2$ built on generator-level (orange line) and detector-level (blue line) events. The black lines define the $1\sigma$ and $2\sigma$ confidence intervals.}
    \label{fig:eftmem}
\end{figure}

\medskip
%\smallskip
\noindent
In the Standard Model Effective Field Theory (SMEFT)~\cite{BUCHMULLER1986621,Hartland:2019bjb}, the effects of new heavy
particles with typical mass scale $M \approx \Lambda$ on SM fields can be parameterized at a lower energy $E \ll \Lambda$ in a model-independent way in terms of a basis of higher-dimensional operators. In this work, we consider the operator
$\mathcal{O}_{Qq}^{11}$, which modifies the coupling between top quarks and light quark--antiquark pair in top quark pair ($t\bar{t}$) production, as displayed in Fig.~\ref{fig:eftmem} (left). The \textsc{MoMEMta} framework is used to estimate the quantity $c_{Qq}^{11}/\Lambda^2$ in a $t\bar{t}$ simulation sample, with $c_{Qq}^{11}$ (referred to as $c$ in the following
for shortness) being the degree of freedom associated to the $\mathcal{O}_{Qq}^{11}$ operator. A fully-leptonic $t\bar{t}$ simulation sample is produced with \textsc{MadGraph\_aMC@NLO} version 2.6.5~\cite{Alwall:2014hca} in the di-muon final state at LO in QCD precision and with corrections up to $1/\Lambda^2$ at the amplitude level,
with the quantity $\Lambda$ set to 1~TeV. The events are then showered with \textsc{Pythia}~8.212~\cite{Sjostrand:2014zea}
and the simulation of particle interactions with the CMS detector is performed with \textsc{Delphes}~3.4.1~\cite{deFavereau:2013fsa}.
The contribution of the SMEFT term to the matrix element $M$ of the process can be broken down into three parts: a SM contribution ($A_{SM}$), a quadratic contribution for the dimension-6 operator only ($A_{quad}$), and an interference term between the two ($A_{int}$). This translates into:
\begin{equation}
    \vert M_{SMEFT} \vert ^2 = \vert M_{SM} + c M_{NP}\vert ^2 = A_{SM} + c A_{int} + c^2 A_{quad}\,.
\end{equation}
With this parametrization, the integral in Eq.~\ref{eq_mem_probability} can be written, for a given $c$, as the sum of three separate integrals:
\begin{equation}
    P(x \vert c) = \frac{1}{\sigma_c} (W_{SM} + c W_{int} + c^2 W_{quad})\,,
\end{equation}

\medskip
%\smallskip
\noindent
with $\sigma_c$ being the visible cross section of the process, in turn similarly parameterized as a function of $c$. The three ME weights are computed with \textsc{MoMEMta} using Gaussian transfer functions on the energies of the visible particles with a standard deviation of 5\% for leptons and 10\% for jets. An additional dimension of integration is introduced in order to handle the combinatorial ambiguity in the assignment between reconstructed final-state b jets and b quarks in the matrix element.

\medskip
%\smallskip
\noindent
A negative log-likelihood function is then built from $P(x \vert c)$ while performing a scan over $c$. The resulting function is shown in Fig.~\ref{fig:eftmem} (right) in blue. The same function excluding the detector effects is also computed and represented in orange.
The estimated quantity $\frac{\hat{c}}{\Lambda^2}$ on events including detector effects is measured to be -0.013~TeV$^{-2}$ with a $2\sigma$ confidence interval [-0.233,~0.210]~TeV$^{-2}$. It is to be noted that in a complete study one would have to take into account also systematic uncertainties that have been neglected here, where only the statistical effect plays a role.

\medskip
\smallskip
\noindent
Finally, two main considerations can be drawn. The two curves in Fig.~\ref{fig:eftmem} show very similar width of the likelihood profiles, proving the strength of the MEM where detector effects are encoded in the computation of the ME weight. Moreover, this method represents a valid alternative to cases where the SMEFT coefficients are estimated one at a time, since in the MEM the maximization of the likelihood can easily be multi-dimensional.
More detail on this study is available in~\cite{Saggio:2019vjo}.

\subsubsection{\texorpdfstring{$\textrm{t}\,\bar{\textrm{t}}\,$}{}H production}

\medskip
%\smallskip
\noindent
The search for Higgs production in association with top quark pairs, $t\bar{t}H$ with $H\rightarrow b\bar{b}$, is a particularly interesting application for the MEM.
It was first studied in~\cite{Artoisenet2013}, and since then has been used extensively by the ATLAS and CMS experiments~\cite{Aad2015a,Aaboud:2017rss,Khachatryan2015,Sirunyan2018,Sirunyan2019}.
The $t\bar{t}H(b\bar{b})$ process has many partons in the final state.
Its main background in final states with at least one charged lepton is $t\bar{t}+b\bar{b}$, which features identical final state partons to the signal.
Discrimination between these processes relies on small differences in kinematics.

\medskip
\smallskip
\noindent
ATLAS combines the MEM with additional multivariate techniques in a search for $t\bar{t}H(b\bar{b})$ with 36 fb$^{-1}$ of data~\cite{Aaboud:2017rss}.
Details of the MEM implementation developed for this analysis and studies of its performance can be found in~\cite{Held:2019gix}.
The following provides a brief summary of the approach chosen for this analysis.
A discriminant, MEM$_{D1}$, is calculated as the logarithm of the ratio of signal and background likelihoods, where the signal is $t\bar{t}H(b\bar{b})$ and the background is $t\bar{t}+b\bar{b}$: MEM$_{D1}= \log_{10}(L_S / L_B)$.
The $t\bar{t}+b\bar{b}$ contribution to the background is dominant in the phase space where the discriminant is used.
Matrix elements are calculated with \textsc{MadGraph5\_aMC@NLO}~\cite{Alwall:2014hca} at leading order; the lack of higher order corrections can at most decrease the performance of the method, but it does not bias the physics. Only gluon-induced Feynman diagrams are considered, which reduces computational time without a significant impact on discrimination power.

\medskip
\smallskip
\noindent
The MEM is used for final states with one charged lepton, where six final state quarks are expected at leading order, alongside the one charged lepton and a neutrino. Directions of all visible partons are assumed to be measured exactly by the ATLAS detector, so the associated transfer function components are $\delta$-distributions. After imposing transverse momentum conservation, seven degrees of freedom remain for the integration. The integration variables are chosen to be the energies of all six final state quarks and the neutrino momentum along the beam direction. The integration itself is performed with \textsc{VEGAS}~\cite{1978JCoPh..27..192L}, based on an framework described in~\cite{Schouten:2014yza}.

\begin{figure}[ht]
 %   \centering
    %\includegraphics[scale=.35]{FIGS/MEM/ttHbb_MEM}
        \includegraphics[scale=.35]{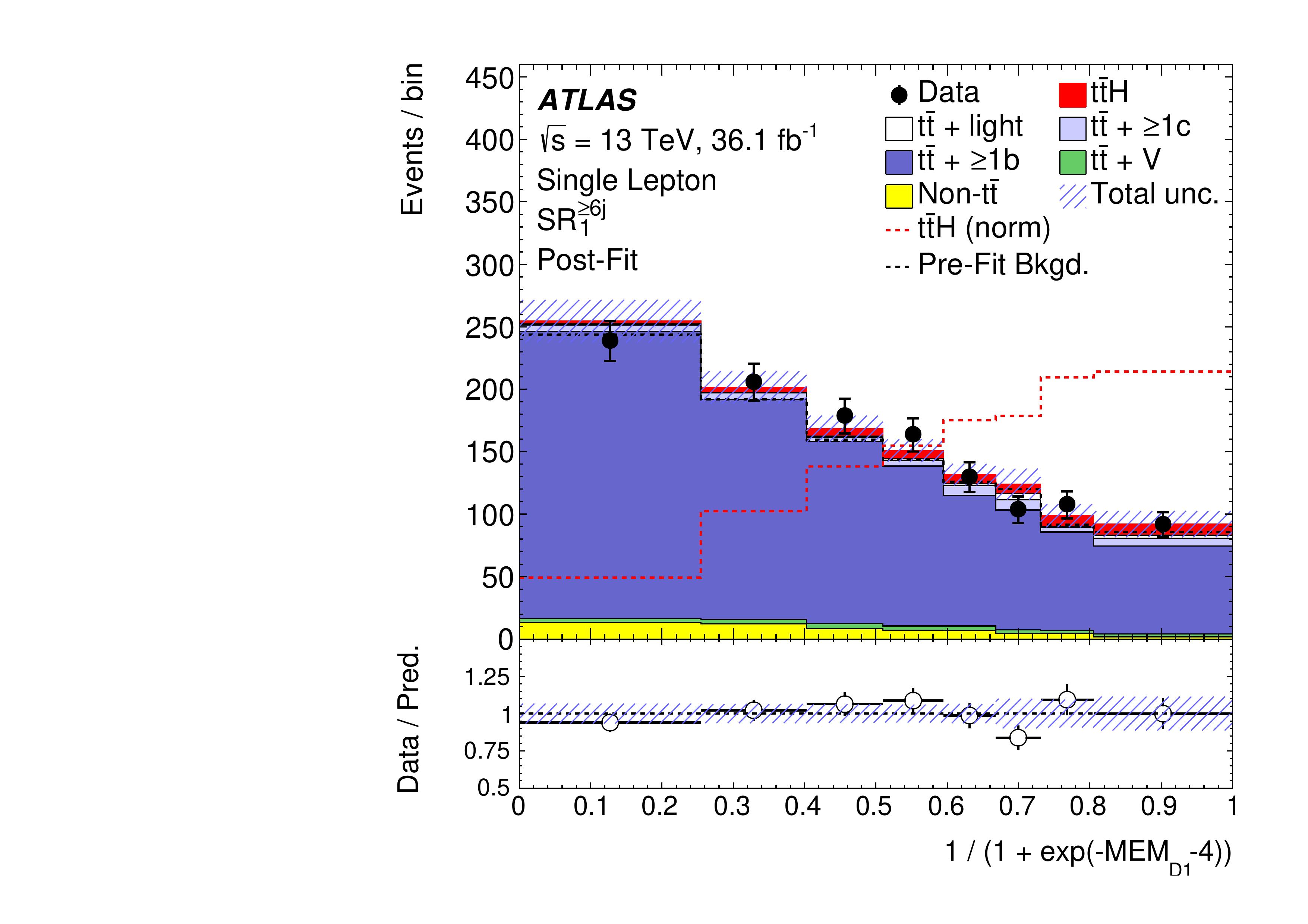}

    \caption{MEM-based discriminant used by ATLAS in the search for $t\bar{t}H(b\bar{b})$. The red dashed histogram shows the normalized density expected from ttH events alone; black points represent ATLAS data, and full histograms the various signal (red) and background contributions~\cite{Aaboud:2017rss}.}
    \label{fig:ttHbb_MEM}
\end{figure}

\medskip
\smallskip
\noindent
Fig.~\ref{fig:ttHbb_MEM} visualizes the MEM$_{D1}$ discriminant, using a sigmoid to map the values into the $[0, 1]$ interval. The data are found in good agreement with the expected distribution, and the discrimination power of the method is seen by comparing the normalized distribution of $t\bar{t}H$ (dashed line) to the background contributions.

\subsection{Summary}

\medskip
%\smallskip
\noindent
The MEM provides a method for evaluating the likelihood of collision events under different hypotheses from first principles. It has applications in parameter measurements and searches for specific collision processes.
The \textsc{MoMEMta} software package provides a flexible implementation of the calculations required. It offers flexibility with its modular structure, and enables use of the MEM for a wide range of applications at the LHC.
The modularity also allows for extensions to handle novel applications, while taking advantage of the optimizations and convenience features provided by \textsc{MoMEMta}.

%\clearpage
\section{New Statistical Learning Tools for Anomaly Detection}\label{sec:anomalyDetection}

\subsection{Overview}

\medskip
%\smallskip
\noindent
Searches for new physics at the LHC proceed by comparing the data collected by the detectors with simulated data sets obtained from software simulations of the underlying physics process, interfaced with a simulation of the detector response. The simulated data describe either the processes predicted by the standard model (\textit{background processes}) or the processes postulated by the particular theoretical extension of the SM under study (\textit{signal processes}). These searches may be broadly categorized into \textit{model-dependent} searches, where the data are compared to both the SM and new physics predictions, and \textit{model-independent} searches, where the data are compared only to the SM in order to search for unexplained deviations from it: these deviations (\textit{anomalies}) may then be attributed to new physics and investigated further. In this context, the term \textit{model-independent} refers to independence with respect to a new physics signal model: the SM background model is always assumed.

\medskip
\smallskip
\noindent
Model-dependent searches are typically approached by producing simulated events for both SM and new physics processes, and hypothesis tests are devised to find which model is favoured by the data. 
In case of model-independent searches, no particular model for the signal processes is assumed: a simulated dataset describing the signal processes cannot therefore be produced. These searches can be approached as anomaly detection problems, where the data are combed to find any observation that is not consistent with the background model. This setting is an example of semi-supervised learning, given that the two data categories involved are a ‘simulated’ dataset, generated by Monte Carlo techniques to represent the known background process and therefore considered as ‘labelled’, and the ‘experimental’ data sample, which is generated by an {\em a-priori} unknown mechanism---possibly comprising contributions from both background and signal processes---and therefore considered as ‘unlabelled’. The presence of a signal in the experimental data is generally inferred through the observation of a significant deviation from the predictions for the background process. 

\medskip
\smallskip
\noindent
Let $Y=(\mathbf{y}_1, \ldots \,, \mathbf{y}_m)', \, \mathbf{y}_l \in \textrm{R}^P,\, l=1, \ldots \, , m$, be the experimental data, with $\mathbf{y}_l$ independent and identically distributed realizations of the random vector with unknown probability density function $f_{BS}$\,$: \textrm{R}^P$\,$\to \textrm{R}$. In addition to the experimental data $Y$, it is possible to generate with the use of Monte Carlo simulations a large sample $X=(\mathbf{x}_1, \ldots \,, \mathbf{x}_n)', \, \mathbf{x}_i \in \textrm{R}^P,\, i=1, \ldots \, , n$. While Monte Carlo simulations in HEP are computationally taxing, a simulated sample of SM events is employed in so many different analyses that the experiments can afford to simulate tens or hundreds of millions of events for SM scenarios: the sample $X$ can therefore be considered arbitrarily large to all practical extents. We assume that the simulated data, as well as the majority of the experimental data, which are known to have been generated by a background process, follow a distribution described 
by the probability density function $f_{B}$\,$: \textrm{R}^P$\,$\to \textrm{R}$. The remaining experimental data have been 
possibly generated by an unknown signal process described by $f_{S}$\,$: \textrm{R}^P$\,$\to \textrm{R}$. In absence of quantum mechanical interference between the signal and the background, the generating mechanism of the experimental data $f_{BS}$ may be thus specified as a mixture model: 

\begin{equation}\label{eq:f_bs}
f_{BS}(\mathbf{y})=(1-\lambda)f_{B}(\mathbf{y})\,+\,\lambda f_{S}(\mathbf{y}),\,\, \lambda \in [0,1)\,,
\end{equation}

\medskip
%\smallskip
\noindent
and the problem may be cast in terms of either parameter estimation, where inference is sought on the value of $\lambda$, or of hypothesis testing, where a simple null hypothesis $\lambda=0$ is tested against a composite alternative $\lambda\neq 0$.

\subsection{Detecting anomalies via hypothesis testing: The Inverse Bagging algorithm}

\medskip
%\smallskip
\noindent
The Inverse Bagging (IB) algorithm~\cite{Vischia2017,Deliverable_4_4,Deliverable_4_5}, developed within the fourth pillar of the AMVA4NewPhysics research program, addresses the problem of anomaly detection by means of hypothesis testing. We formulate the null hypothesis that the processes having generated the experimental data follow the same distribution as the ones corresponding to the simulated data. We then perform a statistical test to quantify how likely it is that the unlabelled data have been indeed generated by the background processes alone.  This algorithm combines hypothesis testing with multiple data sampling, as a means of iterative tests of the properties of the data and the possible presence of unknown signals.

\subsubsection{Hypothesis testing and multiple sampling}

\medskip
%\smallskip
\noindent
Once the null hypothesis is defined as $H_{0}$\,$: f_{BS}(\cdot)=f_{B}(\cdot) \Leftarrow \lambda=0$ and the alternative as $H_{1}$\,$: f_{BS}(\cdot) \neq f_{B}(\cdot) \Leftarrow \lambda\neq 0$, the IB algorithm proceeds by performing a two-sample test on each of $B$ pairs of bootstrap replicas $X_b^{\small{\textrm{*}}}$ and $Y_b^{\small{\textrm{*}}}$, $b = \{1, 2, \ldots \,, B\}$, taken from the data samples $X$ and $Y$, respectively. The size $Q$ of the bootstrap replicas is set to be significantly smaller than the size of the experimental sample under study. The results of the tests are used to classify how anomalous are the individual observations, improving on the insights that may be offered by a standard hypothesis test performed on the original data sets $X$ and $Y$.

\medskip
\smallskip
\noindent
The test statistic associated to the pair  of $X_b^{\small{\textrm{*}}}$ and $Y_b^{\small{\textrm{*}}}$, $\mathbf{T}_l=\left(T_{l1}, \ldots \,, T_{lB_l} \right)'$, $\,l=1, \ldots \,, m$, is considered as part of the useful information associated with each of the observations that are contained in $Y_b^{\small{\textrm{*}}}$. The set of test statistics that concern the same observation, regardless of the specific bootstrap samples they have been computed in, is summarized into an observation score (anomaly detection metric) summarizing how anomalous are the properties of the particular observation. The details of the computation of the anomaly detection metric from the set of test statistic values for each event are given in Section~\ref{sec:ib:resquestions}. The underlying rationale is that if an observation $\mathbf{y}_l$ has been generated by $f_{S}$, then the bootstrap samples that include $\mathbf{y}_l$ will lead to reject $H_0$ more often compared to the samples not including 
it, if the size $Q$ of the sample is small enough. Using small sizes $Q$ would result in samples where an anomaly results more easily in a sizeable deviation from the background distribution. The anomaly detection metric therefore reflects how likely it is for each observation to have been generated by a signal process, and can be hence used for further classification purposes, with observations carrying the most extreme score becoming candidates to be classified as a signal. By defining a sliding-window threshold on the value of the anomaly detection metric for classifying an observation as signal, we obtain a Receiver Operating Characteristic (ROC) curve~\cite{ROC}, which describes the purity of the classifier as a function of its efficiency. 

\begin{figure}[ht]
\centering
  \includegraphics[width=0.49\linewidth,clip]{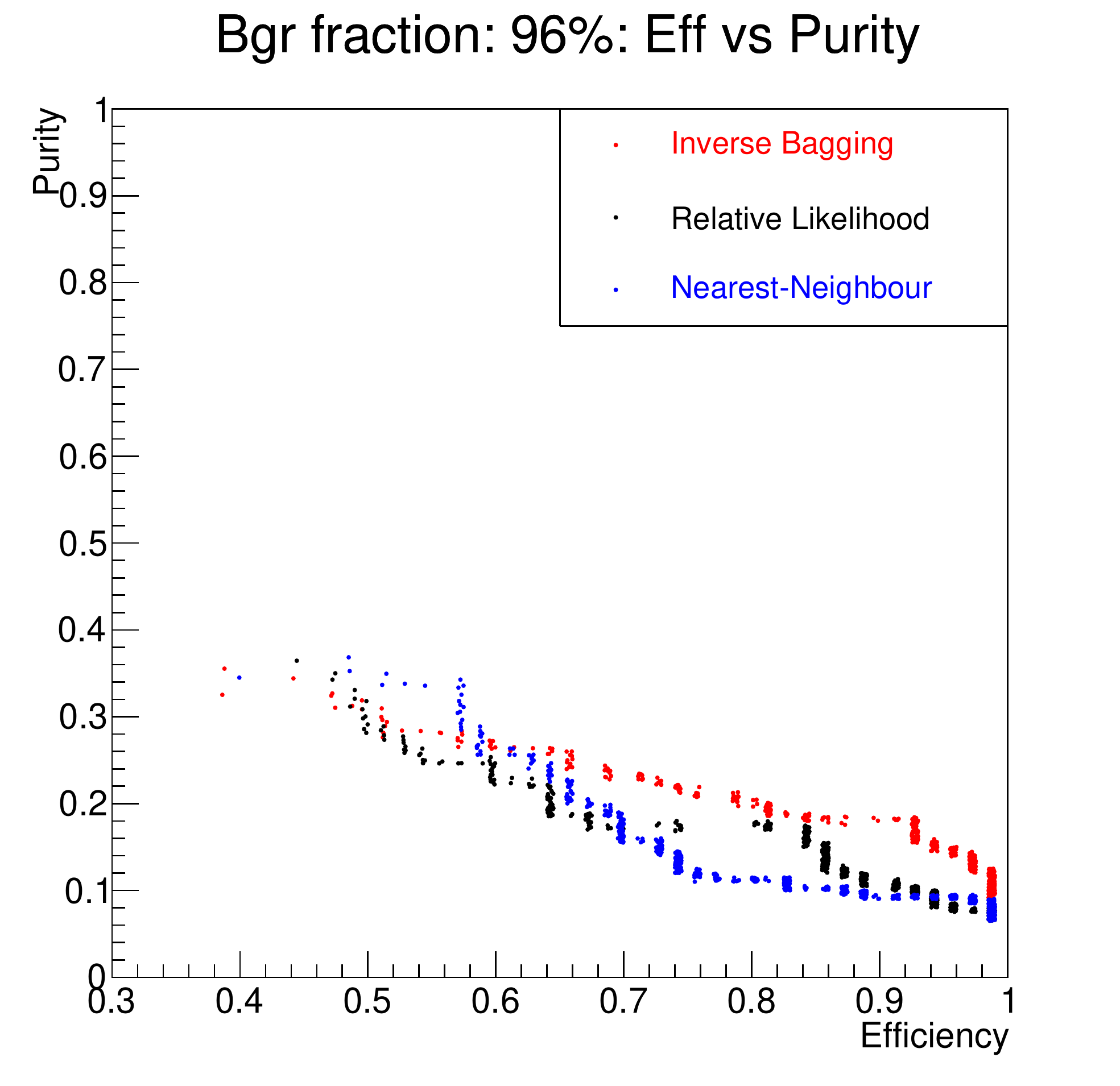}
  \includegraphics[width=0.49\linewidth,clip]{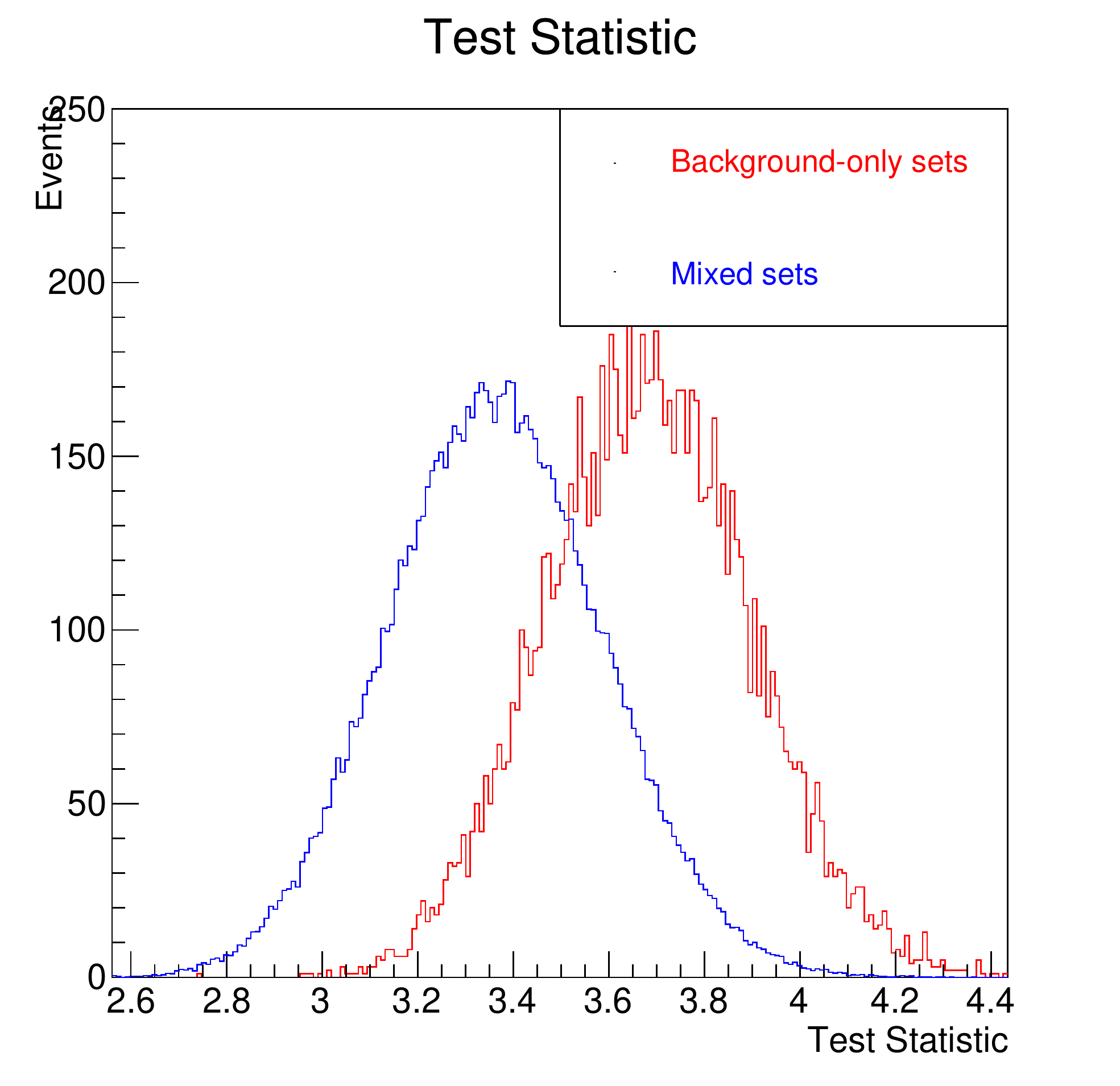}\\

\caption{
{\it Left:} Purity of the discriminators as a function of their efficiency. The simulated data sample is composed by a 96\% background and 4\% signal. {\it Right:} The values of the test statistic (score) for the background-only sets and for the mixed signal+background sets~\cite{Vischia2017}. The performance is calculated on the \textsc{HEPMASS} dataset~\cite{baldietal} as described in the text.
}         
  \label{fig:IB_1}
\end{figure}

\medskip
\smallskip
\noindent
In Refs.~\cite{Vischia2017,Deliverable_4_4,Deliverable_4_5}, the IB algorithm was applied to the \textsc{HEPMASS} dataset~\cite{baldietal}, available at \texttt{http://archive.ics.uci.edu/ml/datasets/HEPMASS}: the dataset consists in a background composed by standard model top quark production and a signal composed by a heavy resonance that decays into a top quark pair that would result in a spectrum different from the SM for any observable involving the full final state (e.g.~the reconstructed visible mass of the top pair decay products).
Refs.~\cite{Vischia2017,Deliverable_4_4,Deliverable_4_5} contain the details on the input features that were used. Fig.~\ref{fig:IB_1} (left) shows a purity versus efficiency curve for the IB algorithm, as well as for two reference classifiers (relative likelihood, and $k$-nearest neighbours) that use only event-based information to classify events. In the considered application, the IB classifier outperforms both. The test statistic used by the IB algorithm has a different distribution for a background-only set of events and for a mixed signal+background set of events ($\lambda=0.04$), as illustrated in  Fig.~\ref{fig:IB_1} (right).

\medskip
\smallskip
             
\subsubsection{Validation of the algorithm and research questions}
\label{sec:ib:resquestions}
\medskip
%\smallskip
\noindent
We took into account the following research questions in order to further validate the algorithm with respect to its original publication:

\begin{description}
\item[Choice of anomaly detection metric]
A key step in the IB algorithm is the score computation, which is our anomaly detection metric. 
A meaningful metric is crucial for inferring the likelihood for each observation to belong to a hypothetical signal. Different metrics may naturally rank each observation in a different way, leading to a different classification as signal- or background-like. Here we consider three methods for anomaly detection metric computation: (a) ‘Test statistic score’, given by the mean of the test statistics $\mathbf{T}_l=\left(T_{l1}, \ldots \,, T_{lB_l} \right)'$ based on the bootstrap samples including $\mathbf{y}_l$:\,\,$R_{Tl}=\frac{1}{B_l} \, \sum_{k=1}^{B_l}T_{lk}\,$, (b) ‘P-value score’, given by
the mean of the p-values $\mathbf{P}_l=\left(p_{l1}, \ldots \,, p_{lB_l} \right)'$, connected with the test  statistics based on the bootstrap samples including $\mathbf{y}_l$:\,\,$R_{Pl}=\frac{1}{B_l} \, \sum_{k=1}^{B_l}p_{lk}\,$, (c) ‘Ok score’, given by the 
proportion of tests based on the bootstrap samples including $\mathbf{y}_l$ and rejected at a given significance 
level $\alpha$\,: $R_{Ok;l}=\frac{1}{B_l} \, \sum_{k=1}^{B_l}1_{\{p_{lk}<a\}}\,$.
\item[Parameters Q and B]
When deploying the IB algorithm, the choice of $Q$ and $B$ is also important. The original formulation of the algorithm requires $Q < m$, as this may imply that a number of bootstrap samples $Y^{\small{\textrm{*}}}$ will include larger proportion of signal observations than $\lambda$, thus increasing the power of the test and making the detection of an hypothetical signal simpler. The choice of $B$ is not independent of $Q$, because the expected number of times each observation $\mathbf{y}_l$ is sampled 
is $E(B_l)=\frac{BQ}{m}$. For a fair comparison, the observation scores are computed based on the same 
number of tests, {\em i.e.} $Q$ and $B$ may vary among the different study cases but should be adjusted accordingly for $E(B_l)$ to be
kept fixed. Moreover, the larger the $B$ values, the more stable the results are expected to be. \item[Performance comparison]
A third significant aspect to take into account is the comparative evaluation of the IB algorithm. For this purpose, we consider an adjustment of the Linear Discriminant Analysis (LDA)~\cite{Izenman2008}, referred to in the
following as ‘LDA score’, suitable for the semi-supervised nature of the problem under study~\cite{kotkowski2019}. The performance of LDA is further compared with that of IB in various scenarios in the context of an improvement of the method~\cite{pietro_vischia_2019_5535703}.
\end{description}

\subsubsection{Simulation settings and results}

\medskip
%\smallskip
\noindent
The research questions described in Section~\ref{sec:ib:resquestions} are explored using either a univariate or multivariate normal distribution for both signal and background, or a multivariate normal distribution for the background and a uniform distribution across a sphere or a hemisphere for the signal.
Early results~\cite{kotkowski2019} suggest that, in the case of univariate and multivariate normal data, a better performance may be achieved when using the test statistic as a score in conjunction with subsampling the data ({\em i.e.} $Q<m$): in particular, the regime $Q \ll m$ seems to be associated with a lower variability of the score, in agreement with the intuition and the studies of Ref.~\cite{Vischia2017}.
For large numbers of bootstrap iterations, $B$, the preliminary results suggest that the classification performance is comparable among different values of $Q$, possibly because by increasing $B$ the variance of the scores converges
to some value that does not depend on $Q$. A small value of $Q$, however, implies lower variability of the scores. A good performance may therefore be obtained without having to resort to a large number of bootstrap samples; early studies seem to confirm this intuition.
The LDA seems favoured against the IB when both signal and background are normally distributed, in line with the assumptions LDA relies upon. However, the preliminary tests suggest that the performance of IB may still be comparable to that of LDA when using small values of $Q$ and selecting the test statistic score: the performance may be even less affected when the normality assumption for each class is removed. Additional tests that use a uniform signal distribution on a sphere and on a hemisphere lead us to conjecture that the IB may have the ability of recognizing both local and global properties of the signal process.

\begin{figure}[ht!]             
\begin{center}                                                      
  \includegraphics[width=0.95\linewidth,clip]{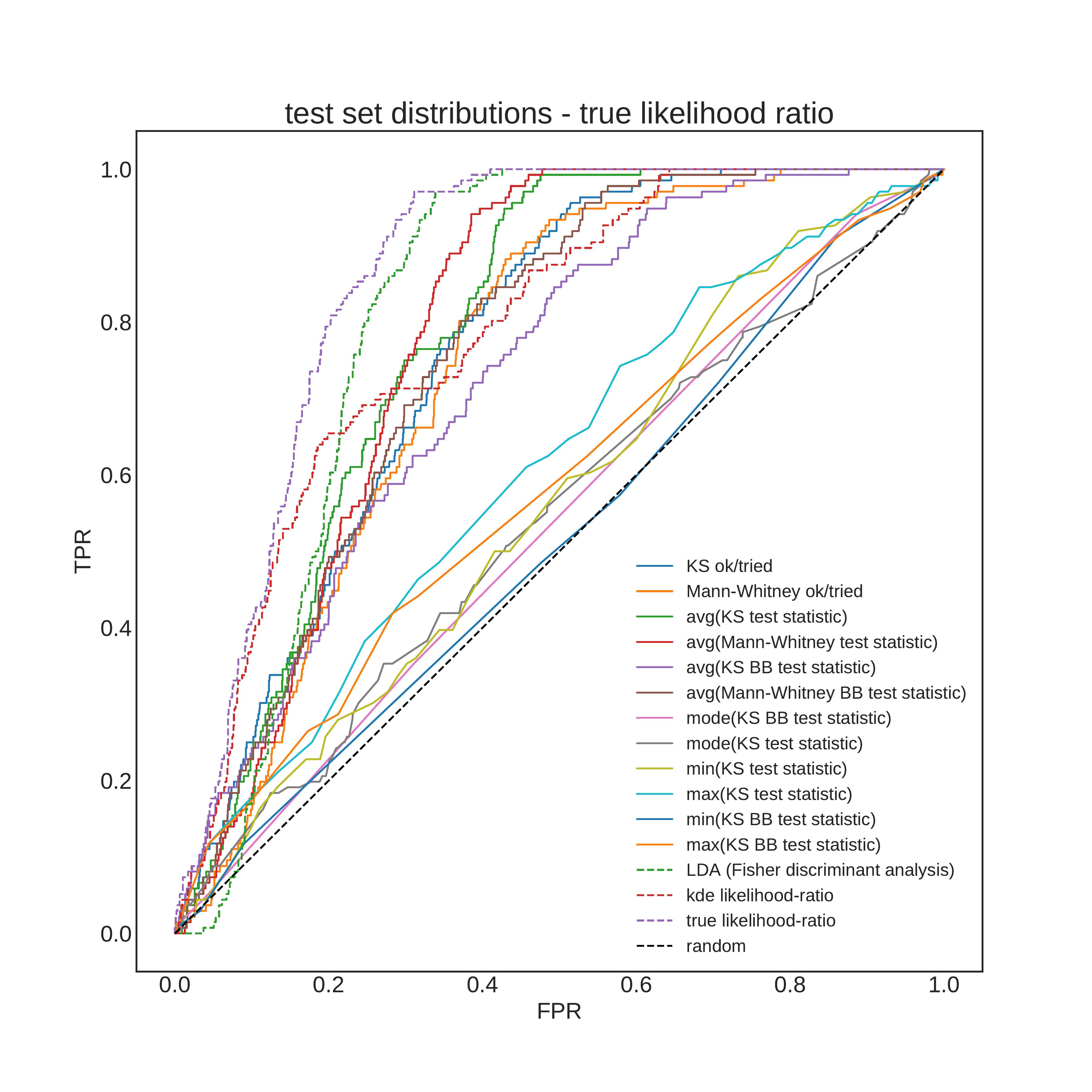}
  \caption{The ROC curve (true positive rate versus false positive rate) for a comparison of different settings of the IB algorithm with several scenarios: the classical likelihood ratio (labelled \textit{true likelihood ratio}) and its KDE approximation (labelled \textit{KDE likelihood ratio}) represent  idealized situations where the signal model is known; a realistic competitor is represented by a semi-supervised LDA approach. The IB algorithm performance is shown for several choices of IB test statistic and of summary statistic for the score, detailed in the text~\cite{pietro_vischia_2019_5535703}.}
  \label{fig:IB_2}
  \end{center}
\end{figure}

\medskip
\smallskip
\noindent
An extensive test~\cite{pietro_vischia_2019_5535703} is performed comparing the IB algorithm to several scenarios, as illustrated in Fig.~\ref{fig:IB_2}. The case where the signal density  is known is represented by two scenarios: a classical likelihood ratio test based on the Neyman--Pearson lemma, and a kernel density estimation performed on the original datasets $X$ and $Y$. A semi-supervised approach---the most realistic competitor to IB---is represented by a semi-supervised LDA. The IB algorithm is represented by several settings, corresponding to different choices of test statistic (Kolmogorov--Smirnov test~\cite{kstest}, Mann--Whitney test~\cite{mannwhitney}) and of score aggregation (minimum, maximum, mode, or median of the set of test statistics). The best setting for the Gaussian mixture in exam ($\lambda=0.06$) appears to be the expected value (mean) of the test statistic. For reference, the performance of a random choice is also shown.

\subsubsection{Possible improvements}

\medskip
%\smallskip
\noindent
Possible improvements to the original IB algorithm include exploring not only additional anomaly detection metric computation methods, using the simple multivariate Gaussian scenarios described {\em supra}, but also a dimensionality reduction approach, which could address the issue of the high-dimensional data with a large number of redundant features that is common in high-energy physics applications. For this approach, standard and more advanced methods of variable sub-sampling (‘max-G-sampling’) are proposed~\cite{Deliverable_4_4} (similarly to~\cite{Casa2018}), as well as methods for adjusting the sampling weights in such a way as to obtain samples more enriched in signal, and to perform tests on the likely more informative variables. ‘G’ in max-G-sampling stands for the
number of test results from which the maximum is taken, and a moderately high G parameter can result to only the highest test statistic values---likely obtained via a feature set rich in informative variables---being saved.
Moreover, given that for the study described {\em supra} uncorrelated generated data were considered, and since it is essential to also validate the IB algorithm performance in presence of correlations affecting the outcome
of the dimensionality reduction technique, a particular transformation of the data intended for its decorrelation is suggested~\cite{Deliverable_4_4}.

\subsubsection{Applications}

\medskip
%\smallskip
\noindent
A preliminary validation of the IB performance on a non-physics-related data set was performed on the well-known spam data~\cite{HastieTrevorTibshiraniRobertFriedman2009}, which are transformed to fit the semi-supervised context of the above described study for anomaly detection; early tests~\cite{kotkowski2019} suggest that the proposed algorithm improvements may have a comparable performance with that of the standard algorithm setting. 
When applied to a common high-energy physics problem so as to be validated on Monte Carlo simulated collision data (which constitutes the main objective of its development) the IB algorithm was found to perform at a similar level as  the LDA score. Also, the performance including the proposed variable max-10-sampling method was comparable with that of the standard approach, and for some scenarios presented a slight improvement.

\subsubsection{Summary}

\medskip
%\smallskip
\noindent
Taking everything into consideration, the IB algorithm exhibits a satisfying performance, achieving its best performance  when the distribution of the signal and background deviates from Gaussianity. The performance of the IB algorithm is particularly good when using small-sized bootstrap samples, both in terms of mean and variance. Different score computation methods have also been tested, with the test statistic appearing as the most effective one for this purpose. At the same time, a reduction of the data dimensionality seems to have the potential to improve the performance of the algorithm; however, further studies are needed on this aspect. Depending on all these factors and on a chosen amount of false-positive-rate, we have observed that the IB algorithm can provide improvements in the true positive rate ranging from 1 to 10\% in some of the considered problems. The improvement is nevertheless problem-dependent, suggesting that the algorithm may be suitable only for a certain class of problems; studies are ongoing to more precisely identify the characteristics of the problems that the IB algorithm can tackle with profit.

\subsection{Detecting anomalies via clustering semi-supervision}

\medskip
%\smallskip
\noindent
With the Inverse Bagging algorithm described {\em supra}, the degree of compatibility between experimental and simulated data is evaluated by means of hypothesis testing. Another approach to anomaly detection, based on a markedly different rationale, consists in semi-supervising learning methods, either by relaxing the assumptions of supervised methods, or by strengthening unsupervised structures via the inclusion of additional available information from simulated data. The latter route has been followed in~\cite{kotkowski2019} and~\cite{Casa2018} where a parametric and, respectively, a non-parametric approach are adopted. Both are summarized {\em infra}.

\subsubsection{The parametric approach}

\medskip
%\smallskip
\noindent
The idea of addressing anomaly detection by semi-supervising parametric clustering was first explored, in the context of new physics searches, in~\cite{vatanen2012semi, kuusela2012semi} under the name of \emph{fixed background model} (FBM). The authors proposed to specify the distributions $f_B$ and $f_S$ themselves in Eq.~\ref{eq:f_bs} within the family of Gaussian mixture models. Parameter estimation was then conducted via maximum likelihood in two steps. First, the background model $\hat f_B$ was obtained based on the background data $X$. Afterwards, keeping the parameters of $\hat f_B$ fixed, the weight $\lambda$ in Eq.~\ref{eq:f_bs} and the parameters characterizing the new possible component $f_S$ were estimated using experimental data $Y$. Maximization of the likelihood function was conducted via a suitable adjustment of the Expectation--Maximization (EM) algorithm~\cite{dempster1977EM}. A goodness-of-fit test served to discard insignificant components and to assure that the whole estimated density was equal to the background component when no signal was detected.

\medskip
\smallskip
\noindent
Due to the curse of dimensionality or numerical issues, in fact, the approach described {\em supra} was found to be sub-optimal or even not liable to be carried out on high-dimensional data. To reduce the data dimensionality while preserving relevant information, penalized methods for variable selection were introduced in the unsupervised framework of mixture modelling in~\cite{pan2007MBclustering, xie2008MBgrouped}, yet with a strong reliance on restrictive assumptions on the clusters shapes. Within the AMVA4NewPhysics program~\cite{kotkowski2019, Deliverable_4_2} we have extended the penalized approach to the FBM to allow for a more flexible modelling without constraining the mixture component covariance matrices, as well as to account for the semi-supervised nature of the anomaly detection problem for new physics searches. The proposed penalized anomaly detection (PAD) method builds on a variant of the EM algorithm, derived in the semi-supervised context to estimate the parameters of a mixture model via the maximization of the penalized likelihood function. 

\medskip
\smallskip
\noindent
The PAD methodology was validated both in the unsupervised setting ({\em i.e.} to estimate the background distribution) and in the semi-supervised anomaly detection setting (to estimate the whole process density) via simulations designed to account for different aspects of the analysis: different implementations for handling variable selection; performance of competing methods; varying configurations of the background and possible signal. The study highlighted a general improvement with respect to the state of the art (see~\cite{kotkowski2019} for details).  

\medskip
\smallskip
\noindent
In the specific context of LHC physics research, the PAD was applied on Monte Carlo data produced to study the experimental signature of processes yielding two energetic jets in the final state. The background data were generated including QCD processes yielding two energetic jets in the final state; these arise when the hard subprocess generates two gluons, two quarks, or a gluon and a quark with large momentum, which then hadronize into the observable pair of energetic jets. The signal data were produced from a heavy resonance (stop quark) postulated in an extension of the Standard Model known as RPV-MSSM~\cite{Barbier:2004ez} with a mass of 1000 GeV, that also leads to two jets in the final state.  Results are presented in Table~\ref{tab:PAD} for varying proportion $\lambda$ of signal events. Since in the presence of imbalanced processes (such as the considered one of signal versus background classification based on mixture models) requires a threshold adjustment that can influence the evaluation, the performance of the method has been measured in terms of Area under the ROC curve (AUC). In terms of AUC, the PAD compares favourably with the FBM model.  

\begin{table}[tb]
	\caption{Summary of the anomaly detection results performed by the PAD and the FBM for MC data with different signal proportions $\lambda.$ For each scenario, $50$ datasets are generated to obtain a mean result with the respective standard deviations presented in brackets. }
\label{tab:PAD}
\centering
 \scriptsize 
\begin{tabular}{cccc}
\hline
Method& $\lambda$ & Average $\hat\lambda$  & Average AUC \\ %\begin{tabular}[r]{@{}l@{}}Average \\misclass.\\ error\end{tabular} & 
		%\begin{tabular}[r]{@{}l@{}}Average percent \\of misclassified\\ sig. observations\end{tabular} & 
	%	\begin{tabular}[c]{@{}l@{}}Average \\adjusted\\ Rand ind. \end{tabular}& 		Average AUC \\ 
\hline
PAD & 0.05 & 0.040(0.012) %& 0.11(0.026) & 0.763(0.216) %& 0.097(0.133) 
& 0.725(0.109) \\ 
PAD & 0.10 & 0.057(0.013) %& 0.10(0.021) & 0.484(0.145)		%& 0.397(0.123)
& 0.818(0.078) \\ 
PAD & 0.15 & 0.086(0.006) %& 0.10(0.009) & 0.402(0.029)		%& 0.507(0.029)
& 0.876(0.017) \\ 
PAD & 0.20 & 0.112(0.006) %& 0.11(0.006) & 0.390(0.023) 		%& 0.513(0.022)
& 0.882(0.012) \\
FBM & 0.05 & 0.025(0.009) %& 0.05(0.002) & 0.904(0.032) 		%& 0.143(0.045)
& 0.708(0.118) \\ 
FBM & 0.10 & 0.046(0.008) %& 0.09(0.002) & 0.872(0.024) 		%& 0.174(0.029)
& 0.764(0.078) \\ 
FBM & 0.15 & 0.070(0.006) %& 0.13(0.003) & 0.845(0.016) 		%& 0.185(0.018)
& 0.771(0.073) \\ 
FBM & 0.20 & 0.096(0.012) %& 0.17(0.003) & 0.818(0.018) 		%& 0.188(0.017)
& 0.780(0.054) \\ 
\hline  
\end{tabular}
\end{table}

\subsubsection{The nonparametric approach}\label{sec:casa_menardi}

\medskip
%\smallskip
\noindent
As an alternative to the parametric approach illustrated {\em supra}, we explored the non-parametric approach to unsupervised learning. From one side, this relies on a more flexible modelling of the processes under analysis, without constraints or specific assumptions about their shape. On the other side, by drawing a correspondence between groups and the modal peaks of the density underlying the observed data, the non-parametric formulation appears particularly consistent with the physical notion of signal, as it is commonly assumed that deviations from the background process manifest themselves as significant peaks in $f_{bs}$, not previously seen in $f_b$. 

\medskip
\smallskip
\noindent
As discussed in detail in~\cite{Casa2018}, two main contributions can be highlighted. Under the assumption that a signal does exist, the main idea is to tune a non-parametric estimate of the density $f_{bs}$, assumed to generate the unlabelled data. This tuning is obtained by selecting the smoothing amount so that the induced partition, where the signal is forced to emerge as a bump in the density distribution, classifies the labelled background data as accurately as possible. The relevance of the forced bump is afterwards tested via the suitable application of a statistical test: if it is deemed to be significant this would provide empirical evidence of a signal, and should then represent the stepping-stone for the possible claim of new physics discovery. As a second side contribution, a variable selection procedure, specifically conceived for the considered framework, is proposed. Here a variable is assumed to be relevant if its empirical distribution shows a changed behaviour in the simulated data with respect to the one in the experimental data, as this difference shall be only due to the presence of a signal, not previously seen in the background density. This idea is pursued by repeatedly comparing the estimated densities on distinct and randomly sampled subsets of variables. Eventually the variables that are more often responsible for a different behaviour of the two marginal distributions are selected.

\begin{figure}[h!]
\begin{center}
\begin{tabular}{cc}
\raisebox{2cm}{\begin{tabular}{rrr}
\hline 
Method & FMI & TPR \\
\hline
NP (2 var) & 0.84& 0.80\\
FBM (6 PC) & 0.77& 0.50\\
FBM (2 var)& 0.78& 0.56\\
\hline
\end{tabular}}&
 \includegraphics[width=.35\textwidth, height = 4cm]{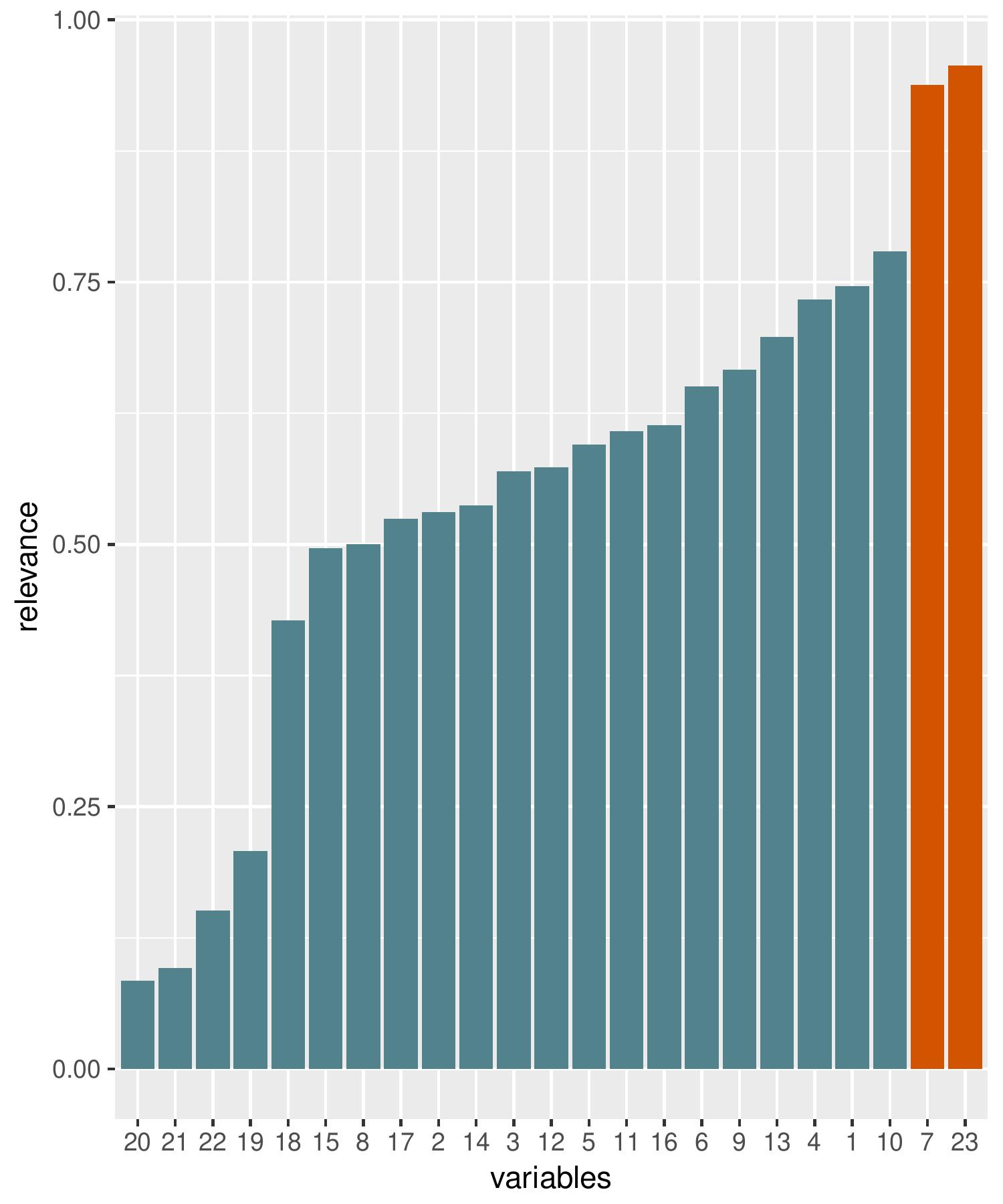}\\
 \includegraphics[scale=.24]{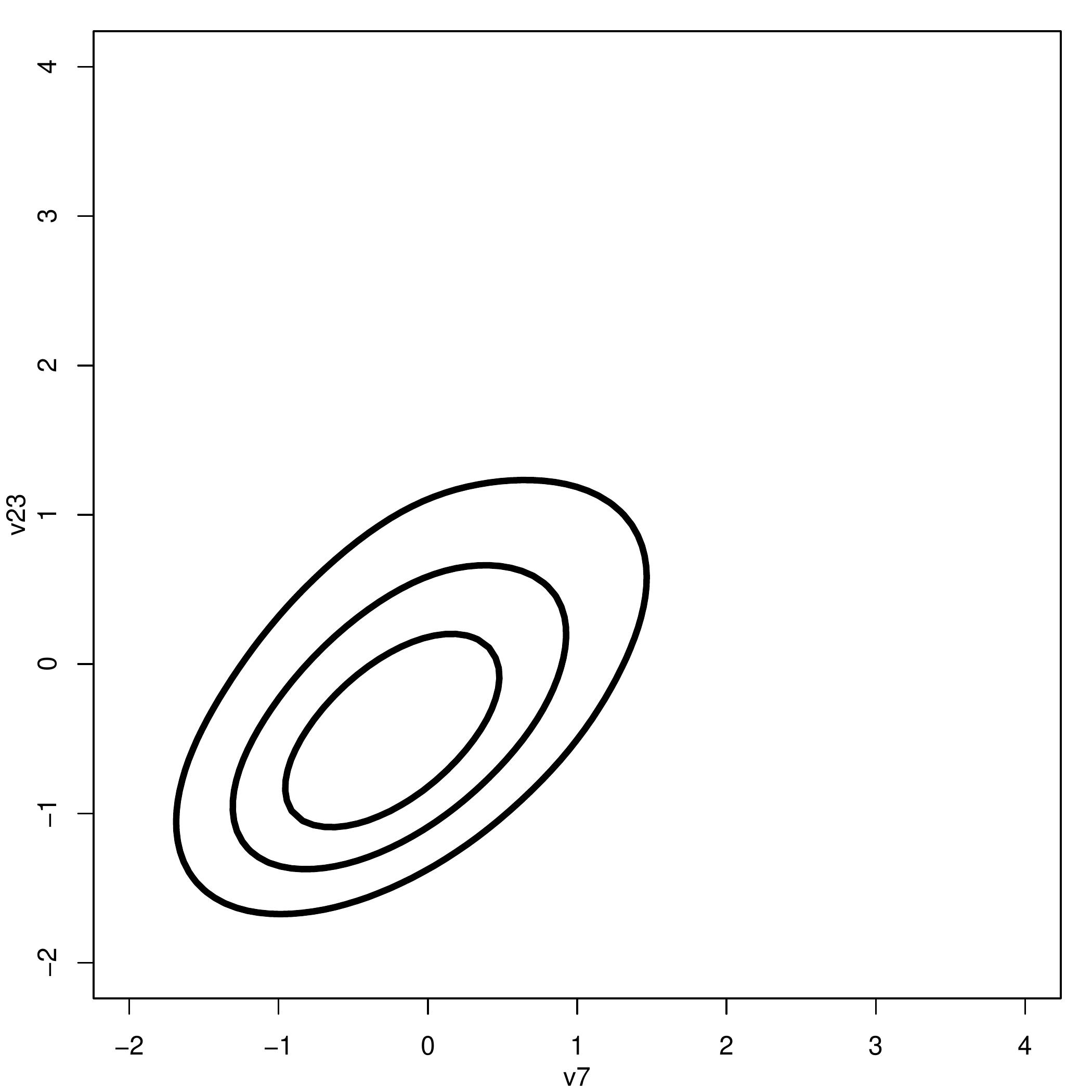}&
    \includegraphics[scale=.24]{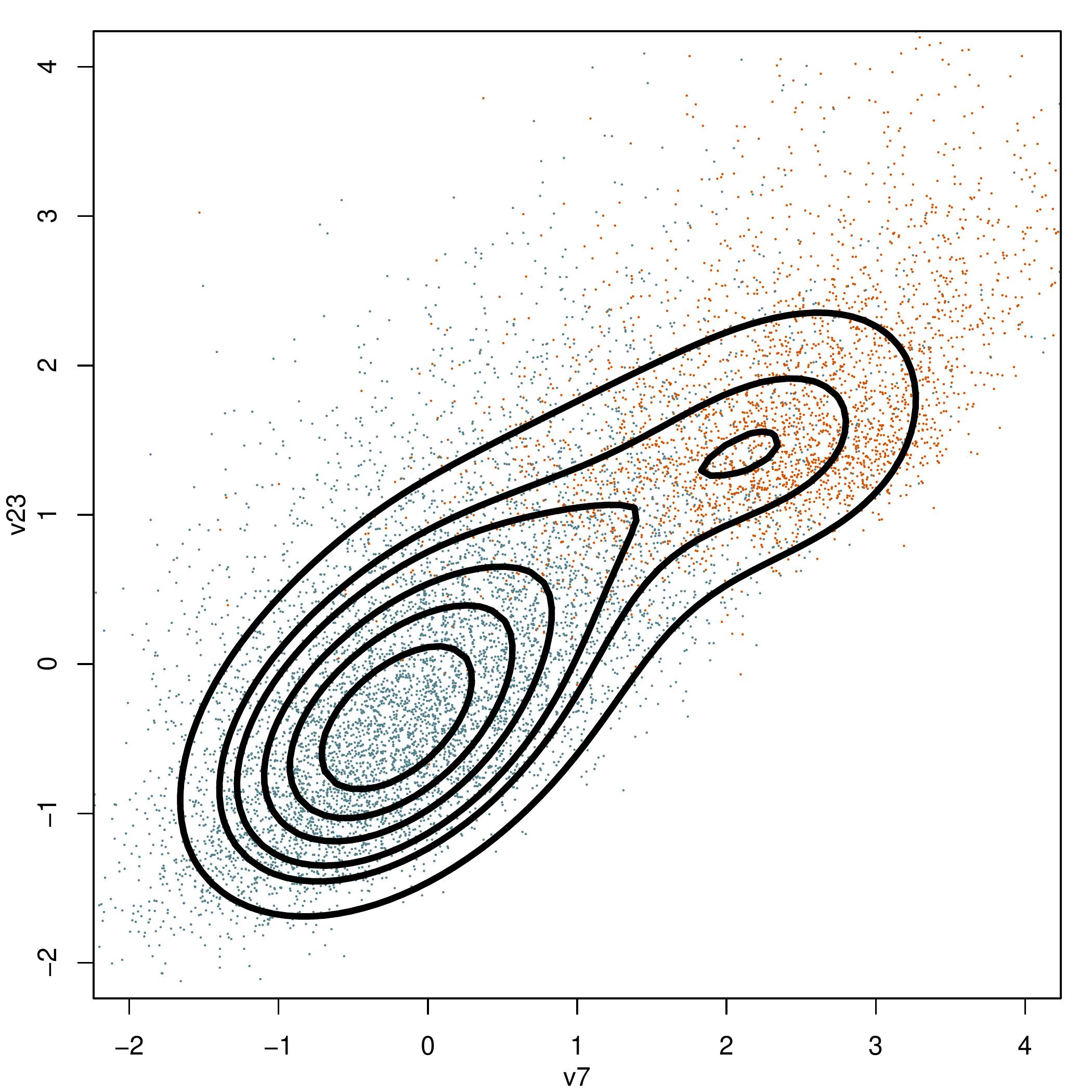}
\end{tabular}
\end{center}
    \caption{Results of the application of the nonparametric (NP) method for anomaly detection~\cite{Casa2018} discussed in Section \ref{sec:casa_menardi}: in the top-left panel, accuracy of classification of signal and background as measured by FMI and TPR, and compared to the parametric competitor Fixed Background Model (FBM) applied to the first 6 principal components the of the data (PC) and to the same 2 variables selected by the nonparametric procedure. In the top right panel relative informativeness of the whole set of variables, as labelled in~\cite{baldietal}. The resulting most relevant variables, in orange, are the combined mass of two bottom quarks with two W-bosons, and the transverse momentum of the leading jet. Bottom panels: contour plot of the density estimate of the two selected variables in the background (left) and overall data including the signal (right), which exhibits as a new peak arising from the background. A subsample of data from background (blue) and signal (orange) has been overimposed.}
    \label{fig:casa_menardi}
\end{figure}

\medskip
%\smallskip
\noindent
The procedure is compared with the Fixed Background Model~\cite{vatanen2012semi, kuusela2012semi} and tested on signal data simulated from a new particle $X$ of unknown mass that decays to a top quark pair $X \to t\bar{t}$ and a background from Standard Model top pair production, identical in its final state to the signal but distinct in the kinematic characteristics because of the lack of an intermediate resonance (see~\cite{baldietal} for a detailed description of the data and their characteristics). A visual glance to the results and the classification accuracy, as measured by the \emph{Fowlkes--Mallows} index (FMI)~\cite[][Ch. 27]{hennig_etal2015} and the true positive rate (TPR), are provided in Fig.~\ref{fig:casa_menardi}.

\subsection{Gaussian processes for modelling invariant mass spectra\label{sec:gausproc}}

\medskip
%\smallskip
\noindent
As already noted {\em supra}, model-independent searches for new physics at the LHC provide a ground for applying semi-supervised methods. One of the most relevant features used in the quest for discovering new physics in a given experimental signature is the invariant mass of the set (or a subset) of objects reconstructed as products of the collision events. The distribution of invariant mass values can be modelled from a Monte Carlo simulation (a labelled dataset) of the Standard Model processes that correspond to the background, and compared with the data obtained at the experiment, which could potentially contain (usually few) signal events among a large amount of background events. Gaussian processes (GP) are a useful tool for regressing the shape of invariant mass spectra and they can be used to disentangle potential signals from the background in a semi-supervised manner.

\subsubsection{Generalities on Gaussian processes}

\medskip
%\smallskip
\noindent
Gaussian processes are a Bayesian inference method in a function space,  defined as ``a collection of random variables, a finite collection of which have a joint Gaussian distribution''~\cite{Rasmussen:2005:GPM:1162254}.
Given a prior distribution over the parameters of a set of functions in the function space and using the likelihood of the observations, we can obtain a posterior distribution over such parameters through Bayes' rule.

\medskip
\smallskip
\noindent
Let $f$ be the function that is regressed and $x$ and $x'$ arbitrary points in the input space $\mathcal X$. The prior on the regression can be noted as
\begin{align}
   f(x) \sim \mathcal{GP}(\mu(x), \Sigma(x, x'))\,,
\end{align}
where $\mathcal{GP}$ is the infinite-dimensional function space associated with the joint Gaussian distribution, from which $f$ is sampled.
Thus, there are two functions that are defined to specify a GP: the \textit{mean} and the \textit{covariance} or \textit{kernel}, respectively,
\begin{align}
   \mu(x) &= \mathbb{E}[f(x)]\,,\\
   \Sigma(x, x') &= \mathbb{E}[(f(x) - \mu(x))(f(x') - \mu(x'))]\,.
\end{align}

\medskip
%\smallskip
\noindent
The mean and covariance functions above, as their names indicate, dictate the mean value and the covariance of the GP distribution for points in the input space.

\medskip
\smallskip
\noindent
For a finite set of points, the prior and posterior are joint Gaussian distributions with as many dimensions as observations, and there is a formalism that allows to predict new output values for arbitrary inputs.
In~\cite{Frate:2017mai}, GPs are used to model the Poisson process corresponding to a binned distribution of events, {\em i.e.} the spectrum. Let the bin centres be denoted by the vector $\bm x = (x_1, \dotsc, x_N)$, and the corresponding observed responses $\bm y = (y_1, \dotsc, y_N)$, the function $f$ is then averaged within the corresponding bin to produce a set of expected counts $\bm{\bar{f}}(\bm x) = (\bar{f}(x_1), \dotsc, \bar{f}(x_N))$. The spectrum is approximated via the product of two multidimensional Gaussians by the probability model:
\begin{equation}
   p(\bm y(\bm x)) = \mathcal N (\bm y | \bm{\bar{f}}(\bm x), \bm{\sigma^2}(\bm x)) \mathcal N (\bm{\bar{f}}(\bm x) | \bm\mu, \bm\Sigma)\,,
\end{equation}
where $\bm \sigma$ are the uncertainties in the values of $\bm{\bar{f}}$, which constitutes a Gaussian approximation of the Poisson noise.
The second factor is an $N$-dimensional Gaussian distribution.
Note that we use the vector $\bm \mu  = (\mu (x_1), \dotsc, \mu (x_N))$ and the matrix $\bm \Sigma$ with $\Sigma_{ij} = \Sigma(x_i, x_j)$, constructed from the mean and covariance functions respectively.

\medskip
\smallskip
\noindent
After some standard algebraic manipulation it is possible to obtain explicit expressions to infer the response of the function $f$ at a new arbitrary input values $\bm{x_\star}$, given $\bm x$ and $\bm y$:
\begin{align}
   \text{\textbf{mean}} (\bm{f_\star}) &= \bm{\mu_\star} + \bm{\Sigma_{\star}}
                                [\bm\Sigma + \bm{\sigma^2}(\bm x)\mathbb{1}]^{-1}(\bm y - \bm\mu), \\
   \label{eq:kernel-inference}
   \text{\textbf{cov}} (\bm{f_\star})&= \bm{\Sigma_{\star\star}} - \bm{\Sigma_\star}
                                        [\bm\Sigma + \bm{\sigma^2}(\bm x)\mathbb{1}]^{-1}
                                        \bm{\Sigma_\star}\,,
\end{align}
where $\bm{f_\star} = \bm{f}(\bm{x_\star})$, $\bm{\bm{\Sigma_\star}} = \bm\Sigma(\bm x, \bm{x_\star})$, and $\bm{\bm{\Sigma_{\star\star}}} = \bm\Sigma(\bm{x_\star}, \bm{x_\star})$; $\bm{\mu_\star}$ is the mean prior corresponding to $\bm{x_\star}$.
Note that the dimension of the GP multivariate Gaussian is extended by the dimension of $\bm{x_\star}$.

\medskip
\smallskip
\noindent
It is possible to initialize the prior mean $\bm\mu$ with a function from domain knowledge but setting it equal to zero is common practice and does not necessarily pose a limitation when using GPs. The most important ingredient to be specified is then the kernel, for which there exist several common choices in the literature ({\em e.g.} the exponential squared or other radial kernels)~\cite{duvenaud2014automatic, Rasmussen:2005:GPM:1162254}; a new kernel could also be crafted for a particular application.\footnote{A set of example kernels, how to compose them, and an explanation on how they can express the structure can be found in~\cite{duvenaud2014automatic} (chapter 2).}
The hyper-parameters in the kernel need to be adjusted as well, which is usually done by finding the maximum log marginal likelihood of the GP:
\begin{align}
\log \mathcal L = - \frac { 1 } { 2 } \log | \bm\Sigma | - ( \mathbf { y } - \bm\mu  ) ^ { T }
\bm\Sigma ^ { - 1 } ( \mathbf { y } - \bm \mu  ) - \frac { N } { 2 } \log 2 \pi\,.
\end{align}

\medskip
%\smallskip
\noindent
The standard algorithm to obtain the value $\bm {f_\star}$, its variance and the likelihood involves solving triangular systems and matrix inversion, as it is implied in the expression for the kernel detailed in Eq.~\ref{eq:kernel-inference}. Details of the algorithm and some optimizations are provided in~\cite{Rasmussen:2005:GPM:1162254}.

\subsubsection{Method overview}

\medskip
%\smallskip
\noindent
Gaussian processes have been used in several contexts in HEP research.
Our work is based on the application in~\cite{Frate:2017mai}, which presents a study on background and signal distributions modelling in the dijet mass spectrum. The goal of the method is to accommodate the background and identify a signal component on the data, if present.

\medskip
\smallskip
\noindent
We start with the following prescriptions for the mean and kernel functions:

\begin{align}
   \mu(x) &= 0\,, \\
   \label{eq:bkg-kernel}
   \Sigma_B \left( x , x ^ { \prime } \right) &= A \, \, \exp  
   \left( \frac { d - \left( x + x ^ { \prime } \right) } { 2 a } \right)
   \sqrt { \frac { 2 l ( x ) l \left( x ^ { \prime } \right) }
   { l ( x ) ^ { 2 } + l \left( x ^ { \prime } \right) ^ { 2 } } }
   \exp \left( \frac { - \left( x - x ^ { \prime } \right) ^ { 2 } } { l ( x ) ^ { 2 }
   + l \left( x ^ { \prime } \right) ^ { 2 } } \right)\,, \\
   \label{eq:sigkernel}
   \Sigma_S (x, x') &= C \,\, \exp \left( - \frac { 1 } { 2 } \left( x - x ^ { \prime } \right) ^ { 2 } / k ^ { 2 } \right)
   \exp \left( - \frac { 1 } { 2 } \left( ( x - m ) ^ { 2 } + \left( x ^ { \prime } - m \right) ^ { 2 } \right) / t ^ { 2 } \right)\,,
\end{align}
where $l(x)=bx+c$ is a linear function. For a description of all kernel hyper-parameters in Eqs.~\ref{eq:bkg-kernel} and~\ref{eq:sigkernel}, and a motivation for their functional forms, see the cited study.
We refer to the kernel hyper-parameters collectively as $\theta_B = \{A, a, b, c, d\}$ and $\theta_S = \{C, m, t, k\}$, respectively.

\medskip
\smallskip
\noindent
We present procedures that can operate using two or three steps as follows:

\paragraph*{First step} 

A GP fit is performed on a pure background distribution ({\em i.e.} from simulation) using the background kernel $\Sigma_1=\Sigma_B$. We obtain a model for the background distribution and find the corresponding hyper-parameters of the first kernel, $\theta_1$.

\paragraph*{Second step} 

The optimized hyper-parameters of the background fit are kept fixed and used in performing another fit using the kernel $\Sigma_{12} = \Sigma_1 + \Sigma_2$, where the second component in the sum is of the signal kind $\Sigma_2=\Sigma_S$ (Eq.~\ref{eq:sigkernel}); thus a new GP model is obtained, including the parameters corresponding to the new component, $\theta_2$.

\paragraph*{Third step} 

A further optimization is performed where a new signal kernel $\Sigma_S$ is added, $\Sigma_{123} = \Sigma_{12} + \Sigma_{3}$, and the parameters of $\Sigma_{12}$ ({\em i.e.} $\theta_{12} = \theta_1 \cup \theta_2$) are kept fixed. From this, the parameters of the last component ($\theta_3$) can be extracted.

%\vspace{2em}

\medskip
\medskip
\smallskip
\noindent
In the case of the two-step procedure, the background kernel is used in a first step; then, in a second step, a signal kernel identifies a concentrated excess or deficit centred at $m$ with width $t$.
For the three-step procedure, the first two steps are used to accommodate a background with a turn-on; then a third step performs the signal detection. This procedure can be considered as semi-supervised: we first model the background-only invariant mass distribution, {\em i.e.} the labelled dataset, and then find a model for the unlabelled data distribution where signal could appear.

\subsubsection{Modelling the dijet invariant mass: QCD background and artificial signal}

\medskip
%\smallskip
\noindent
For testing the two-step procedure, we used as toy dataset a simulated dijet invariant mass spectrum following the event selections described in~\cite{Aaboud:2018ufy}.

\medskip
\smallskip
\noindent
Before reporting the results of the GP optimization, we describe the injection of signal in the invariant mass spectrum. The starting point is the simulated background dataset we generate pseudo-data from, where the signal can be injected.
Since the spectrum ranges over several orders of magnitude in the number of events, we prescribe the amplitude of the injected signal via a quantity defined from a signal-over-background ratio ($R$). We calculate the amplitude of the Gaussian signal from the mean and width: $R$ is the ratio of signal to background events that should be in a window constructed from the interval given by the mean and width. The number of events taken into account in a given window is given by the bin counts of the distribution contained in the window; bins whose centres are outside the range (mean $\pm$ width/2) are not counted:
\begin{align}
  R = \frac{\text{Injected signal events in the window}}{\text{Background events in the window}}\,.
\end{align}

\medskip
%\smallskip
\noindent
Analogously, the extraction of the $R$ (and hence the amplitude) of the signal comes from the
same ratio within a window defined by the mean and width of the extracted signal from the parameters of the signal kernel.
We inject signals with different possible combinations of values for a Gaussian distribution, namely the mass (3, 3.5, 4, 4.5, and 5 TeV), the width (150, 300, and 450 GeV), and the $R$ ratio (0.1, 0.2, 0.3, and 0.4), totalling 60 values.
The values were chosen to cover a wide range of the spectrum and different intensities of the signal hypotheses.

\medskip
\smallskip
\noindent
In Table~\ref{table:Rvals} we provide the number of events injected that correspond to different values of the ratio $R$, to give a sense of the mapping between the two quantities.

\begin{table}[ht]
\centering
   \begin{tabular}{c | c c c c} 
   \hline
   $R$ & 0.1 & 0.2 & 0.3 & 0.4 \\ 
   \hline
   $\#$ injected events & 460 $\pm$ 30 & 920 $\pm$ 40 & 1380 $\pm$ 50 & 1840 $\pm$ 70 \\
   \hline
  \end{tabular}
  \caption{Number of injected signal events within a window for values of R, for a signal of 3 TeV mass and 150 GeV width. The values and errors obtained are the mean and standard deviation of the distribution of values obtained after repeating the injection in 100 background toys.}
  \label{table:Rvals}
\end{table}

\begin{figure}[ht]
\centering
\includegraphics[width=0.7\textwidth]{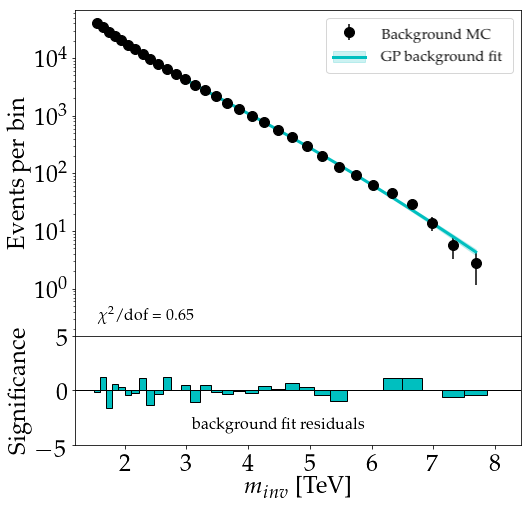}
\caption{Top panel: invariant mass spectrum showing a GP fit for the background. Bottom panel: per-bin significance of the deviation between the number of events and the fit.}
\label{fig:GP-Bkg}
\end{figure}

\medskip
\smallskip
\noindent
We present in Fig.~\ref{fig:GP-Bkg} a first fit (first step) performed on the pure background simulated data.\footnote{In this and subsequent plots of this section, the displayed significances correspond to ``signed z-values only if $p$-value $<$ 0.5'', as defined in~\cite{2012EPJP..127...25C}, where the $p$-value is calculated assuming that each bin count follows a Poisson distribution.}
The posterior GP background mean, the data and their respective errors are displayed.
We can observe that the GP fit is able to accommodate the distribution with a smooth model whose bin-wise discrepancies with data are small.
To measure such discrepancies, the $\chi^2$ divided by the number of degrees of freedom is calculated.

\begin{figure}[ht]
\centering
\includegraphics[width=0.7\textwidth,height=0.7\textwidth]{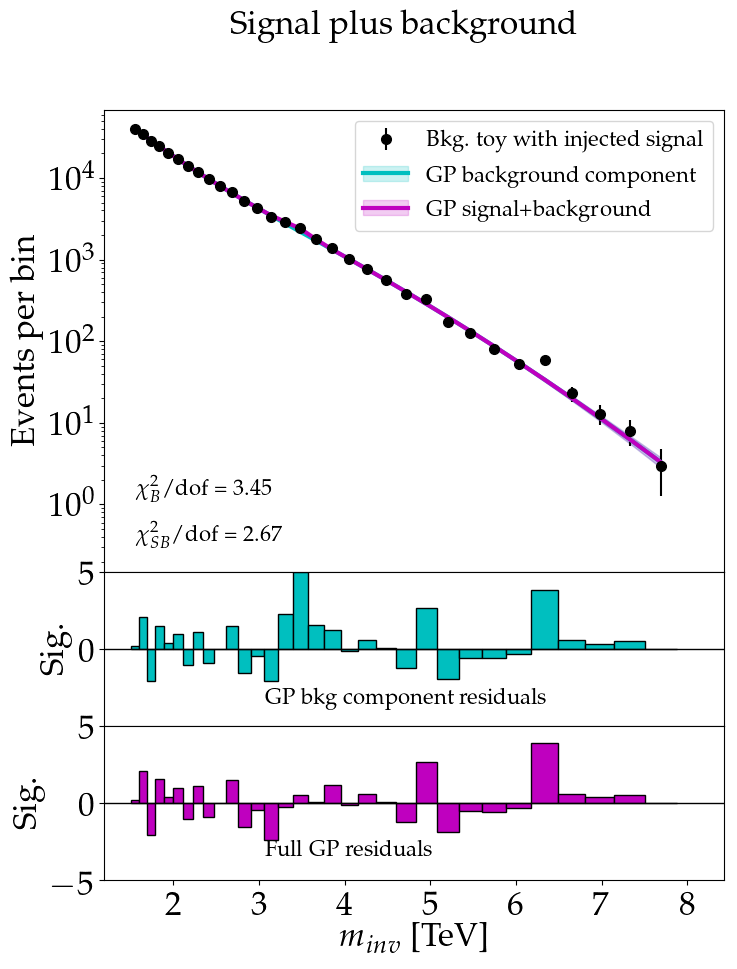}\\
\caption{Top panel: invariant mass spectrum displaying a GP background fit, with event counts for a background toy with signal injected centred at 3.5 TeV with a width of 150 GeV and R of 0.1, and a signal plus background fit. Solid coloured lines represent the GP fit components. Middle and bottom panels: per-bin significance of the discrepancy between the event counts and indicated fits.}
\label{fig:GP-SigBkg}
\end{figure}

\medskip
\smallskip
\noindent
Fig.~\ref{fig:GP-SigBkg} shows a similar plot for the second step in the presence of an injected signal. By observing the residuals we verify that the GP can identify the injected signal: the middle panel clearly reveals a localized set of discrepant bins in the region of the injected signal that the background-only component of the fit does not capture; whereas the bottom panel shows that the injected signal has been incorporated in the full fit.

\begin{figure}[ht]
\centering
\includegraphics[width=0.8\textwidth]{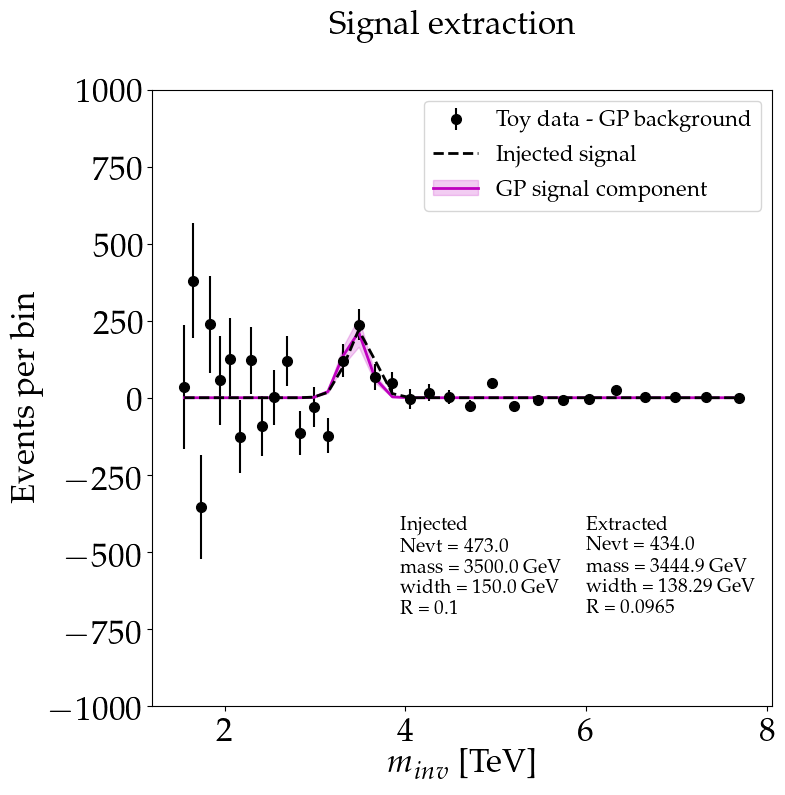}
\caption{Residual plot corresponding to Fig.~\ref{fig:GP-SigBkg}. The GP signal component (solid magenta) and the signal injected (dashed black line) are displayed as well as a subtraction of the toy data set with a signal injected minus the background GP fit (black dots with error bars). Injected and extracted signal values are shown.}
\label{fig:GP-Sig}
\end{figure}

\medskip
\smallskip
\noindent
A corresponding graph where signal identification becomes more evident is presented in Fig.~\ref{fig:GP-Sig}. In the case shown, the GP signal component correctly locates the injected signal and is able to recover relevant parameters (strength, width, mean).
However, when we approach the faintest or widest signals, the method becomes prone to mis-identification. The method is also used in the absence of the signal when performing the second step, in order to observe to which extent spurious signals are detected.

\newpage
\begin{figure}[ht]
\centering
\includegraphics[width=0.9\textwidth,height=0.8\textwidth]{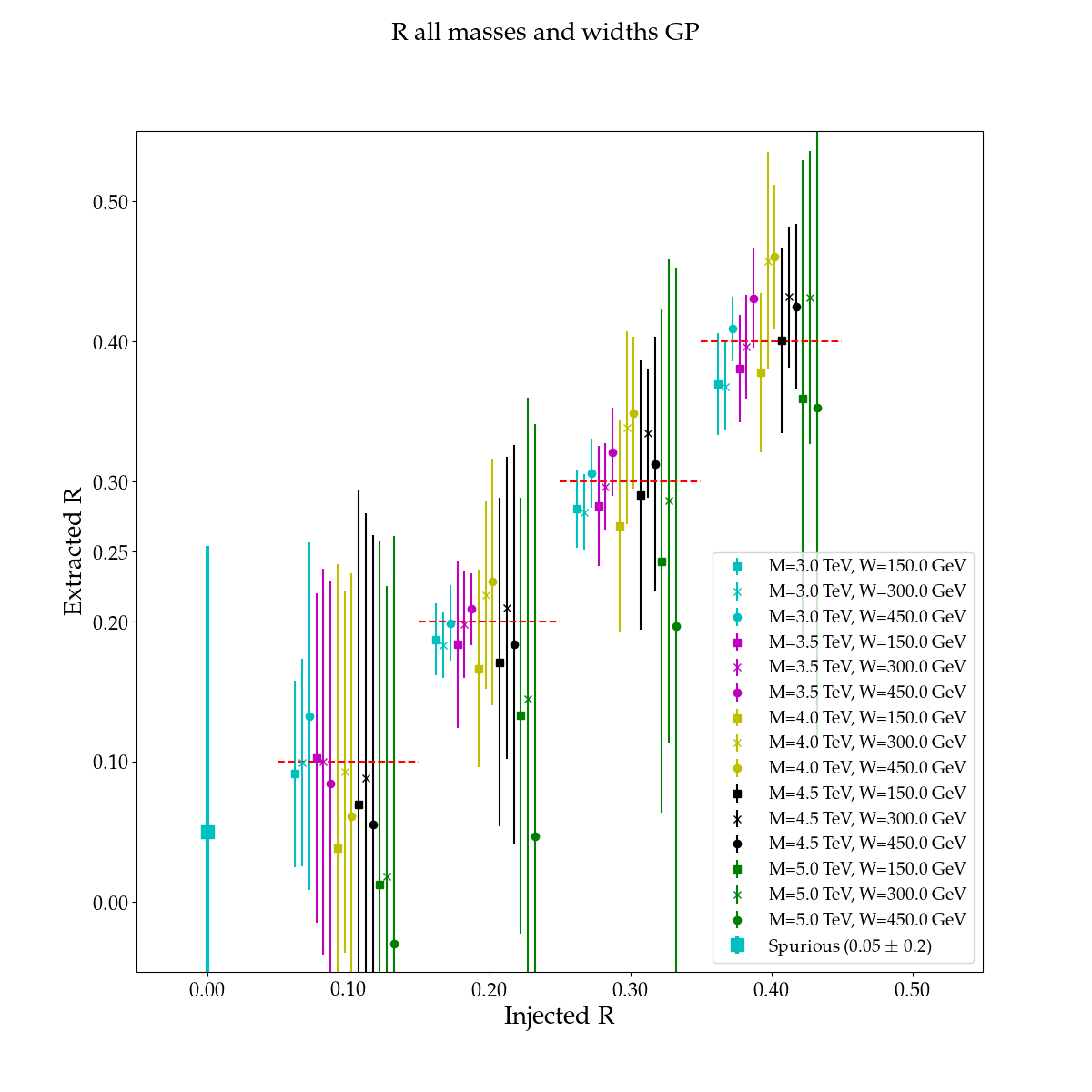}
\caption{Linearity plots for the R values of the signal injected in the dijet spectrum; each plot corresponds to indicated mass M and width W (both in GeV). The values of the width are the same for each plot column, and those of the mass are fixed through each plot row. Points and error bars (means and rms) are calculated from the distribution of extracted width values. A dashed red $x = y$ line is plotted for reference.}
\label{fig:GP-R-Op1}
\end{figure}

\medskip
\smallskip
\noindent
Good agreement is generally found when checking the linearity between injected and extracted $R$ values in Fig.~\ref{fig:GP-R-Op1}. For every signal hypothesis, a sampling of the Gaussian signal is performed and injected 100 times on different background pseudo-data, allowing the extraction of a distribution of 100 values for each signal parameter and for each hypothesis. Despite the good agreement, we observe that the higher mass hypotheses tend to be poorly identified; this is due to the specific binning of this spectrum, where bin widths increase approximately from 90 GeV at lower masses to 360 GeV at the right-end of the spectrum. Even if $R$ is by construction a more consistent parameter to prescribe signal strengths in different regions of the spectrum than, for example, a fixed number of events, it still can lead to undesirably faint signals at high mass values, where the event counts are small.

\medskip
\smallskip
\noindent
The value presented as the upper error (mean+rms) of the spurious signal detection ($R =$ 0.25) serves as an indication of the ability of the method to detect genuine signals.
As we stated before, GPs are in general tools flexible enough to model the distribution without prescribing a mean function, and in terms of signal extraction they do not yield a significant gain in the extraction power.

\subsubsection{Further studies}

\medskip
%\smallskip
\noindent
An application of the three-step procedure described {\em supra} was used to accommodate backgrounds with turn-ons (first two steps) and signals (last step). Here we give a general outline on this use case; further detail can be found in~\cite{jimenezModel2019}.

\medskip
\smallskip
\noindent
The motivation for using a three-step procedure arose in an attempt of modelling the invariant mass spectrum of top quark pairs ($t\bar t$) to search for a $Z'$-boson decaying to $t\bar t$.
That spectrum covers a mass range from 0.5 to 2 TeV  
with a turn-on for lower values (roughly below 0.8 TeV) followed by a decay, the latter being somewhat similar to the dijet spectrum.
In an effort to use the two-step procedure in this case, we observed that the signal kernel was not able to identify injected signals in the second step, but always captured the turn-on part instead. Thus, the first and second step (with $\Sigma_1$ and $\Sigma_{12}$ respectively) were used on the background-only sample, successfully accommodating the background spectrum; a third step (using $\Sigma_{123}$) was then applied for signal extraction.

\medskip
\smallskip
\noindent
The testing of the method was similar to that performed in the dijet case. After fitting the background spectrum, we generate and use background pseudo-data to inject $Z'$ signals corresponding to simulated resonance hypotheses at two mass points, {\em i.e.} 750 GeV and 1250 GeV. The last step of the procedure is then applied 100 times, where we obtain the signal parameters ($\theta_3$). 

\medskip
\smallskip
\noindent
As reported in~\cite{jimenezModel2019}, signal extraction is achieved with moderate success. Two main limiting factors are the faintness of the nominal simulated signals used, and in some cases an observed shift of the GP signal component towards the turn-on point. Concerning the first issue, we tested the extraction where the nominal simulated signal was amplified by different factors. For the shift in the signal component, we instead limited the $\theta_3$ parameters in the optimization to lay outside the turn-on region, a procedure that helps overcome the issue; still, in the case of spurious detection (no signal injected), the signal component would mostly tend to find spurious signals near the lower bound allowed for the location ({\em i.e.} $m$ in $\theta_3$).
This and other limitations reported in the study need further work in order to enable the described method to be applicable to resonance searches in the $t \bar t$ mass spectrum.

\subsubsection{Conclusions}

\medskip
%\smallskip
\noindent
We explored different aspects of GP approaches that are able to perform background modelling and signal detection, without any prior information on the mean, {\em i.e.} $\mu(x)=0$.
The two-step procedure was able to detect signals with different widths, intensities, and locations in the dijet invariant mass spectrum.
We use $R$, a ratio of events defined within a window to measure the intensity of the signal; since the spurious detection leads to identifying signals up to 0.25, we use this value as an indication of the faintest signal that the method is capable to identify.
We also described the application of the GP method in three steps in the more challenging scenario of the search for a resonant signal in the $t\bar t$ invariant mass spectrum: the first two steps model the background, including the turn-on region, and the third step is used for extracting the signal.

\medskip
\smallskip
\noindent
The GP methods that we used provide a way towards alternative background modelling and signal identification techniques. The procedures presented here lend themselves to further improvement: in particular the performance of the method applied to the dijet mass spectrum could be further improved with a better definition of the injected amplitude and, in the case of the top quark pair spectrum, the three-step procedure showed some limitations in modelling the background turn-on and capturing the signal.
Finally, the application of the method to model-independent searches could be better served by more versatile methods, which could be studied by exploring other kernels and by avoiding reliance on a complex iterative fitting procedure.

\section{Similarity Search for the Fast Simulation of the ATLAS Forward Calorimeter \label{s:caloshowers}}

\subsection{Overview}

\medskip
%\smallskip
\noindent
The physics measurements of the experiments at the LHC strongly rely on detailed and precise simulations to predict the complex response of the detectors. These simulations are essential for the comparison between experimental results and theoretical predictions. The detailed simulation of the interaction and propagation of the particles through the detector is typically performed with the \textsc{Geant4}~\cite{geant4} simulation toolkit, and it demands very large computing resources. The development of innovative computational and algorithmic techniques is fundamental to cope with the complexity of the simulation, which is further expected to increase in the next decade, with the increment of the volume of data collected at the LHC.

\subsection{Fast simulation with frozen showers \label{sec:fast_sim}}

\medskip
%\smallskip
\noindent
 {In the ATLAS experiment, the simulation chain is managed by Athena~\cite{Athena, Athena_repo}, an open software used for the simulation of the particle interaction and propagation in the detector, as well as the physics object reconstruction.} The ATLAS experiment makes large use of fast simulation techniques to reduce the computational resources required by the entire simulation chain. A prominent example is the \emph{FastCaloSim} module~\cite{FastCaloSim} {in Athena}, which uses a parameterized model of the calorimeter response in the barrel. 

\medskip
\smallskip
\noindent
When it comes to simulating the response of a calorimeter, specifically built with very dense material to entirely absorb the energy of the particles that traverse it, the requirements of computational resources could increase exponentially, up to $\sim70\%$ of the total CPU time dedicated to the entire simulation process. This is especially relevant for the simulation of the ATLAS forward calorimeter (FCal)~\cite{FCal, FCalPerformance}. The FCal is a sampling calorimeter that covers the most forward region of the ATLAS detector, in the range of pseudo-rapidity $3.1 < \vert \eta \vert < 4.9$. Due to the close position to the beamline, energy and density of particles reaching the FCal are very high. The harsh environment is reproduced by Monte Carlo simulations, responsible for the propagation of high-energy particles that enter the calorimeter and form showers of secondary particles inside it. The FCal consists of three consecutive modules along the beamline. The first module, the closest to the interaction point, is made of copper and is designed to absorb the majority of the electromagnetic showers. The following modules, made with tungsten, are designed to absorb the hadronic contribution to the particle showers. Since the total energy of a showering particle is predominantly deposited in the sampling calorimeter through a large number of soft electrons and photons, the work presented here focuses on the fast simulation of the FCal response in the electromagnetic module.

\begin{figure}
\centering
\includegraphics[width=1.\textwidth]{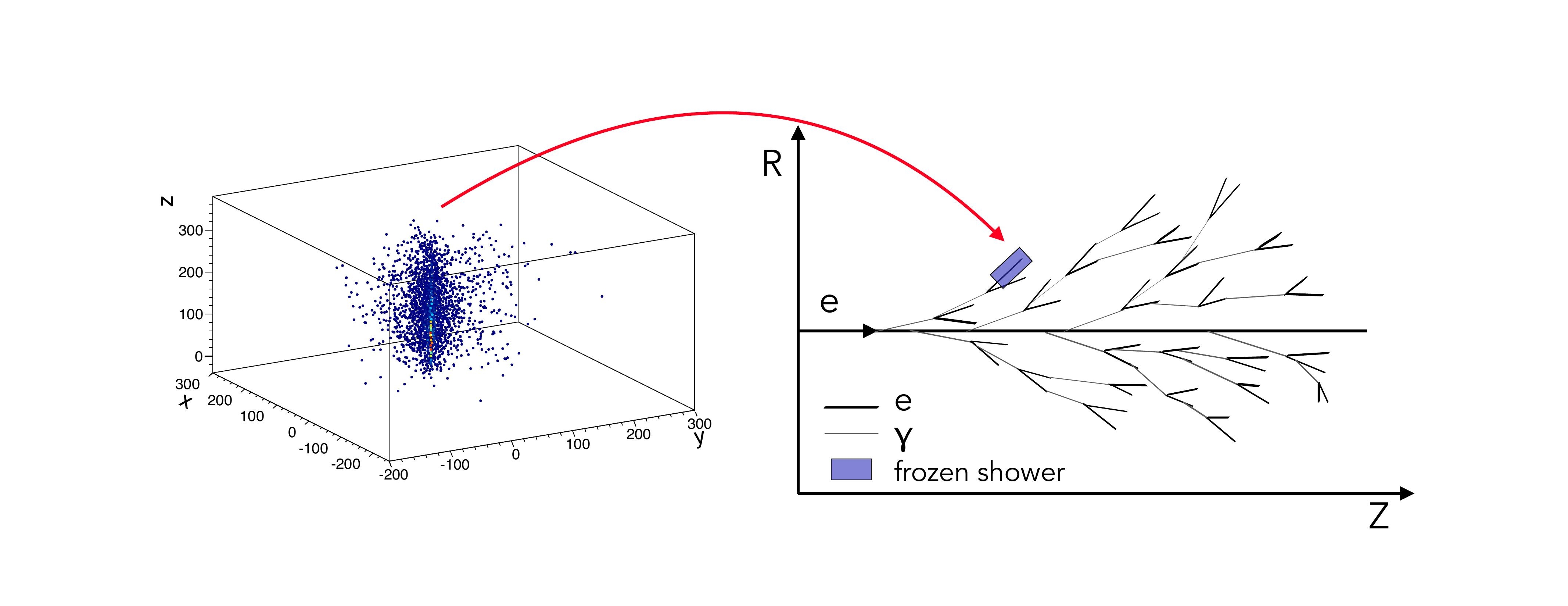}
\caption{Sketch of the particle showers. A shower collected in the frozen library (left) is used to replace a low-energy particle produced in the simulation (right).}
\label{fig:showers_scheme}
\end{figure}

\medskip
\smallskip
\noindent
In ATLAS, the simulation of the FCal detector already makes use of fast techniques aimed at reducing the complexity of the response. The FCal fast simulation is based on a \emph{frozen shower library} approach~\cite{Barberio_2008}, which consists of storing pre-simulated electromagnetic showers initiated by high-energy particles entering the front face of the calorimeter, subsequently used to quickly model the detector response. The frozen shower substitution is illustrated in Fig.~\ref{fig:showers_scheme}. 

\medskip
%\smallskip
\noindent
The generation of the frozen library requires a preliminary simulation of the low-energy particles obtained with the propagation in the FCal of the particles arising from simulated collision events. In the work presented here, LHC-produced top quark pair events $pp \to t\bar{t} +X$ have been used. When secondary particles with energy below a given threshold are produced, typically electrons below 1 GeV and photons below 10~MeV, the corresponding kinematic information is saved in the frozen library, together with the collection of hits corresponding to the subsequent shower produced by them. The frozen shower is then used in the fast simulation of the FCal response. When a low-energy particle is produced in the FCal, the simulation (performed with \textsc{Geant4}) is suspended, the particle is associated to one entry in the library and the corresponding electromagnetic shower, whose energy is scaled to match the energy of the particle produced, is sampled.

\medskip
\smallskip
\noindent
The ATLAS FCal calorimeter consists of an absorber matrix (made with copper, in the case of the electromagnetic module) instrumented with cylindrical electrodes parallel to the beamline. Each electrode is composed by a rod (anode) placed inside a tube (cathode). The small gap in between is filled with liquid argon, used as the active material of the sampling calorimeter. The distance of the particle to the active material is an important parameter to determine the characteristics of the originated shower.  

\medskip
\smallskip
\noindent
The frozen library currently used by ATLAS for the FCal fast simulation contains pre-simulated showers parameterized in bins of pseudorapidity $\eta$ and $d$, the distance of the low-energy particle to the centre of the closest rod. The energy of the particle is also stored. During the simulation, the matching between the simulated particle and one entry of the library is performed by selecting the closest $d$ and $\eta$ bin and then finding the closest energy available within the bin. 

\medskip
\smallskip
\noindent
The energy resolution of the detector response obtained with the fast simulation should always be able to reproduce the result obtained with a standard simulation (full simulation). The fast simulation approach currently used in ATLAS strongly reduces the computational requirements of the simulation. However, it does not reproduce well the energy resolution provided by the full simulation, thus requiring some tuning of the parameters in the library prior to its usage.

\subsection{Similarity search}

\medskip
%\smallskip
\noindent
The new proposed strategy~\cite{Tosciri:2739405} employs \emph{similarity search} techniques to select the most suitable particle--shower pair in the library. Such techniques constitute a branch of machine learning and include \emph{clustering} and \emph{indexing} methods that enable quick and efficient searches for vectors similar to each other~\cite{Babenko2}. These methods are usually employed in applications such as image features search and document retrieval in large scale databases, for example in search engines. A new version of the library is also used. The stored showers are keyed by 6-dimensional parameter vectors containing the angular and spatial information of the low-energy particles, as well as their energy.

\medskip
\smallskip
\noindent
Similarity search techniques can be described as approximate nearest neighbour searches to find similar vectors. A prediction for a query vector is made from data, by measuring the distance with the instances in the dataset and selecting the most similar ones. 
The similarity can be defined using different metrics, one of the most common being the Euclidean distance $D(\mathbf{p}, \mathbf{q})$ between two vectors $\mathbf{p}$ and $\mathbf{q}$, defined as
\begin{equation}
%\begin{split}
D(\mathbf{p}, \mathbf{q}) = \sqrt{(q_{1}-p_{1})^{2} + (q_{2}-p_{2})^{2} + ... + (q_{d}-p_{d})^{2}} =  \sqrt{\sum_{i=1}^{d}(q_{i}-p_{i})^{2}}\,.
%\end{split}
\label{eq:euclidean}
\end{equation}

\medskip
%\smallskip
\noindent
In the frozen library context, given a query vector containing the kinematic information related to the simulated low-energy particle, the search is performed to find the most similar vector of kinematic variables collected in the library. The search is not performed exhaustively in the whole dataset, but is based on approximation methods that often involve the use of clustering algorithms to limit the portion of the dataset considered in the search. This involves pre-processing the dataset with an indexing procedure, which learns a data structure from the dataset. 

\medskip
\smallskip
\noindent
In this work, the \emph{Facebook AI Similarity Search} (Faiss)~\cite{Faiss} package, developed by the Facebook Artificial Intelligence researchers to efficiently retrieve multimedia documents, has been used to construct the index and perform the search within the frozen library. Several indexing options to build a structure from the dataset are available with Faiss. One of the most common structures is called \emph{inverted index}~\cite{Babenko} and is based on the space partition into Voronoi regions defined by a \emph{K}-means clustering algorithm. Each vector is encoded into the corresponding region (\emph{code}) and saved in an array (the inverted index). Given a query vector, the search is then performed on the indexed dataset, based on the similarity with the centroids of the clusters. The vectors contained in each queried cluster are decoded sequentially and compared to the query.  
The Faiss package also provides a method called \emph{Product Quantizer}~\cite{PQ} to compress the vectors and perform a faster search at the cost of accuracy. The Product Quantizer is particularly useful when dealing with high-dimensional vectors and very large datasets ($>10^{6}$ elements). Another interesting method is the so-called \emph{Hierarchical Navigable Small World graphs} (HNSW)~\cite{Malkov}. The HNSW consists of a multi-layer structure built from the dataset, where the vectors are connected through links, based on the similarity metric, and organized over different layers according to the length of the links. Then the search is performed across the layers, starting from the region characterized by the longest links. The HNSW is considered state-of-the-art among similarity search algorithms and provides a good trade-off between speed and accuracy both in high- and low-dimensional data.

\subsection{Validation and results \label{sec:validation}}

\medskip
%\smallskip
\noindent
To validate the new approach, a library containing $\sim10^{5}$ showers has been generated with electrons of energy below 1 GeV. Then, high-energy electrons are generated at the interaction point with discrete energies of $[100, 200, 300, 400, 500]$ GeV and pseudorapidity in the range $3.35 < \vert \eta \vert < 4.60$, and propagated using the fast simulation with the new implementation. 

\begin{figure}
\centering
\includegraphics[width=0.75\textwidth]{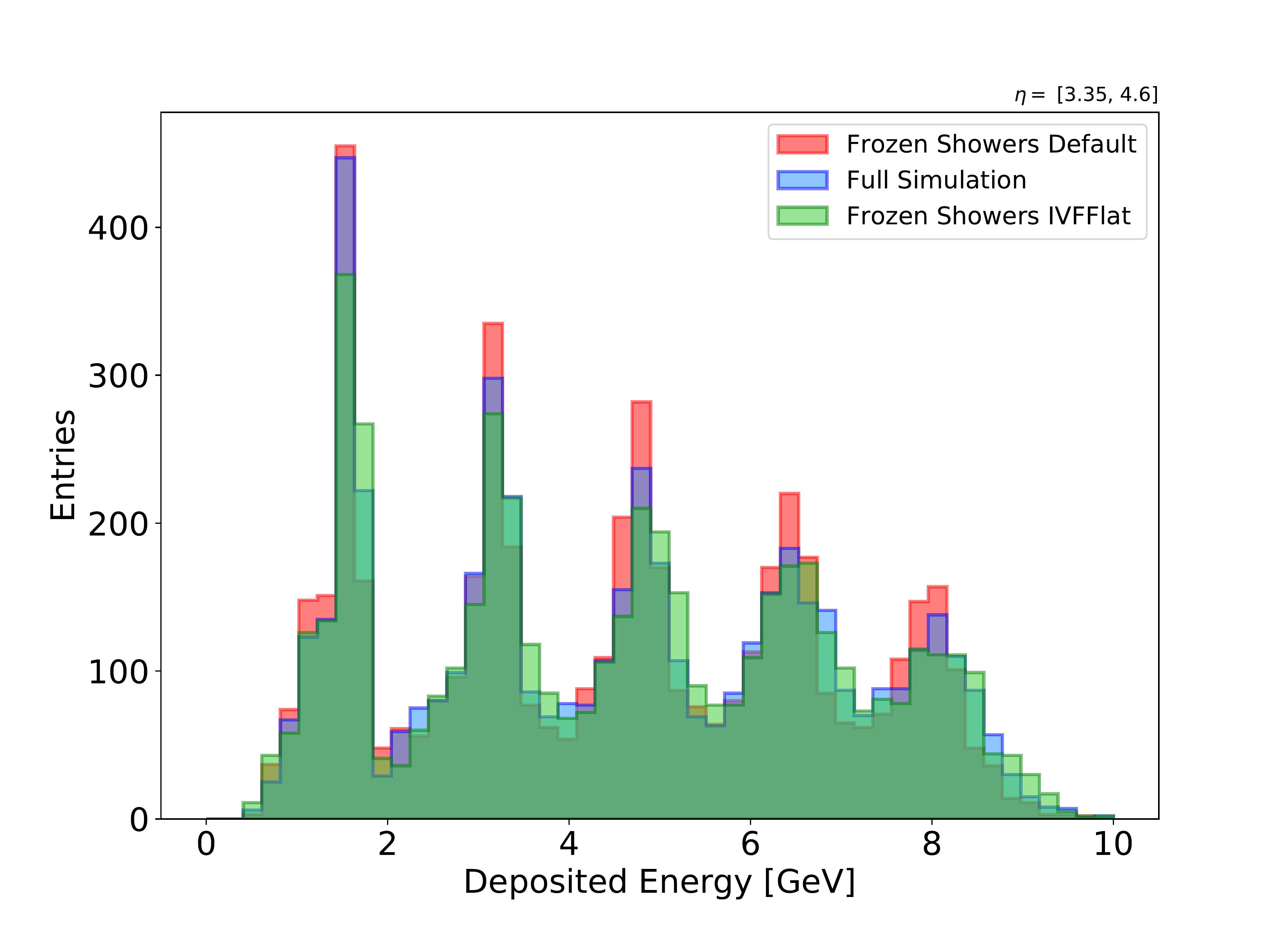}
\caption{Distribution of the energy deposited by the showers originated by the low-energy electrons in the FCal calorimeter.}
\label{fig:DepE}
\end{figure}

\medskip
\smallskip
\noindent
The processing time is evaluated as the average of the CPU time spent executing the propagation of a high-energy electron of fixed initial energy. This time includes the fast simulation of $\sim10^{4}$ low-energy showers produced in the propagation.
The detector resolution is measured as the ratio $\sigma_{E}/{E}$ between the standard deviation and the mean of the distribution of the energy deposited by the low-energy electrons. This distribution is shown in Fig.~\ref{fig:DepE} for the fully \textsc{Geant4}-based simulation (blue), the default library (red), and the new library (green).

\medskip
%\smallskip
\noindent
Different indexing methods have been tested and compared on the basis of CPU time and resolution response. On the frozen shower library, all the tested methods show optimal results in terms of resolution response. The method based on the HNSW structure has been chosen as a benchmark, based on a slightly better performance in terms of CPU requirements.

\begin{figure}
  \centering
  \includegraphics[width=0.7\textwidth]{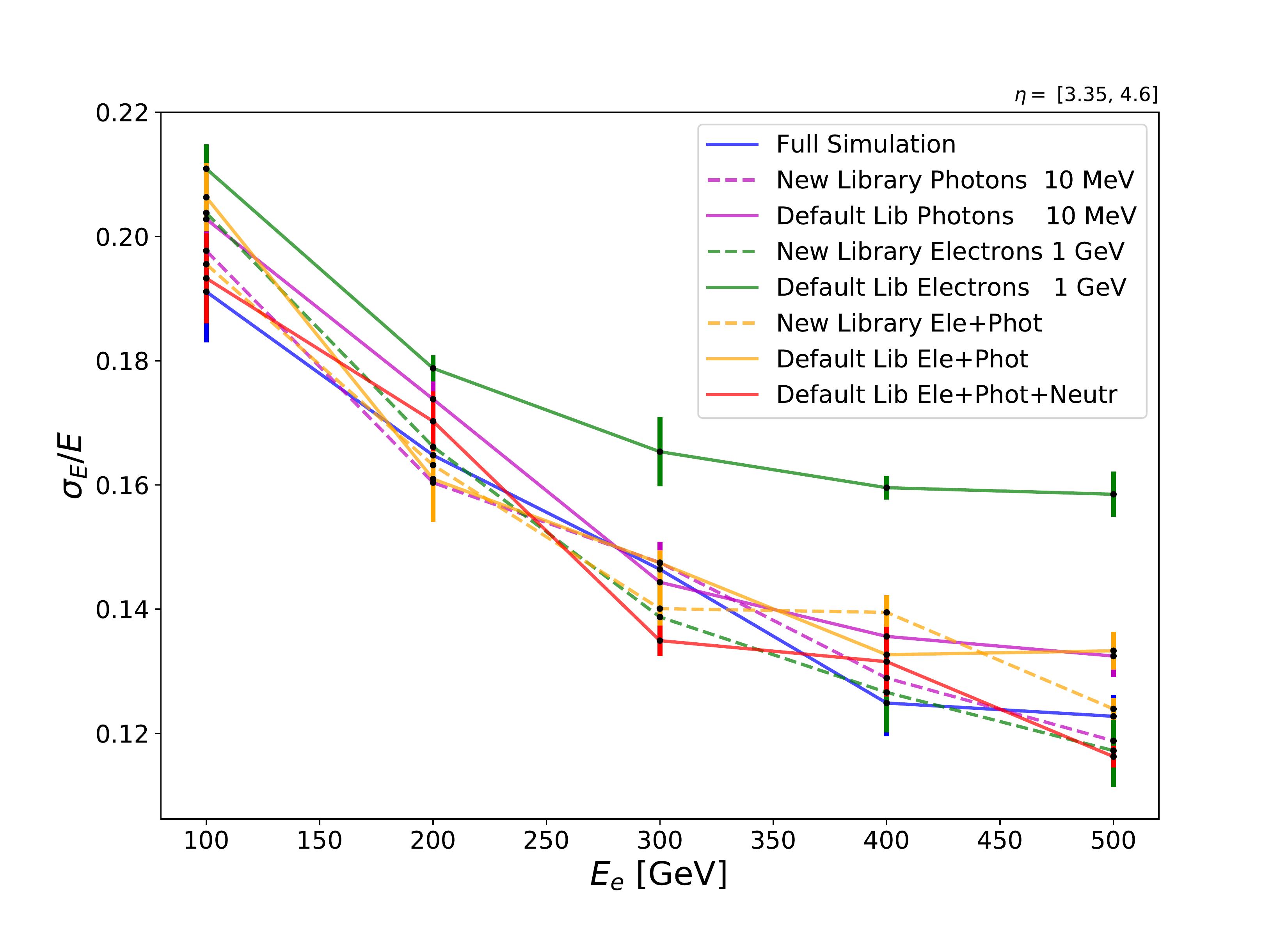}
  \caption{Resolution response obtained with the new (dashed lines) and the old (full lines) libraries.}
  \label{fig:final_result_Res}
\end{figure}

\medskip
\smallskip
\noindent
A specific library has to be generated for all the particles that require a fast simulation. During simulation, the different libraries can either be used individually or in combination. For comparison, a library containing showers originated from photons with an energy threshold of 10~MeV was also produced. Fig.~\ref{fig:final_result_Res} shows the final comparison of the detector resolution response obtained with different versions of the default libraries (continuous curves) and the new libraries (dashed curves). 

\begin{figure}
  \centering
  \includegraphics[width=0.7\textwidth]{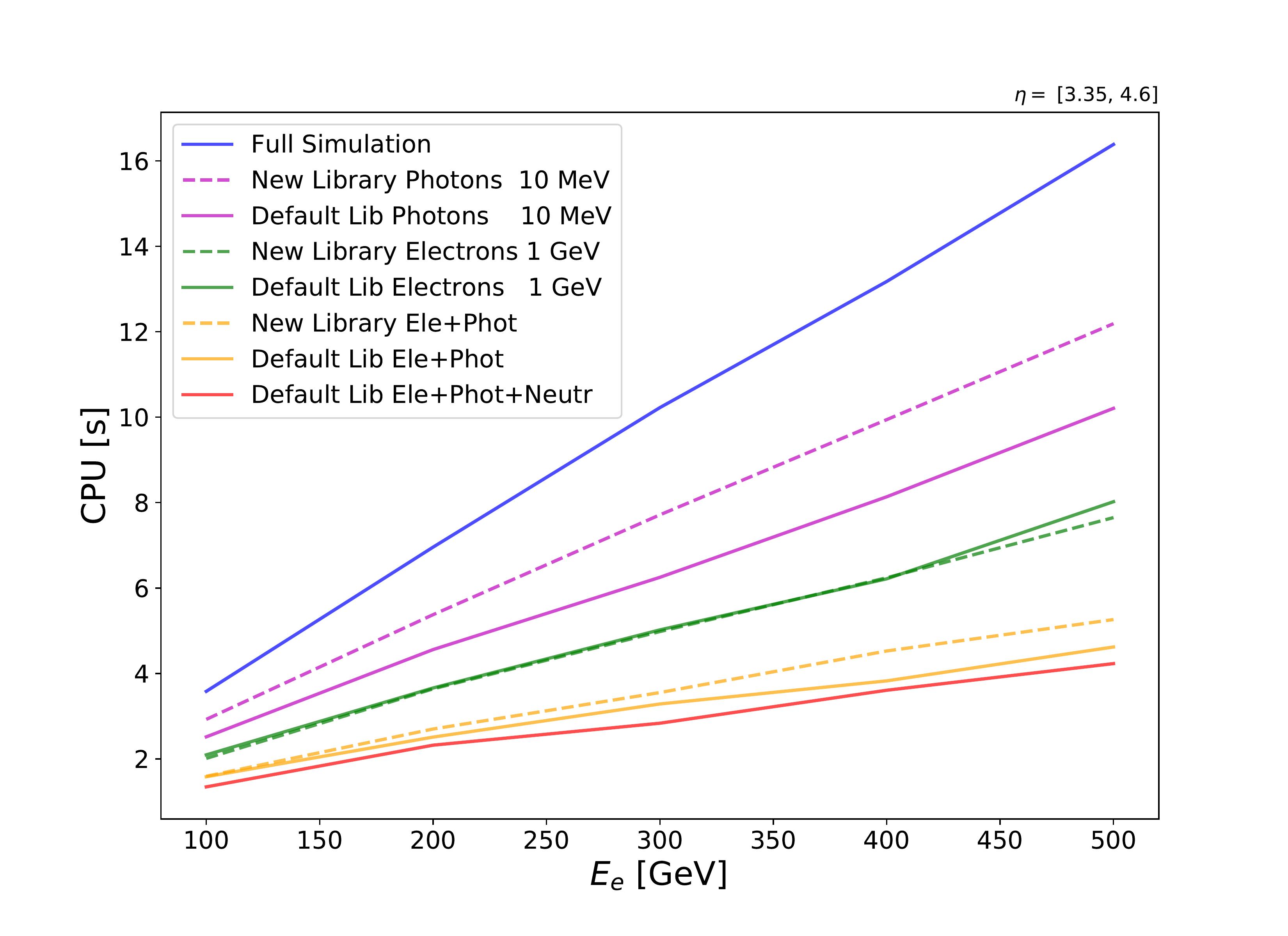}
  \caption{CPU response obtained with the new (dashed lines) and the old (full lines) libraries.}
  \label{fig:final_result_CPU}
\end{figure}

\medskip
%\smallskip
\noindent
The electron and photon libraries have been tested both individually and in combination in the default and new implementations. The resolution obtained with the full simulation (blue) is in general well reproduced, except for the default electron library (green) that presents some discrepancies. Also, for the default library, a previous tuning of the library parameters was needed. The new approach provides optimal results in terms of detector resolution response.
Fig.~\ref{fig:final_result_CPU} shows the final comparison of the CPU response. 

\medskip
%\smallskip
\noindent
The electron libraries (green) provide a significant gain (up to $\sim50\%$) with respect to the full simulation, with comparable results for the two different implementations. The photon library used in the new implementation appears to be less efficient than the default one. In general, the photon library provides a smaller gain to the simulation than the electron library. This is due to the smaller threshold used for the generation of the photon libraries. When the electron and photon libraries are used in combination, the simulation is significantly accelerated (up to $\sim70\%$ ), and again provides results comparable between the two approaches, as shown by the yellow curves. The fact that the different performance obtained with the photon libraries is not reflected in the combined results can be related again to the different thresholds used for the library generation, which makes the electron library much more likely to be used during simulation. 

 \begin{figure}
  \centering
  \includegraphics[width=0.8\textwidth]{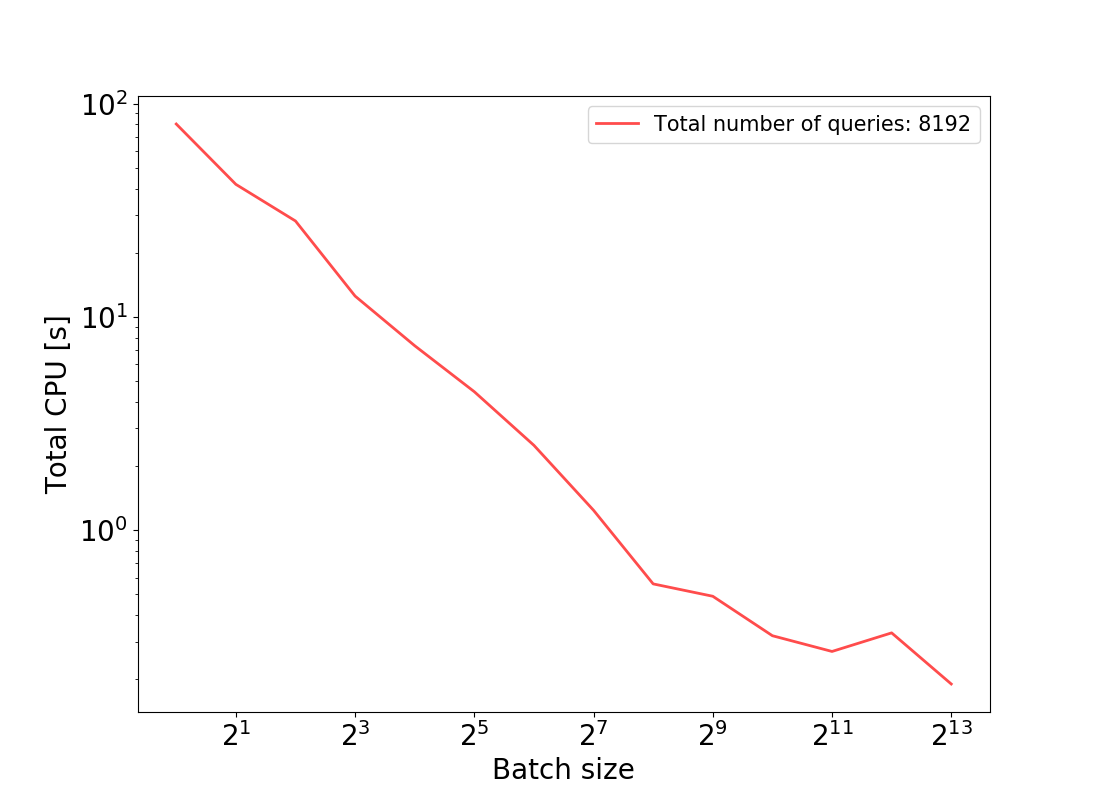}
  \caption{Faiss search in batches: total CPU vs batch size.}
  \label{fig:batch}
\end{figure}

\subsection{Conclusions and prospects} 

\medskip
%\smallskip
\noindent
The proposed approach for a new fast simulation of the ATLAS FCal was fully implemented within the ATLAS simulation framework. The new method based on similarity search techniques provides optimal results in terms of detector resolution response, outperforming the default approach, which needs to be adjusted before usage. A significant gain in CPU requirements with respect to the full simulation, amounting to an improvement of about 70\%, is also obtained as a comparable result to what obtained with the default approach. 
CPU acceleration could be further enhanced with a library sampling performed in batches. When this work was performed, the batching functionality could not be enabled, as a multi-threading processing model was not supported by the ATLAS simulation framework. However, the Faiss implementation is optimized for batch searches. 

\medskip
\smallskip
\noindent
The potential improvement due to batch searches was tested in a standalone mode, using two separate datasets, one as the library, and the other as a collection of query vectors. Fig.~\ref{fig:batch} shows that the CPU time decreases significantly as the batch size increases. This suggests that the fast simulation implemented in ATLAS would greatly benefit from batch searches.

\section{Towards the Full Optimization of Physics Measurements}\label{sec:mode}

\subsection {Introduction} 

\medskip
%\smallskip
\noindent
In the previous sections we have described a number of applications to LHC research of advanced multivariate methods, to which AMVA4NewPhysics provided a significant contribution. The produced examples provide a clear indication of the true revolution that the process of inference extraction from complex collider data withstood in the course of the past decade. In a situation where computer science is moving forward very quickly towards artificial intelligence applications in all domains of science and technology, as well as elsewhere, one is bound to ask what may the next step be in HEP research.

\medskip
\smallskip
\noindent
We offer a partial answer to the above question in this section, where we summarize the result of studies aimed at achieving the full end-to-end optimization of physics measurements based on injecting information on the final analysis goal in the dimensionality reduction procedures applied to collider data through supervised classification.

\subsection {The misalignment problem}

\medskip
%\smallskip
\noindent
The typical workflow of the measurement of a physical quantity from data collected by particle collisions at the LHC involves the construction of an informative summary statistic---usually one-dimensional---from multidimensional high-level features of observed data events, and the extraction of inference on the parameter(s) of interest by a likelihood fit or some other statistical procedure. The high-level features, usually individual four-momenta of identified relevant particles, are themselves the result of a large compression procedure that starts with millions of digital readouts of as many electronic channels; this procedure is generically addressed as event reconstruction. 

\medskip
\smallskip
\noindent
Event reconstruction, of which we have discussed some aspects in the previous Sections (Section~\ref{sec:btagging}, Section~\ref{s:caloshowers}) is a very complex task, and  in ATLAS and CMS the development of the relevant software requires an investment of hundreds of person-years; it involves thorough calibrations, tunings of data collection procedures, modelling of detector response, and a number of related studies. We will not consider these procedures further in this section, other than noting that they necessarily target, as their optimization goal, suitable simplified surrogates of the true goals of the experiment. The complexity of the whole set of procedures that convert a raw dataset of digitized detector readouts into the measurement of a fundamental physical quantity prevent a direct optimization of the ultimate goal of experimentalists, which is usually the minimization of the total uncertainty on the measurement of a set of parameters of interest.  For example, the electromagnetic calorimeter of an LHC experiment is naturally built to improve as much as possible the energy resolution and identification capabilities of energetic, isolated photons (see {\em supra}, Section~\ref{s:caloshowers}). Those figures of merit are manageable optimization proxies for the discovery reach of the experiment to high-mass resonances decaying into pairs of photons, which is much harder to study and maximize directly, but in general they are not monotonous functions of the latter quantity, which is dependent on additional parameters and conditions.

\medskip
\smallskip
\noindent
The above mismatch between desirable goals and optimization proxies constitute what could be called a \emph{misalignment problem}. Realigning our construction and reconstruction strategies to our experimental goals is a topic that will require our attention in the forthcoming years, leveraging differentiable programming tools that already enable the investigation of full end-to-end optimization strategies for particle reconstruction. A few efforts have started to consider the design of experiments as the subject of a full optimization~\cite{MODE,MODENPNI,blackbox,lhcbcalo}, and there is hope that even full-blown particle detectors of the scale of ATLAS or CMS, which are among the most complex instruments ever built by human beings, may one day be assisted in their design by artificial intelligent tools leveraging automatic differentiation. 

\medskip
\smallskip
\noindent
A misalignment problem of smaller scale, yet by itself often one of quite large impact, exists also in the smaller-scale data compression step that starts from the $O(50)$ high-level event features resulting from event reconstruction, and ends in a one-dimensional summary statistic on which experimental physicists base the extraction of information on the quantity (or quantities) of interest. The latter is usually the mass of a particle, or the cross section of an interesting phenomenon, and/or some other fundamental parameter of the underlying theory. Whatever the measurement goal may be, the workflow of the final inference step often involves the use of supervised classification to construct a discriminating variable that effectively separates the wanted signal from all competing backgrounds. The classification task may be performed by training a neural network, as exemplified in Section~\ref{sec:higgsml}, {\em supra}; the network learns from simulated samples of signal and background processes, and produces an effective summary which, in absence of systematic effects, is often close to optimal in performance. In this context, an optimal classifier is one monotonous with the likelihood ratio, which by the Neyman--Pearson lemma is the most powerful test statistic to separate two simple hypotheses~\cite{NPLemma}. The produced summary statistic is optimal in the absence of systematic effects, but it is in general not optimal when model uncertainties or other biases have to be taken into account. The neural network, oblivious of those systematic uncertainties, produces a summary that does not guarantee the optimality of the final inference. In statistical terms, the summary is not {\em sufficient}: it does not make optimal use of the input information in the task of estimating the parameter of interest.

\subsection {INFERNO}

\medskip
%\smallskip
\noindent
Recognizing the misalignment problem typically encountered in data analyses that perform a dimensionality reduction step from high-level features to a neural network classifier output, we designed an algorithm that promises full end-to-end optimality and approximate sufficiency of the constructed summary statistic. The algorithm, called INFERNO~\cite{inferno} (an acronym of {\em Inference-Aware Neural Optimization}), exploits \texttt{TensorFlow} libraries to construct a fully differentiable pipeline that includes a neural network (used for the dimensionality-reduction step) as well as a model of the inference-extraction procedure (a binned likelihood fit to the network outputs). In INFERNO, nuisance parameters affecting the background shape are introduced in the model, and the loss function of the neural network is retro-fitted with knowledge of the size of the full uncertainty expected on the final parameter of interest. The latter results from the inclusion in the model of all statistical as well as systematic effects, and is estimated from the inverse of the Hessian matrix of the likelihood fit employed for the parameter estimation step. Stochastic gradient descent then allows the network to learn a data representation that minimizes the variance of the parameter of interest.

\begin{figure}[ht]
\includegraphics[width=\textwidth]{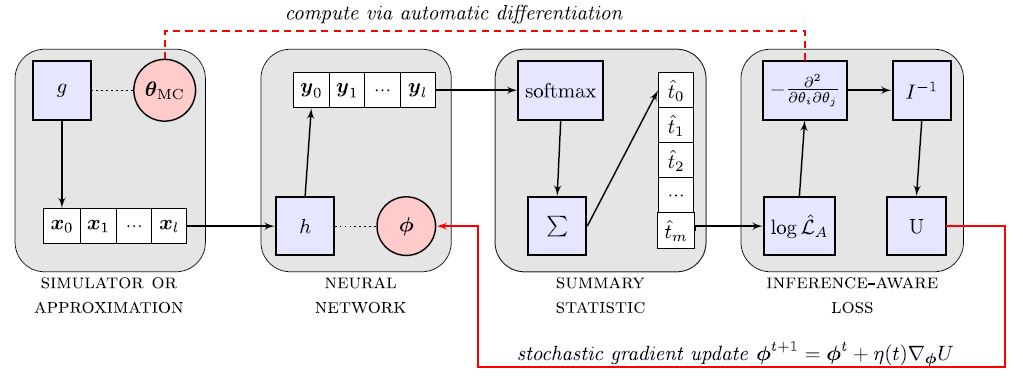}
\caption{\em Pipeline of INFERNO. Data produced by a simulator (left box) is fed to a neural network (second box from left). The network output is processed through a softmax operator to make it differentiable, and constitutes the summary statistic (third box from left) employed by a likelihood maximization (right box) to perform inference on the parameter of interest and related nuisances $\theta$, and obtain an estimate of the associated variance. The variance is fed back into the loop such that the network is capable of optimizing its parameters $\phi$ to produce the smallest uncertainty on the parameter of interest given nuisance parameters~\cite{inferno}. }
\label{f:inferno}
\end{figure}

\medskip
\smallskip
\noindent
The algorithm was originally implemented with TensorFlow 1.0~\cite{tensorflow} and is freely available~\cite{tf1_inferno}; it has also recently implemented within TF2.0~\cite{tf2_inferno} and PyTorch~\cite{pytorch_inferno}. A sketch of the software pipeline which INFERNO is based on is shown in Fig.~\ref{f:inferno}.

\subsection {Outlook}

\medskip
%\smallskip
\noindent
INFERNO is innovative and powerful, but its application is not always straightforward as it expects the user to produce a suitable differentiable model of the effect of nuisance parameters; so far the algorithm has only been proven to work---quite effectively---when applied to synthetic datasets. Its application to a full-blown LHC data analysis is in progress at the time of writing; we fully expect that the solution of the misalignment problem discussed {\em supra}, provided by INFERNO, will yield much higher sensitivity to physics analyses. Following INFERNO, a few similar attempts have recently been published, see {\em e.g.}~\cite{wunsch}. We direct readers interested in more detail on this area of research to a recent review of methods that address the incorporation of nuisance parameters in supervised classification for HEP research~\cite{dorigo-decastro}.

\medskip
\medskip
%\clearpage
\section{Concluding Remarks}\label{sec:summary}

\medskip
%\smallskip
\noindent
The present document summarizes the most significant studies carried out by members of the AMVA4NewPhysics ITN in the years 2015--2019 within the context of searches and measurements of the Higgs boson and of LHC searches for new physics phenomena in datasets collected during Run 1 and Run 2 by the ATLAS and CMS experiments at the CERN LHC. The common denominator of the reported studies is the attempt to go beyond current data analysis practice in HEP experiments, exploiting state-of-the-art techniques from computer science and developing entirely new methods to extract the maximum amount of information from the invaluable datasets of proton--proton collisions.

\medskip
\smallskip
\noindent
The effort that brought these studies to the fore could not have been undertaken without the resources made available by the European Union, as highlighted {\em infra}. However, the breadth of results and the significant stride they provide in the direction of more effective data analysis in fundamental physics research would have been impossible without the cohesive motivation of the young participants of the AMVA4NewPhysics training network. For that reason, it is necessary to stress here that an additional untold success of the network's action has been to provide the spark for that collaborative effort. 

\medskip
\medskip
%\clearpage
\section*{Acknowledgements}

\medskip
%\smallskip
\noindent
This work is part of a project that has received funding from the European Union’s Horizon 2020 research and innovation programme under grant agreement $\textrm{N}^o\,\, 675440$.

\medskip
\medskip
\medskip

\begin{figure}[ht!]
\begin{center}
\includegraphics [width=8cm]{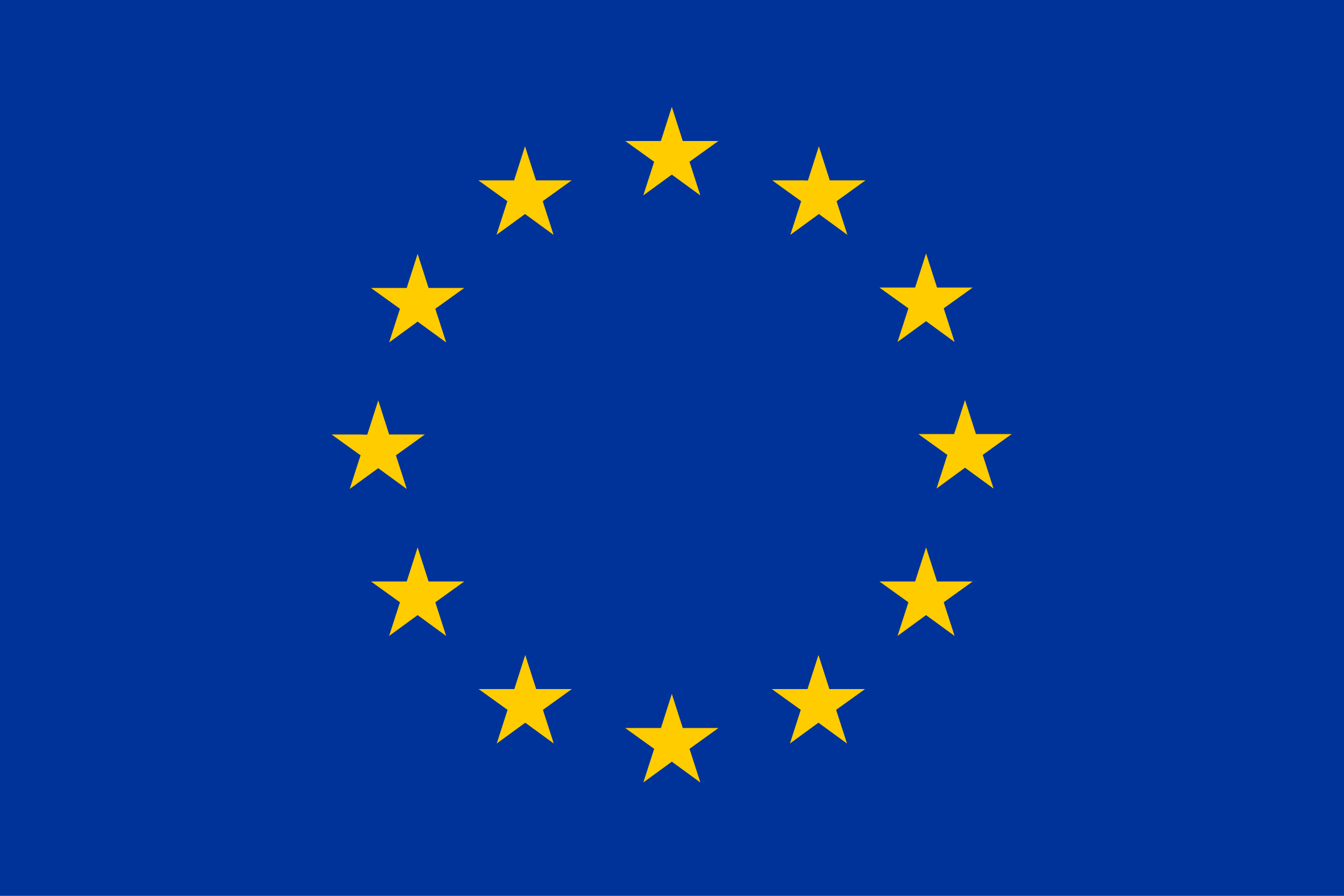}
\end{center}
\end{figure}

%\clearpage
%%%%%%%%%%%%%%%%%%%%%%%%%%%%%%%%%%%%%%%%%%%%%%%%%%%%%%%%%%%%%%%

\section*{ORCID iDs}

\medskip
\medskip

\begin{table}[ht]
    \small 
    \begin{tabular}{@{}l l@{}}
        Anna Stakia                 &\href{https://orcid.org/0000-0001-6277-7171}{0000-0001-6277-7171}\\ \cline{1-2}
        Tommaso Dorigo              &\href{https://orcid.org/0000-0002-1659-8727}{0000-0002-1659-8727}\\ \cline{1-2}
        Giovanni Banelli            &\href{https://orcid.org/0000-0001-5490-3430}{0000-0001-5490-3430}\\ \cline{1-2}
        Daniela Bortoletto          &\href{https://orcid.org/0000-0002-1287-4712}{0000-0002-1287-4712}\\ \cline{1-2}
        Alessandro Casa             &\href{https://orcid.org/0000-0002-2929-3850}{0000-0002-2929-3850}\\ \cline{1-2}
        Pablo de Castro             &\href{https://orcid.org/0000-0002-4828-6568}{0000-0002-4828-6568}\\ \cline{1-2}
        Christophe Delaere          &\href{https://orcid.org/0000-0001-8707-6021}{0000-0001-8707-6021}\\ \cline{1-2}
        Julien Donini               &\href{https://orcid.org/0000-0002-8998-0839}{0000-0002-8998-0839}\\ \cline{1-2}
        Livio Finos                 &\href{https://orcid.org/0000-0003-3181-8078}{0000-0003-3181-8078}\\ \cline{1-2}
        Michele Gallinaro           &\href{https://orcid.org/0000-0003-1261-2277}{0000-0003-1261-2277}\\ \cline{1-2}
        Andrea Giammanco            &\href{https://orcid.org/0000-0001-9640-8294}{0000-0001-9640-8294}\\ \cline{1-2}
        Alexander Held              &\href{https://orcid.org/0000-0002-8924-5885}{0000-0002-8924-5885}\\ \cline{1-2}
        Fabricio Jiménez Morales    &\href{https://orcid.org/0000-0002-8401-8501}{0000-0002-8401-8501}\\ \cline{1-2}
        Grzegorz Kotkowski          &\href{https://orcid.org/0000-0003-4779-6766}{0000-0003-4779-6766}\\ \cline{1-2}
        Seng Pei Liew               &\href{https://orcid.org/0000-0003-2419-2505}{0000-0003-2419-2505}\\ \cline{1-2}
        Fabio Maltoni               &\href{https://orcid.org/0000-0003-4890-0676}{0000-0003-4890-0676}\\ \cline{1-2}
        Giovanna Menardi            &\href{https://orcid.org/0000-0003-0429-3034}{0000-0003-0429-3034}\\ \cline{1-2}
        Ioanna Papavergou           &\href{https://orcid.org/0000-0002-7992-2686}{0000-0002-7992-2686}\\ \cline{1-2}
        Alessia Saggio              &\href{https://orcid.org/0000-0002-7385-3317}{0000-0002-7385-3317}\\ \cline{1-2}
        Bruno Scarpa                &\href{https://orcid.org/0000-0002-9628-5164}{0000-0002-9628-5164}\\ \cline{1-2}
        Giles Chatham Strong        &\href{https://orcid.org/0000-0002-4640-6108}{0000-0002-4640-6108}\\ \cline{1-2}
        Cecilia Tosciri             &\href{https://orcid.org/0000-0001-6485-2227}{0000-0001-6485-2227}\\ \cline{1-2}
        Jo\~{a}o Varela             &\href{https://orcid.org/0000-0003-2613-3146}{0000-0003-2613-3146}\\ \cline{1-2}
        Pietro Vischia              &\href{https://orcid.org/0000-0002-7088-8557}{0000-0002-7088-8557}\\ \cline{1-2}
        Andreas Weiler              &\href{https://orcid.org/0000-0002-5605-8258}{0000-0002-5605-8258}\\ \cline{1-2}
    \end{tabular}
\end{table}
\clearpage

%%%%%%%%%%%%%%%%%%
%\section*{References}
\bibliography{MAIN.bib}
% \input{main.bbl} %<-- works here, not on elsevier interface
%\bibliography{BIBS.bib}

\end{document}